\documentclass[twoside,a4paper]{amsart}

\usepackage{xcolor}
\definecolor{webgreen}{rgb}{0,.5,0}
\definecolor{webbrown}{rgb}{.6,0,0} 
\definecolor{RoyalBlue}{cmyk}{1, 0.50, 0, 0}
\usepackage[colorlinks=true, breaklinks=true, urlcolor=webbrown, linkcolor=RoyalBlue, citecolor=webgreen,  backref=page]{hyperref}
\usepackage{epsfig, graphicx, subfigure}
\usepackage{verbatim, setspace,pifont}
\usepackage{amsmath, amssymb}

\newcommand{\R}{{\mathbb R}}
\newcommand{\N}{{\mathbb N}}
\newcommand{\C}{{\mathbb C}}

\newcommand{\be}{\beta}

\newcommand{\ga}{\gamma}
\newcommand{\Ga}{\Gamma}

\newcommand{\la}{\lambda}
\newcommand{\ep}{\varepsilon}

\newcommand{\re}{{\mathsf{Re}}\,}
\newcommand{\im}{{\mathsf{Im}}\,}

\renewcommand{\deg}{\mathsf{deg}}
\renewcommand{\arg}{\mathsf{arg}}

\newcommand{\rhy}   {\textnormal{RHP}-${\boldsymbol Y}$}
\newcommand{\rht}   {\textnormal{RHP}-${\boldsymbol T}$}
\newcommand{\rhs}   {\textnormal{RHP}-${\boldsymbol S}$}
\newcommand{\rhr}   {\textnormal{RHP}-${\boldsymbol R}$}
\newcommand{\rhm}   {\textnormal{RHP}-${\boldsymbol M}$}
\newcommand{\rhp}   {\textnormal{RHP}-${\boldsymbol P}_e$}

\newcommand{\rhpc}   {\textnormal{RHP}-${\boldsymbol P}_c$}

\newcommand{\rhpsi}   {\textnormal{RHP}-${\boldsymbol \Psi}$}

\newtheorem{lemma}{Lemma}[section]
\newtheorem{theorem}[lemma]{Theorem}
\newtheorem{definition}[lemma]{Definition}
\newtheorem{proposition}[lemma]{Proposition}
\newtheorem{corollary}[lemma]{Corollary}
\newtheorem{remark}[lemma]{Remark}

\numberwithin{equation}{section}

\begin{document}

\title{Topological Expansion in the Complex Cubic Log-Gas Model. One-Cut Case}

\author{Pavel  Bleher}
\address{Department of Mathematical Sciences, Indiana University-Purdue University Indianapolis, 402~North Blackford Street, Indianapolis, IN 46202, USA}
\email{pbleher@iupui.edu}

\author{Alfredo Dea\~no}
\address{School of Mathematics, Statistics and Actuarial Science,  University of Kent,  Canterbury, Kent, CT2 7NF, United Kingdom}
\email{A.Deano-Cabrera@kent.ac.uk}

\author{Maxim Yattselev}
\address{Department of Mathematical Sciences, Indiana University-Purdue University Indianapolis, 402~North Blackford Street, Indianapolis, IN 46202, USA}
\email{maxyatts@iupui.edu}

\dedicatory{Dedicated to David Ruelle and Yakov Sinai}

\begin{abstract}
We prove the topological expansion for the cubic log-gas partition function
\[
Z_N(t)= \int_\Gamma\cdots\int_\Gamma\prod_{1\leq j<k\leq N}(z_j-z_k)^2
\prod_{k=1}^Ne^{-N\left(-\frac{z^3}{3}+tz\right)}\mathrm dz_1\cdots \mathrm dz_N,
\]
where $t$ is a complex parameter and $\Gamma$ is an unbounded contour on the complex plane extending 
from $e^{\pi \mathrm i}\infty$ to $e^{\pi \mathrm i/3}\infty$. 
The complex cubic log-gas model exhibits two phase regions on the complex $t$-plane, with one cut and two cuts, separated by analytic critical arcs of the two types
of phase transition: split of a cut and birth of a cut. The common point of the critical arcs is a tricritical point of
the Painlev\'e I type. In the present paper we prove the topological expansion for $\log Z_N(t)$ in the one-cut phase region. The proof is based on the Riemann--Hilbert approach to semiclassical asymptotic expansions for the associated orthogonal polynomials and the theory of
$S$-curves and quadratic differentials. \\
\end{abstract}

\keywords{Log-gas model, partition function, topological expansion, equilibrium measure, S-curve, quadratic differential, orthogonal polynomials, non-Hermitian orthogonality, Riemann--Hilbert problem, nonlinear steepest descent method.}
\subjclass[2010]{33C47, 30E15, 31A25,  15B52} 

\thanks{The work of the first author (P.B.) is supported in part by the National Science Foundation (NSF) Grants DMS-1265172 and DMS-1565602. P.B. also gratefully acknowledges support from the Simons Center for Geometry and Physics, Stony Brook University, at which some of the research for this paper was performed. The second author (A.D.) acknowledges financial support from projects MTM2012-36732-C03-01 and MTM2012-34787 from the Spanish Ministry of Economy and Competitivity. The research of the third author (M.Y.) is supported by a grant from the Simons Foundation, CGM-354538.}

\maketitle

\setcounter{tocdepth}{3}

\section{Introduction}\label{s:intro}

The main goal of this work is to analyze the topological expansion in the cubic log-gas model with a general complex coupling constant
and semiclassical asymptotics of related orthogonal polynomials. 
The partition function of the cubic log-gas model is given as
\begin{equation}\label{ZN}
Z_N(u)=\int_\Ga\ldots\int_\Ga \prod_{1\leq j<k\leq N}(z_j-z_k)^2\,
\prod_{j=1}^N e^{-N \left(\frac{z_j^2}{2}-uz_j^3\right)}\mathrm dz_1\ldots \mathrm dz_N,
\end{equation}
where $u>0$ is a coupling constant and the contour of integration $\Ga$ goes from $e^{\pi \mathrm i}\infty$ to $e^{\pi \mathrm i/3}\infty$.
This work is a continuation of the works of Bleher and Dea\~no \cite{BD1,BD2}.
\begin{center}
\begin{figure}[h]\label{Gamma_t1}
\begin{center}
\scalebox{0.35}{\includegraphics{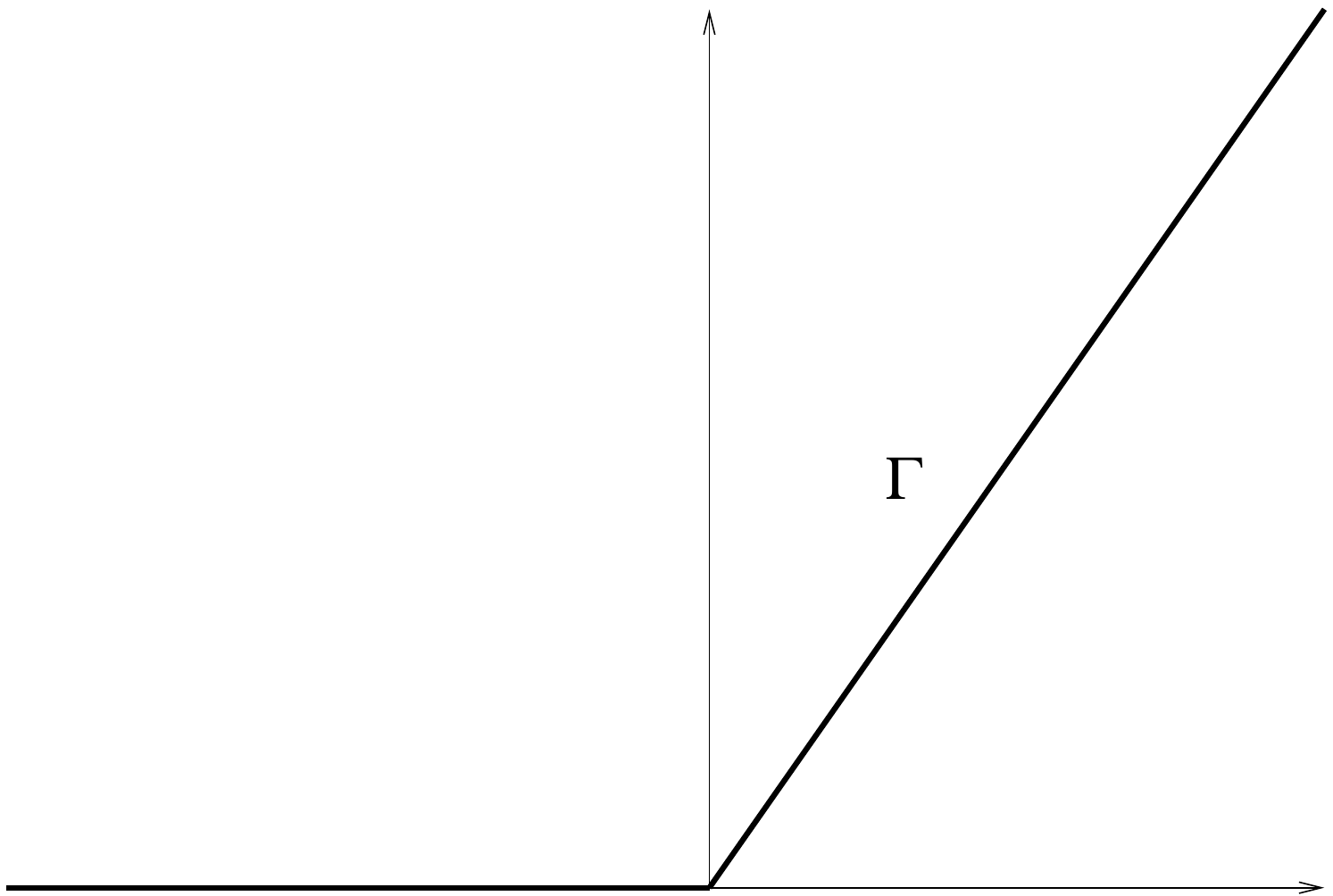}}
\end{center}
  \caption[sectors ]{The  contour  $\Ga$ of integration.}
 \end{figure}
\label{Gammat1}
\end{center}
As proven in  \cite{BD1}, the free energy of the cubic log-gas model,
\begin{equation}\label{free}
F_N(u):=\frac{1}{N^2}\ln \frac{Z_N(u)}{Z_N(0)}\,,
\end{equation}
admits an asymptotic expansion as $N\to\infty$ in powers of $\frac{1}{N^2}\,$,
\begin{equation}\label{top1}
F_N(u)\sim \sum_{g=0}^{\infty}\frac{ F^{(2g)}(u)}{N^{2g}},
\end{equation}
for any $u$ in the interval $0\le u< u_c$, where
\begin{equation}\label{uc}
u_c=\frac{3^{1/4}}{18}
\end{equation}
is a critical point.
In addition, the functions $F^{(2g)}(u)$ admit an analytic continuation to the disk $|u|<u_c$ on the complex plane, 
and if we expand them in powers of $u$,
\begin{equation}\label{top2}
F^{(2g)}(u)=\sum_{j=1}^\infty \frac{f^{(2g)}_{2j}u^{2j}}{(2j)!},
\end{equation}
then the coefficient $f^{(2g)}_{2j}$ is a positive integer number that counts the number 
of $3$-valent connected graphs with $2j$ vertices on a Riemann surface of genus $g$. Asymptotic expansion 
\eqref{top1} is called the {\it topological expansion}. For more details on this aspect of the theory, 
we refer the reader to the classical papers of Bessis, Itzykson and Zuber \cite{BIZ}, 
Br\'{e}zin, Itzykson, Parisi and Zuber \cite{BIPZ}, the monograph of Forrester \cite[Section 1.6]{Forrester}, the works of 
Mulase \cite{Mulase}, Di Francesco \cite{DiFrancesco}, Ercolani and McLaughlin \cite{EMcL,EMcLP}, and references therein, or the very readable introduction by Zvonkin \cite{Zvonkin}.

As shown in \cite{BD1}, the coefficients $f^{(2g)}_{2j}/(2j)!$ of power series \eqref{top2}
behave, when $j\to\infty$, as
\begin{equation}\label{gg1}
\begin{aligned}
\frac{f^{(2g)}_{2j}}{(2j)!} &=\frac{K_{2g}  j^{\frac{5g-7}{2}}}{u_c^{2j}}\left(1+\mathcal{O}(j^{-1/2})\right),
\quad K_{2g}>0.
\end{aligned}
\end{equation}
This implies that  $u_c$ is the radius of convergence of power series \eqref{top2}. In fact, $u=u_c$ is a singular point
of the functions $\eqref{top2}$. The topological expansion in a neighborhood of the critical point
$u_c$ has been obtain in the work of Bleher and Dea\~no \cite{BD2}. This topological expansion is
closely related to the Painlev\'e I equation. The relation to the Painlev\'e I equation can be
already seen in asymptotic formula \eqref{gg1}. Namely, if we rescale the coefficients $K_{2g}$ in 
\eqref{gg1}, by introducing the coefficients
\begin{equation}\label{gg2}
C_{2g}=\frac{\Gamma\left(\frac{5g-1}{2}\right)u_c^{g}K_{2g}}{6\cdot 3^{1/4}}\,,
\end{equation}
and consider the following generating function:
\begin{equation}\label{genfun}
y(t)=\sum_{g=0}^{\infty} C_{2g}t^{\frac{1-5g}{2}}\,,
\end{equation}
then $y(t)$ solves the Painlev\'e I differential equation,
\begin{equation}\label{ode}
y''(t)=a_0y^2(t)-a_1t,
\end{equation}
with $a_0=2^{\frac{5}{2}}3^{\frac{9}{4}}\,$, $a_1=2^{\frac{3}{2}}3^{-\frac{5}{4}}\,$ (see  \cite{BE,Eyn,BD1}).
  
It is noteworthy that the key ingredient in the proof of topological expansion \eqref{top1} in \cite{BD1} is  the derivation of semiclassical asymptotic formulae for the recurrence coefficients $\ga_n^2$, $\be_n$
of the corresponding monic orthogonal polynomials $P_n(z)=z^n+\ldots$. The orthogonality condition is stated on the contour $\Ga$:
\begin{equation}\label{op_1}
\int_{\Gamma} P_n(z)z^k w(z)\mathrm dz=0, \quad k=0,1,\ldots,n-1;\qquad w(z)=e^{-N\left(\frac{z^2}{2}+uz^3\right)} 
\end{equation}

Namely, as proven in \cite{BD1},
for any $u$ such that $0\le u<u_c$, there exists $\ep>0$ such that as $N,n\to \infty$ with $1-\ep\le \frac{n}{N}\le 1+\ep$, 
the recurrence coefficients $\ga_n^2$ and $\be_n$ admit the asymptotic expansions in powers of $\frac{1}{N^2}$:
\begin{equation}\label{asympgb}
\left\{
\begin{aligned}
\gamma_n^2 & \sim \sum_{k=0}^{\infty}\frac{1}{N^{2k}}g_{2k}\left(\frac{n}{N},u\right),\\
\beta_n & \sim \sum_{k=0}^{\infty}\frac{1}{N^{2k}}
b_{2k}\left(\frac{n}{N}+\frac{1}{2N},u\right),
\end{aligned}
\right.
\end{equation}
where the functions $g_{2k}(s,u)$, $b_{2k}(s,u)$, $k=0,1,\ldots$,  do not
depend on $n$ and $N$ and are analytic in $s$ at $s=1$.

In the paper \cite{BD2} this asymptotic expansion is extended to the double scaling asymptotic expansion 
of the recurrence coefficients at the critical point $u_c$.
In the double scaling regime we set 
\begin{equation}\label{nNv}
\frac{n}{N}=1+vN^{-4/5}.
\end{equation}
where $v\in\mathbb{R}$ is a scaling variable. Then as proven in \cite{BD2}, at $u=u_c$ the recurrence coefficients $\ga_n^2$ and $\be_n$ admit the asymptotic expansions in powers of $N^{-2/5}$ as $N\to\infty$:
\begin{equation} \label{dsc}
\left\{
\begin{aligned}
\ga_{n}^2&\sim \ga^2_c+\sum_{k=1}^{\infty}\frac{1}{N^{2k/5}}\, {p}_{2k}(v),\\
\be_{n}& \sim \be_c+\sum_{k=1}^{\infty}\frac{1}{N^{2k/5}}\,{q}_{2k}(\tilde{v}),
\end{aligned}
\right.
\end{equation}
where the functions ${p}_{2k}(v)$, ${q}_{2k}(\tilde{v})$ are expressed in terms of the Boutroux tritronqu\'ee solution
to Painlev\'e I and $\tilde v=v+\frac{N^{-1/5}}{2}\,$. As shown in \cite{BD2}, expansions \eqref{asympgb}
and \eqref{dsc} can be extended for large $N$ to $u$ in overlapping intervals, $[0,u_c-N^{-0.79}]$ for \eqref{asympgb}
and $[u_c-N^{-0.65},u_c]$ for \eqref{dsc}, and this can be used to obtain the double scaling asymptotic formula for
the partition function. 

Namely, let $u-u_c=C\la N^{-\frac{4}{5}}$, where $C=2^{-\frac{12}{5}}3^{-\frac{7}{4}}$ and
$\la$ is a complex scaling variable in the double scaling regime. Then for $\la$ outside of a neighborhood
of the poles of the Boutroux tritronqu\'ee solution to Painlev\'e I $y(\la)$,
the partition function $Z_N(u)$ can be written as
\begin{equation}\label{ZNu}
Z_N(u)=Z_N^{\rm reg}(u)Z_N^{\rm sing}(\lambda)
\left(1+\mathcal{O}(N^{-\varepsilon})\right),\quad \varepsilon>0,
\end{equation}
where the regular factor is
\begin{equation}\label{ZNreg}
Z_N^{\rm reg}(u)=e^{N^2 [a+b(u-u_c)+c(u-u_c)^2]+d},
\end{equation}
with some explicit constants $a,b,c,d$, and the singular factor is
\begin{equation}\label{ZNsing}
Z_N^{\rm sing}(\lambda)=e^{-Y(\lambda)},
\end{equation}
where $Y(\lambda)$ is a solution of the differential equation
\begin{equation}\label{Yy_ODE}
Y''(\lambda)=y(\lambda),
\end{equation}
with  the boundary condition
\begin{equation}\label{asymp_Y}
Y(\lambda)=\frac{2\sqrt{6}}{45}(-\lambda)^{5/2}-\frac{1}{48}\log(-\lambda)+\mathcal{O}((-\lambda)^{-5/2}), \qquad \lambda\to-\infty.
\end{equation} 
Asymptotic formula \eqref{ZNu}
is used in \cite{BD2} to prove the conjecture of David \cite{Dav1,Dav2} that the poles of $y(\la)$
give rise to zeros of $Z_N(u)$.

This work is a continuation of  \cite{BD1,BD2}. The main goal of it is to investigate the topological expansion
of the cubic log-gas model for complex values of $u$. Formula \eqref{ZN} is not very convenient for this purpose 
because the contour of integration $\Ga$ should be rotated to secure the convergence of the integral. Instead, let us
make the change of variables in \eqref{ZN},
\begin{equation}\label{wz}
z_j=(3u)^{-1/3}\zeta_j+\frac{1}{6u}\,,
\end{equation}
where we assume that $u>0$ and $(3u)^{-1/3}>0$. Then 
\begin{equation}\label{wz1}
\frac{z_j^2}{2}-uz_j^3-\frac{1}{108u^2}=-\frac{\zeta_j^3}{3}+t\zeta_j,
\end{equation}
where
\begin{equation}\label{tu}
t=\frac{1}{4(3u)^{4/3}}\,,
\end{equation}
and with the help of the Cauchy theorem, formula \eqref{ZN} can be reduced to
\begin{equation}
\label{cm0}
Z_N(u) =C_N \int_\Gamma\cdots\int_\Gamma\prod_{1\leq j<k\leq N}(\zeta_j-\zeta_k)^2
\prod_{k=1}^Ne^{-N\left( -\frac{\zeta_k^3}{3}+t\zeta_k\right)}\mathrm d\zeta_1\cdots \mathrm d\zeta_N,
\end{equation}
where $C_N$ is an explicit constant. Now the integral converges for any complex $t$.

The primary interest of the present study is the asymptotic analysis of the partition function
\begin{equation}
\label{cm1}
Z_N(t) := \int_\Gamma\cdots\int_\Gamma\prod_{1\leq j<k\leq N}(z_j-z_k)^2\prod_{k=1}^Ne^{-NV(z;t)}
\mathrm dz_1\cdots \mathrm dz_N,
\end{equation}
with respect to the  parameter $t\in \C$ for the case when
\begin{equation}\label{cm2}
V(z;t) = -\frac{z^3}{3}+tz, \quad t\in\C,
\end{equation}
where $\Ga$ is an unbounded smooth contour such that for any parametrization $z(s)$, $s\in\R$, of $\Ga$ there exists $\epsilon\in(0,\pi/6)$ and $s_0>0$ for which
\begin{equation}
\label{cm2b}
\left\{
\begin{array}{ll}
|\arg(z(s))-\pi/3|\leq \pi/6-\epsilon, & s\geq s_0, \medskip \\
|\arg(z(s))-\pi|\leq \pi/6-\epsilon, & s\leq -s_0,
\end{array}
\right.
\end{equation}
where $\arg(z(s))\in[0,2\pi)$. The above conditions ensure that the partition function $Z_N(t)$ is finite and due to analyticity of the integrand does not depend on a particular $\Ga$ satisfying \eqref{cm2b}. Hence, we shall denote by $\mathcal T$ the collection of all such contours.

We analyze the partition function via the corresponding monic orthogonal polynomials 
\begin{equation}
\label{cm3}
\int_{\Ga} z^kP_n(z;t,N)e^{-NV(z;t)} \mathrm dz=0,\quad k\in\{0,\ldots, n-1\}.
\end{equation} 
Due to the non-Hermitian character of the above relations, it might happen that polynomial satisfying \eqref{cm3} is non-unique. In this case we understand by $P_n(z;t,N)$ the monic polynomial of the smallest degree (such a polynomial is always unique). One way of connecting $Z_N(t)$ to $P_n(z;t,N)$ is via {\it three term recurrence relation}. More precisely, it is known that
\begin{equation}
\label{cm4}
zP_n(z;t,N) = P_{n+1}(z;t,N)+\be_n(t,N) P_n(z;t,N)+\ga_n^2(t,N) P_{n-1}(z;t,N),
\end{equation}
granted all the polynomials in \eqref{cm4} have prescribed degrees, where
\begin{equation}
\label{cm5}
\left\{
\begin{array}{lll}
\ga_n^2(t,N) & = & h_n(t,N)/h_{n-1}(t,N), \medskip \\
h_n(t,N) & = & \displaystyle \int_{\Ga} P_n^2(z;t,N)e^{-NV(z;t)} \mathrm dz.
\end{array}
\right.
\end{equation} 
Observe that if $P_n(z;t,N)=P_{n+1}(z;t,N)$ with both polynomials having degree $n$, then $h_n(t,N)=0$ and $h_{n+1}(t,N)=\infty$. More generally, it holds that $h_n(t,N)$ is a meromorphic function of $t$ and so is $\gamma_n^2(t,N)$. It is further known that the recurrence coefficients $\ga_N^2(t,N)$ satisfy the {\it Toda equation},
\begin{equation}
\label{cm6}
\frac{\partial^2 F_N(t)}{\partial t^2}=\ga_N^2(t,N), \qquad F_N(t)=\frac{1}{N^2}\log Z_N(t).
\end{equation}
Another way of connecting $Z_N(t)$ to orthogonal polynomials is through the formula
\[
Z_N(t) = N!\prod_{n=0}^{N-1}h_n(t,N),
\]
where $h_n(t,N)$ are given in \eqref{cm5}. However, we shall not elaborate on this approach.

The structure of the paper is as follows:
\begin{itemize}
\item In Sections \ref{s:Sprop} and \ref{s:Structure} we describe equilibrium measures and corresponding $S$-curves for the cubic model under consideration. This leads
us to a precise description of the phase diagram of the cubic model on the complex $t$-plane. 
\item In Section \ref{s:main} we present the main results of the paper: the topological expansion in the one-cut phase region and the
asymptotic expansion of the orthogonal polynomials and their recurrence coefficients.
\item In Section \ref{s:pr-geom} we obtain various results about the detailed structure of the $S$-curves and critical graphs of the 
quadratic differential.
\item In Section \ref{s:g} we evaluate the $g$-function and its asymptotic behavior at singular points.
\item In Section \ref{s:aa} we apply the Riemann--Hilbert approach to derive the asymptotic behavior of the 
orthogonal polynomials and their recurrence coefficients. 
\item And finally, in Section \ref{s:ae} we prove the topological expansion in the one-cut phase region. 
\end{itemize}

\section{Equilibrium Measures and S-Property}\label{s:Sprop}

It is well understood that the zeros of polynomials satisfying \eqref{cm3} asymptotically distribute as a certain \emph{weighted equilibrium measure} on an \emph{S-contour} corresponding to the weight function \eqref{cm2}. In this section we discuss these notions in greater detail. Our consideration will use the recent works of Huybrechs, Kuijlaars, and Lejon \cite{HKL} and Kuijlaars and Silva~\cite{KS}. Let us start with some
definitions.

\begin{definition} Let $V$ be an entire function. The logarithmic energy in the external field $\re V$ of a measure $\nu$ in the complex plane
is defined as
\[
E_V(\nu)=\iint \log \frac{1}{|s-t|}\mathrm d\nu(s) \mathrm d\nu(t)+\int \re V(s)\mathrm d\nu(s).
\]
The equilibrium energy of a contour $\Ga$ in the external field $\re V$ is equal to
\begin{equation}
\label{em1}
\mathcal E_V(\Ga)=\inf_{\nu\in \mathcal M(\Ga)} E_V(\nu),
\end{equation}
where $\mathcal M(\Ga)$ denotes the space of Borel probability measures on $\Ga$.
\end{definition}

When $\re V(s)-\log|s|\to+\infty$ as $\Ga\ni s\to\infty$, there exists a unique minimizing measure for \eqref{em1}, which is called the {\it weighted equilibrium measure} of $\Ga$, say $\mu_\Ga$, in the external field $\re V$, see \cite[Theorem I.1.3]{SaffTotik} or \cite{HKL}. We shall use this definition in the case of the cubic polynomial \eqref{cm2} and $\Ga\in\mathcal T$. The support of $\mu_\Ga$, say $J_\Ga$, is a compact subset of $\Ga$. The equilibrium measure  $\mu=\mu_\Ga$ is characterized by the Euler--Lagrange variational conditions:
\begin{equation}
\label{em2}
2U^\mu(z)+\re V(z)\;
\left\{
\begin{aligned}
&= \ell,\qquad z\in J_\Ga,\\
&\ge \ell,\qquad z\in \Ga\setminus J_\Ga,
\end{aligned}
\right.
\end{equation}
where $\ell=\ell_\Ga$ is a constant, the Lagrange multiplier, and
\[
U^\mu(z)=-\int\log|z-s|d\mu(s)
\]
is the logarithmic potential of $\mu$, see \cite[Theorem~I.3.3]{SaffTotik}. Any $\Ga\in\mathcal T$ can be used to define $Z_N(t)$ in \eqref{cm1}, nevertheless, it is well understood in the theory of non-Hermitian orthogonal polynomials, starting with the works of Stahl \cite{St85,St85b,St86} and Gonchar and Rakhmanov \cite{GRakh87} that one should use the contour whose equilibrium measure has support  \emph{symmetric} (with the \emph{S-property}) in the external field $\re V$. We make this idea precise in the following definition.
\begin{definition} 
The support $J_\Ga$ has the S-property in the external field $\re V$, if it consists of a finite number of open analytic arcs and their endpoints, and  on each arc it holds that
\begin{equation}
\label{em4}
\frac{\partial }{\partial n_+}\,\big(2U^{\mu_\Gamma}+\re V\big)=
\frac{\partial }{\partial n_-}\,\big(2U^{\mu_\Gamma}+\re V\big),
\end{equation}
where $\frac{\partial }{\partial n_+}$ and  $\frac{\partial }{\partial n_-}$ are the normal derivatives from the $(+)$- and $(-)$-side of $\Ga$. We shall say that a curve $\Ga\in\mathcal T$ is an S-curve in the field $\re V$, if $J_\Ga$ has the S-property in this field.
\end{definition}

It is also understood that geometrically $J_\Ga$ is comprised of \emph{critical trajectories} of quadratic differentials. Recall that if $Q$ is a meromorphic function, a \emph{trajectory} (resp. \emph{orthogonal trajectory}) of a quadratic differential $-Q(z)\mathrm dz^2$ is a maximal regular arc on which
\[
-Q(z(s))\big(z^\prime(s)\big)^2>0 \quad \big(\text{resp.} \quad -Q(z(s))\big(z^\prime(s)\big)^2<0\big)
\]
for any local uniformizing parameter. A trajectory is called \emph{critical} if it is incident with a \emph{finite critical point} (zero or a simple pole of $-Q(z)\mathrm dz^2$) and it is called \emph{short} if it is incident only with finite critical points. We designate the expression \emph{critical (orthogonal) graph of $-Q(z)\mathrm dz^2$} for the totality of the critical (orthogonal) trajectories $-Q(z)\mathrm dz^2$.

The following theorem is a specialization to $V(z;t)$ of  \cite[Theorem~2.3]{KS}.

\begin{theorem}
Let $V(z;t)$ be given by \eqref{cm2}.
\label{fundamental} 
\begin{enumerate}
  \item There exists a contour $\Ga_t\in\mathcal T$ such that
\begin{equation}
\label{em3}
\mathcal E_V(\Ga_t)=\sup_{\Ga\in\mathcal T} \mathcal E_V(\Ga).
\end{equation}
  \item The equilibrium measure $\mu_t:=\mu_{\Ga_t}$ is the same for every $\Ga_t$ satisfying \eqref{em3}. The support $J_t$ of $\mu_t$ has the S-property in the external field $\re V(z;t)$.
  \item The function
\begin{equation}
\label{em5}
Q(z;t)=\left(\frac{V'(z;t)}{2}- \int \frac{\mathrm d\mu_t(s)}{z-s}\right)^2,\quad z\in \C\setminus J_t,
\end{equation}
is a polynomial of degree 4.
\item The support $J_t$ consists of short critical trajectories of the quadratic differential $-Q(z;t)\mathrm dz^2$ that connect simple zeros of $Q(z;t)$ and the equation
\begin{equation}
\label{em6}
\mathrm d\mu_t(z)=-\frac{1}{\pi i}\,Q_+^{1/2}(z;t)\mathrm dz, \quad z\in J_t,
\end{equation}
holds on each such critical trajectory, where $Q^{1/2}(z;t)=\frac12z^2+\mathcal{O}(z)$ as $z\to\infty$.
\end{enumerate}  
\end{theorem}

Much information on the structure of the critical graphs of a quadratic differential can be found in the excellent monographs \cite{Jenkins,Pommerenke,Strebel}. Since $\deg\, Q(z)=4$, $J_{\Ga_t}$ consists of one or two arcs, corresponding (respectively) to the cases where $Q(z)$ has two simple zeros and one double zero, and the case where it has four simple zeros. In this paper we study the case of a single arc and investigate the two-cut case in a later publication. In the next section we discuss which values of $t$ correspond to the one-cut case and describe the geometry of the critical graphs in more detail.

\section{Structure of $\Ga_t$}\label{s:Structure}

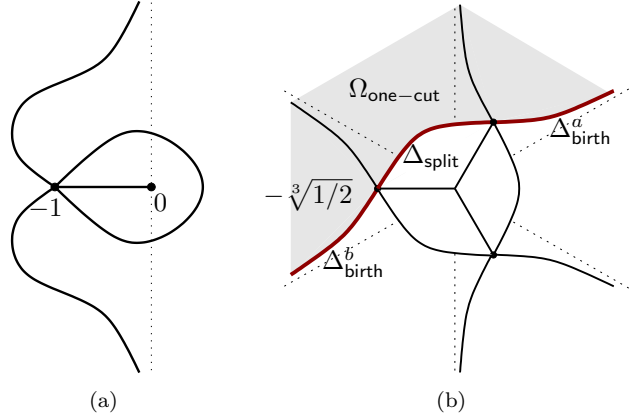
\begin{figure}[ht!]
\centering
\subfigure[]{\includegraphics[scale=.5]{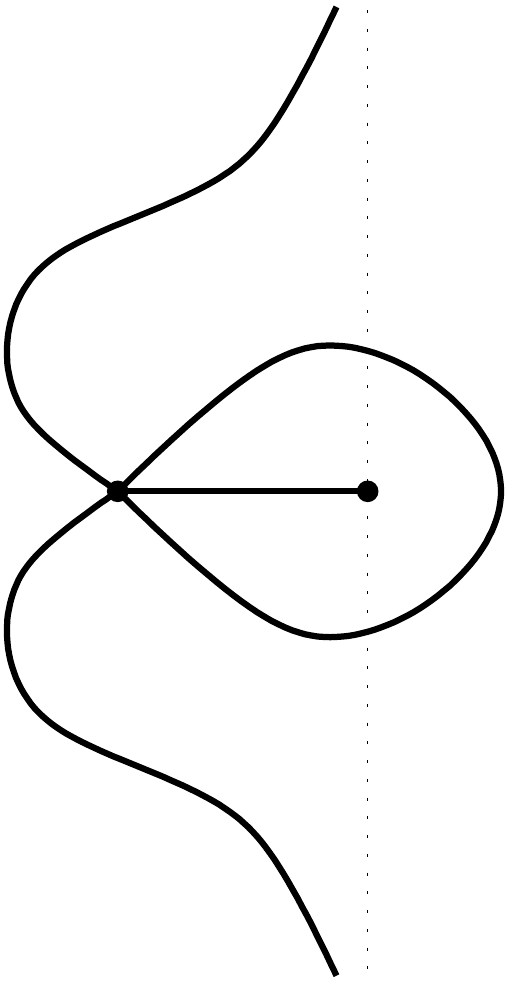}}
\begin{picture}(0,0)
\put(-23,62){$0$}
\put(-70,60){$-1$}
\end{picture}
\quad\quad
\subfigure[]{\includegraphics[scale=.8]{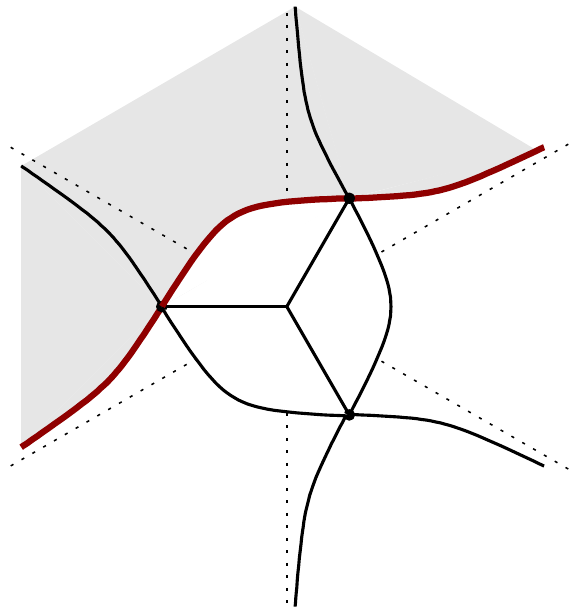}}
\begin{picture}(0,0)
\put(-142,65){$-\sqrt[3]{1/2}$}
\put(-35,90){$\Delta^a_\mathsf{birth}$}
\put(-90,80){$\Delta_\mathsf{split}$}
\put(-120,40){$\Delta^b_\mathsf{birth}$}
\put(-110,105){$\Omega_\mathsf{one-cut}$}
\end{picture}
\caption{\small schematic representation of (a)  the critical graph $\mathcal C$; (b) the set $\Delta$ (solid lines) and the domain $\Omega_\mathsf{one-cut}$ (shaded region).}
\label{fig:loops}
\end{figure}

\begin{figure}[ht!]
\centering
\includegraphics[scale=1.2]{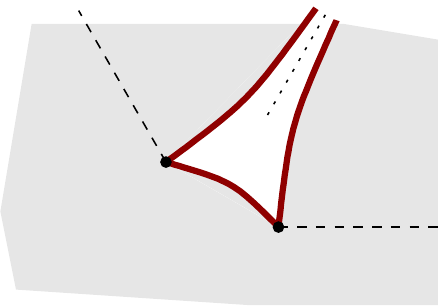}
\begin{picture}(0,0)
\put(-64,18){$t_\mathsf{cr}$}
\put(-140,47){$e^{2\pi\mathrm i/3}t_\mathsf{cr}$}
\put(-53,55){$C_\mathsf{birth}^b$}
\put(-95,77){$C_\mathsf{birth}^a$}
\put(-92,33){$C_\mathsf{split}$}
\put(-20,18){$S$}
\put(-153,80){$e^{2\pi\mathrm i/3}S$}
\put(-135,15){$O_\mathsf{one-cut}$}
\end{picture}
\caption{\small Domain $O_\mathsf{one-cut}$ (shaded region); $\partial O_\mathsf{one-cut}$ consisting of the open bounded arc $C_\mathsf{split}$, two open semi-unbounded arcs $C_\mathsf{birth}^a$ and $C_\mathsf{birth}^b$, and two points $t_\mathsf{cr}$ and $e^{2\pi\mathrm i/3}t_\mathsf{cr}$; the semi-unbounded open horizontal rays $S$ and $e^{2\pi\mathrm i/3}S$ (dashed lines).}
\label{fig:C-curves}
\end{figure}
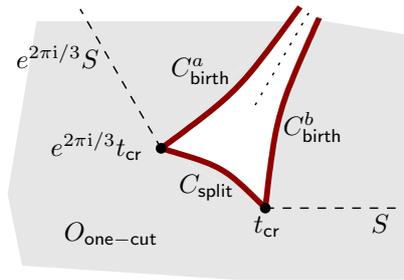

The structure of $\Ga_t$ and its dependence on $t$ has been heuristically described in \cite{AMAM1,AMAM2}. Our goal here is to provide rigorous mathematical justifications for this description, when $J_t$ consists of a single arc. Clearly, in this case $Q(z;t)$ should be of the form
\begin{equation}
\label{ts1}
Q(z;t)=\frac14(z-a(t))(z-b(t))(z-c(t))^2.
\end{equation}
It follows from  \eqref{em5} in conjunction with \eqref{cm2} that
\begin{equation}
\label{ts2}
Q(z;t)=\left(\frac{-z^2+t}{2}-\frac{1}{z}+\mathcal O(z^{-2})\right)^2=\frac{(z^2-t)^2}{4}+z+C.
\end{equation}
Thus, by equating the coefficients in \eqref{ts1} and \eqref{ts2}, we obtain a system of equations
\begin{equation}
\label{ts3}
\left\{
\begin{aligned}
&a+b+2c=0,\\
&ab+c^2+2(a+b)c=-2t\,,\\
& 2abc+(a+b)c^2=-4.
\end{aligned}
\right.
\end{equation}
By setting $x:=(a+b)/2$ and eliminating the product $ab$ from the second and third relations in \eqref{ts3}, we get that
\begin{equation}
\label{ts4}
x^3-tx-1 = 0.
\end{equation}
To study the solutions of \eqref{ts4}, denote by $\mathcal C$ the critical graph of an auxiliary quadratic differential
\begin{equation}
\label{aux-d}
-(1+1/s)^3 \mathrm ds^2,
\end{equation}
see Figure~\ref{fig:loops}(a). We show in Section~\ref{s:pr-geom} that $\mathcal C$ consists of 5 critical trajectories emanating from $-1$ at the angles $2\pi k/5$, $k\in\{0,\ldots,4\}$, one of them being $(-1,0)$, other two forming a loop crossing the real line approximately at $0.635$, and the last two approaching infinity along the imaginary axis without changing the half-plane (upper or lower).
Given $\mathcal C$, define
\[
\Delta:=\big\{x:~2x^3\in\mathcal C\big\}.
\]
Further, put $\Omega_\mathsf{one-cut}$ to be the shaded region on Figure~\ref{fig:loops}(b) and set
\[
\partial \Omega_\mathsf{one-cut} = \Delta^b_\mathsf{birth}\cup \big\{-2^{-1/3}\big\} \cup \Delta_\mathsf{split} \cup \big\{e^{\pi\mathrm i/3}2^{-1/3}\big\} \cup \Delta^a_\mathsf{birth},
\]
where $\Delta_\mathsf{split}$ connects $-2^{-1/3}$ and $e^{\pi\mathrm i/3}2^{-1/3}$, $\Delta^b_\mathsf{birth}$ extends to infinity in the direction of the angle $7\pi/6$ while $\Delta^a_\mathsf{birth}$ extends to infinity in the direction of the angle $\pi/6$. Finally, let $t(x):=(x^3-1)/x$ and set
\begin{equation}
\label{ts5}
\left\{
\begin{array}{l}
t_\mathsf{cr} := 3\cdot2^{-2/3} = t\big(-2^{-1/3}\big), \medskip \\ 
O_\mathsf{one-cut}:=t(\Omega_\mathsf{one-cut}), \medskip \\
C_\mathsf{split} := t\big(\Delta_\mathsf{split}\big), \quad C^b_\mathsf{birth} := t\big(\Delta^b_\mathsf{birth}\big), \quad C^a_\mathsf{birth} := t\big(\Delta^a_\mathsf{birth}\big), \medskip \\
S:=(t_\mathsf{cr},\infty), \quad e^{2\pi\mathrm i/3}S := \big\{z:~e^{-2\pi\mathrm i/3}z\in S\big\},
\end{array}
\right.
\end{equation}
see Figure~\ref{fig:C-curves}. 
The function $t(x)$ is holomorphic in $\Omega_\mathsf{one-cut}$ with non-vanishing derivative there. It maps $\Omega_\mathsf{one-cut}$ onto $O_\mathsf{one-cut}$ in a one-to-one fashion. Hence, the inverse map $x(t)$ exists and is holomorphic. Altogether, the following proposition holds.

\begin{proposition}
\label{prop:x}
There exists a holomorphic branch $x(t)$ of \eqref{ts4} that maps $O_\mathsf{one-cut}$ conformally onto $\Omega_\mathsf{one-cut}$. The function $x(t)$ possesses analytic continuations across each of the arcs $C_\mathsf{split}$, $C_\mathsf{birth}^a$, and $C_\mathsf{birth}^b$. The functions
\begin{equation}
\label{ts6}
\left\{
\begin{array}{lll}
a(t) &:=& x(t)-\mathrm i\sqrt2/\sqrt x(t), \medskip \\
b(t) &:=& x(t)+\mathrm i\sqrt2/\sqrt x(t), \medskip \\
c(t) &:=& -x(t),
\end{array}
\right.
\end{equation}
are holomorphic in $O_\mathsf{one-cut}$, where $\sqrt x(t)$  is the branch holomorphic in $O_\mathsf{one-cut}$ satisfying $\sqrt x(0)=e^{\pi\mathrm i/3}$. It is a matter of a routine verification to check that they also satisfy \eqref{ts3}.
\end{proposition}

Below, we adapt the following convention: $\Ga(z_1,z_2)$ (resp. $\Ga[z_1,z_2]$) stands for the trajectory or orthogonal trajectory (resp. the closure of) connecting $z_1$ and $z_2$, oriented from $z_1$ to $z_2$, and $\Ga\big(z,e^{\mathrm i\theta}\infty\big)$ (resp. $\Ga\big(e^{\mathrm i\theta}\infty,z\big)$) stands for the orthogonal trajectory ending at $z$, approaching infinity at the angle $\theta$, and oriented away from $z$ (resp. oriented towards $z$).\footnote{This notation is unambiguous as the corresponding trajectories are unique for polynomial differentials as follows from Teichm\"uller's lemma, see \eqref{teichmuller} further below.}

\begin{figure}[ht!]
\subfigure[]{\includegraphics[scale=.2]{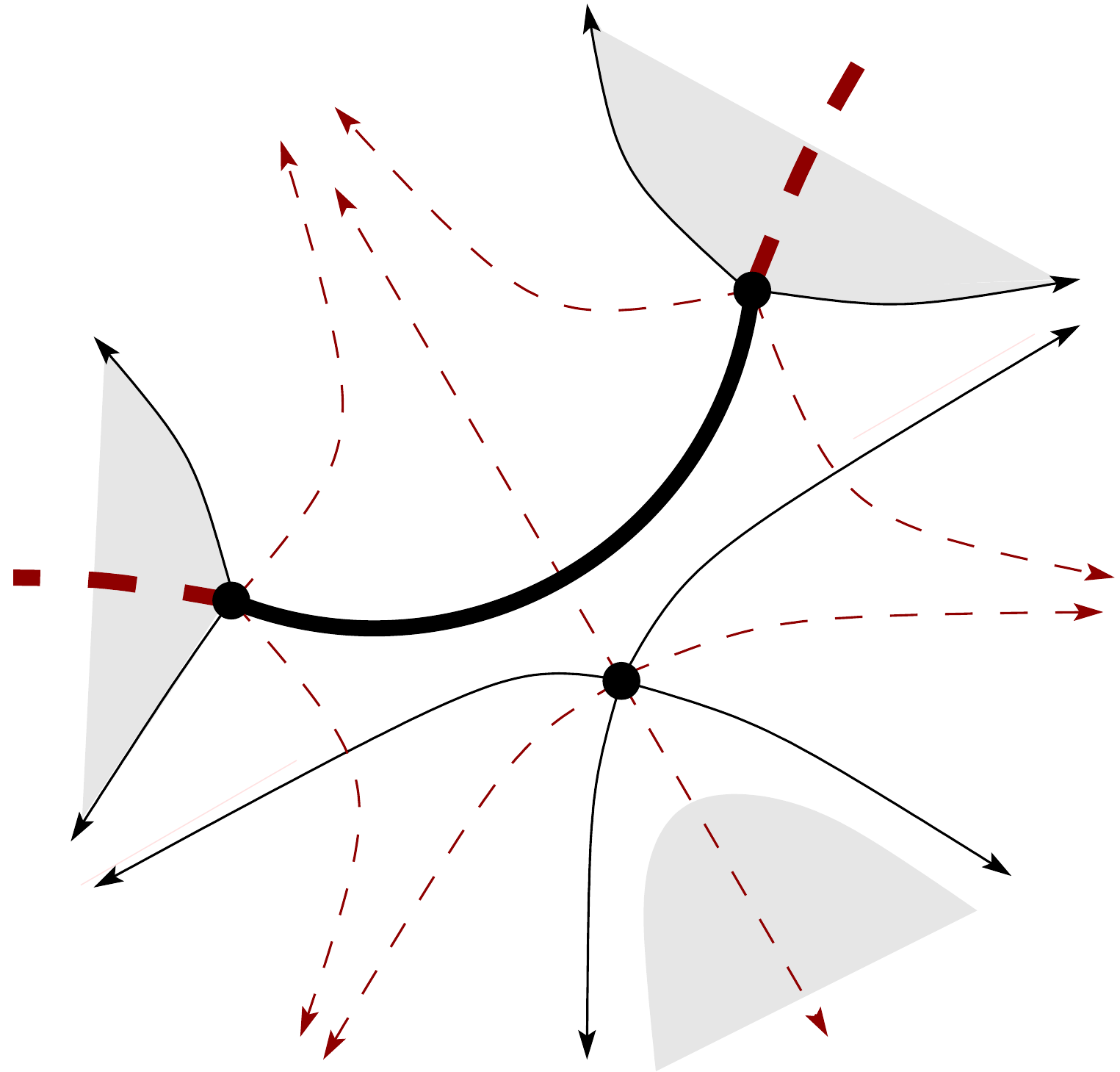}}
\subfigure[]{\includegraphics[scale=.2]{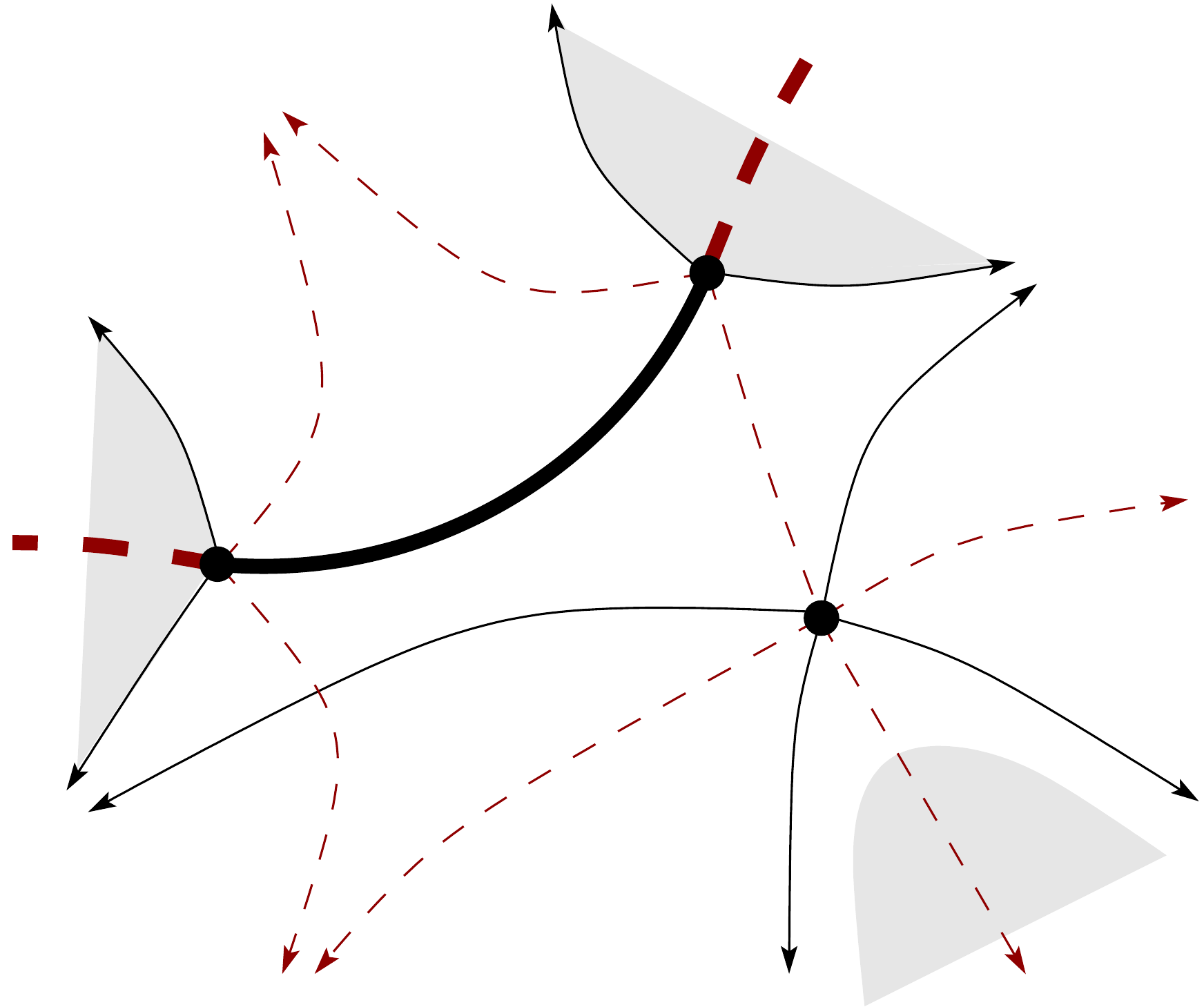}}
\subfigure[]{\includegraphics[scale=.19]{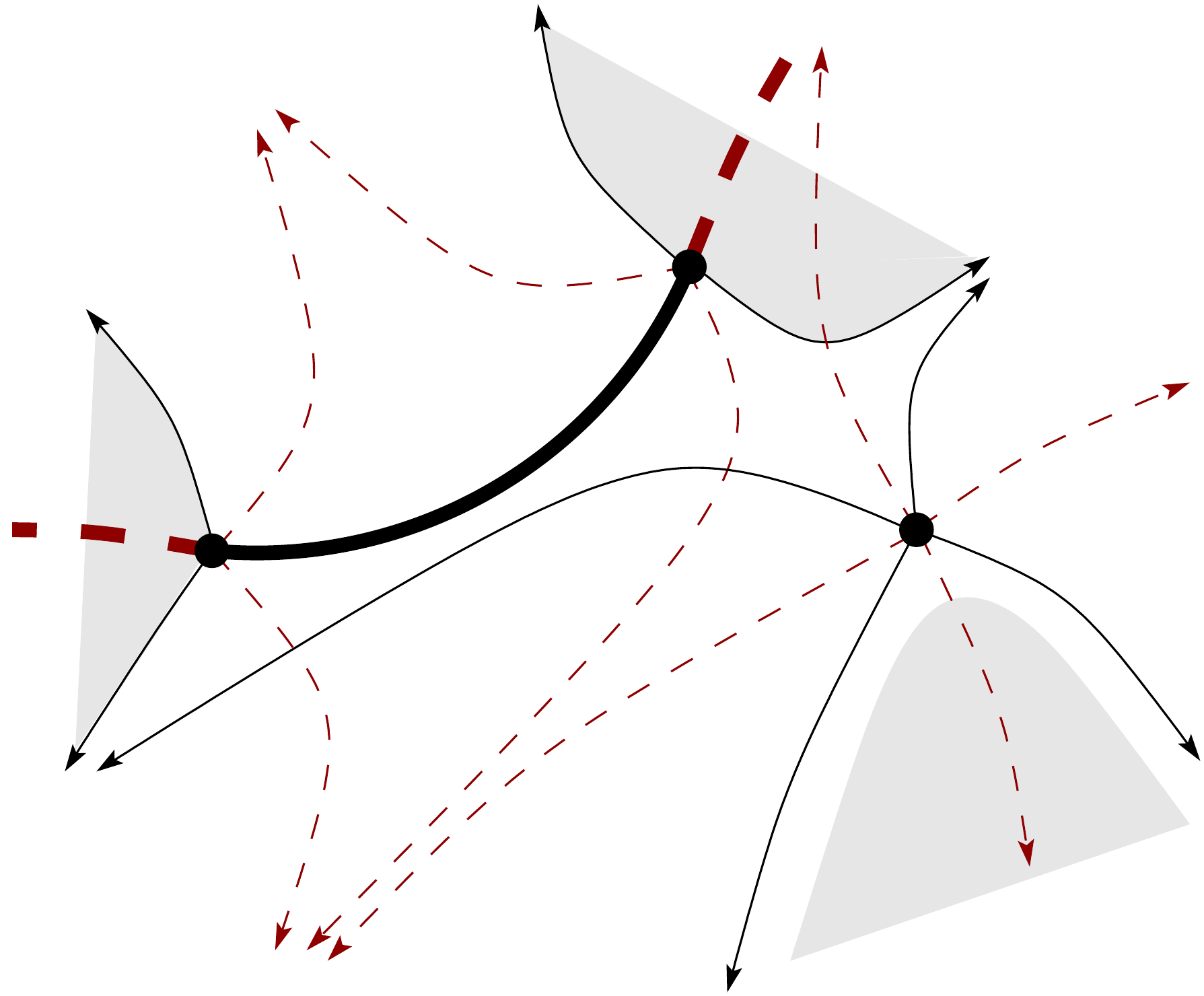}}
\subfigure[]{\includegraphics[scale=.18]{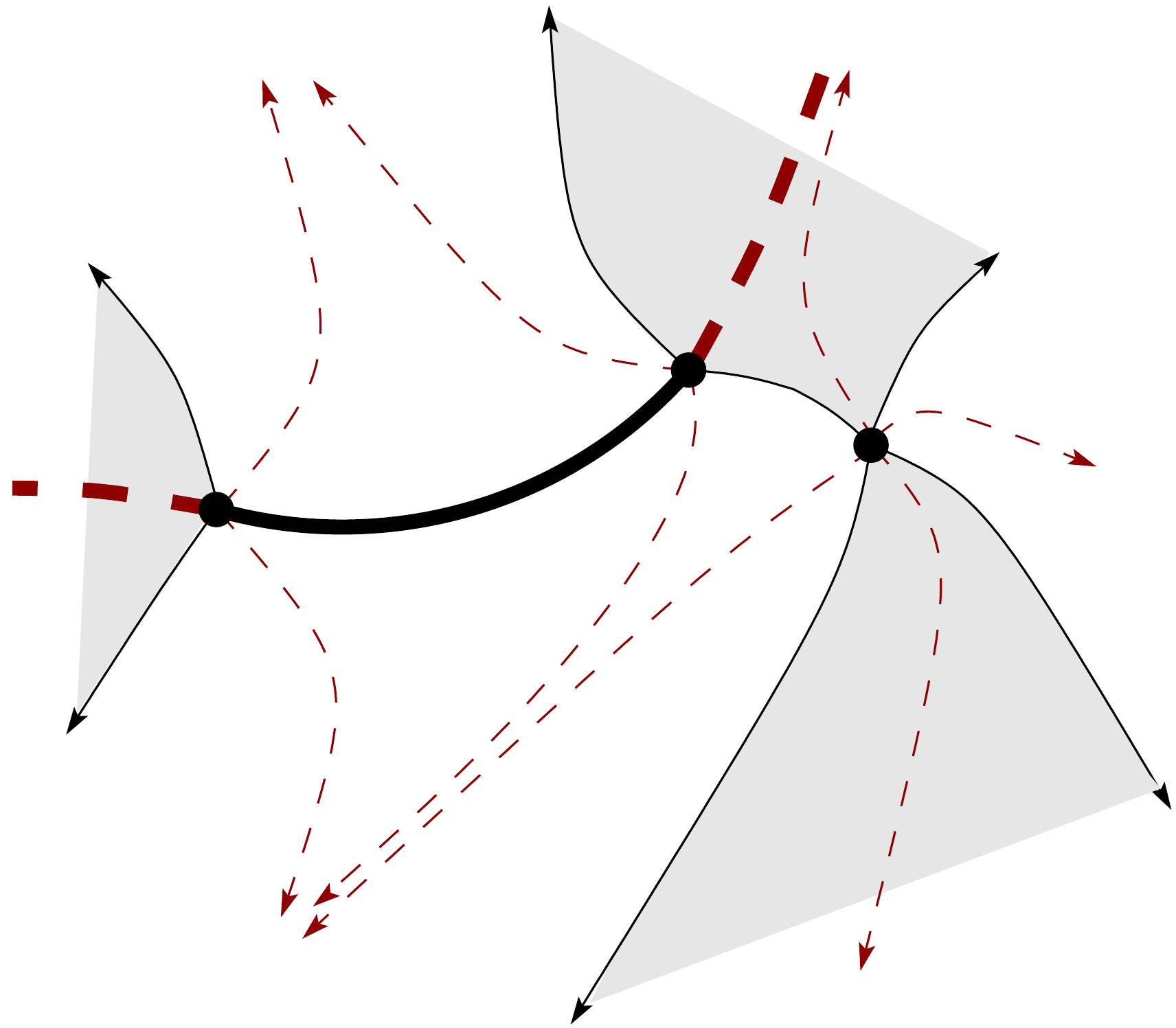}}\newline
\subfigure[]{\includegraphics[scale=.21]{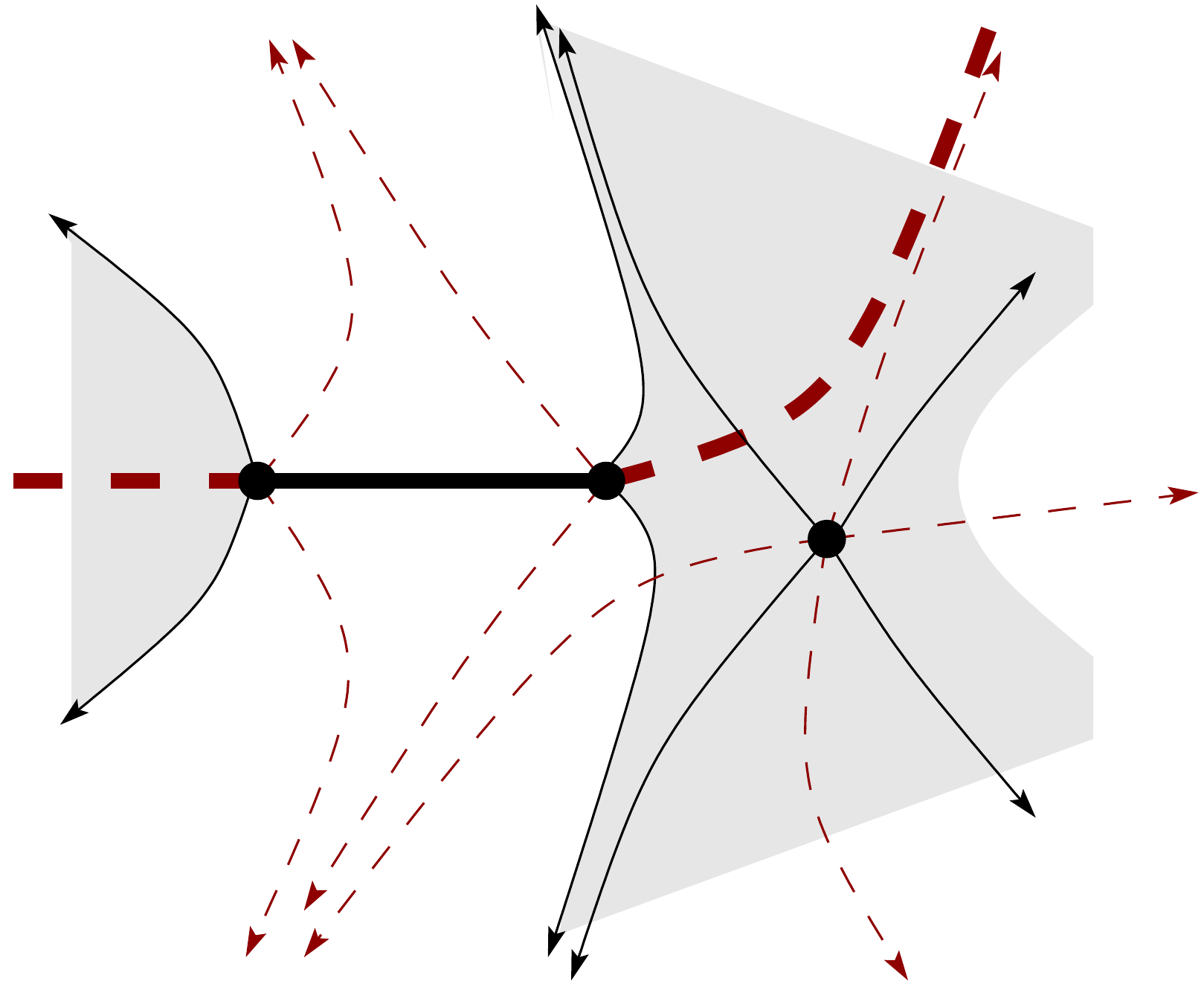}}
\subfigure[]{\includegraphics[scale=.21]{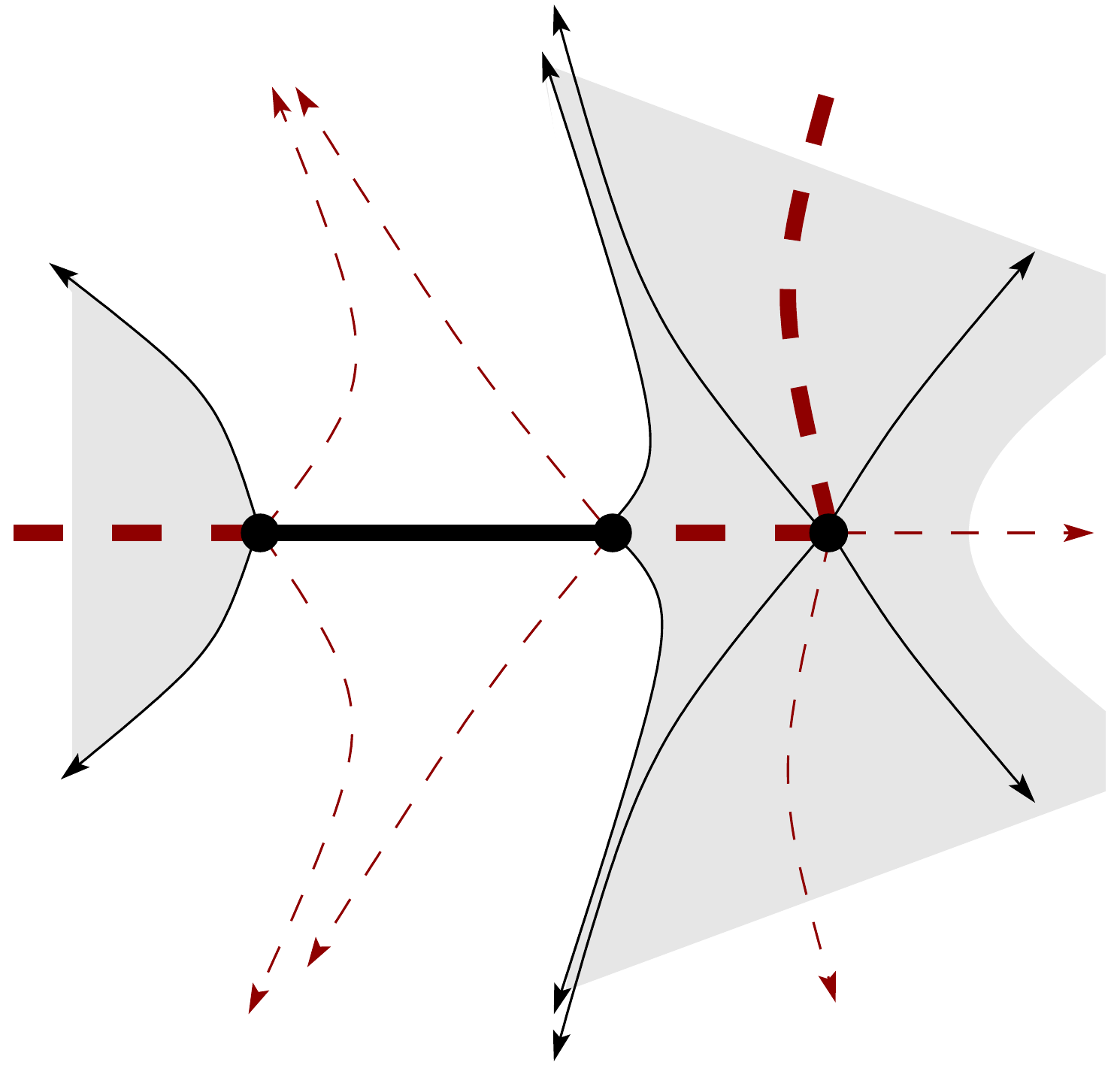}}
\subfigure[]{\includegraphics[scale=.21]{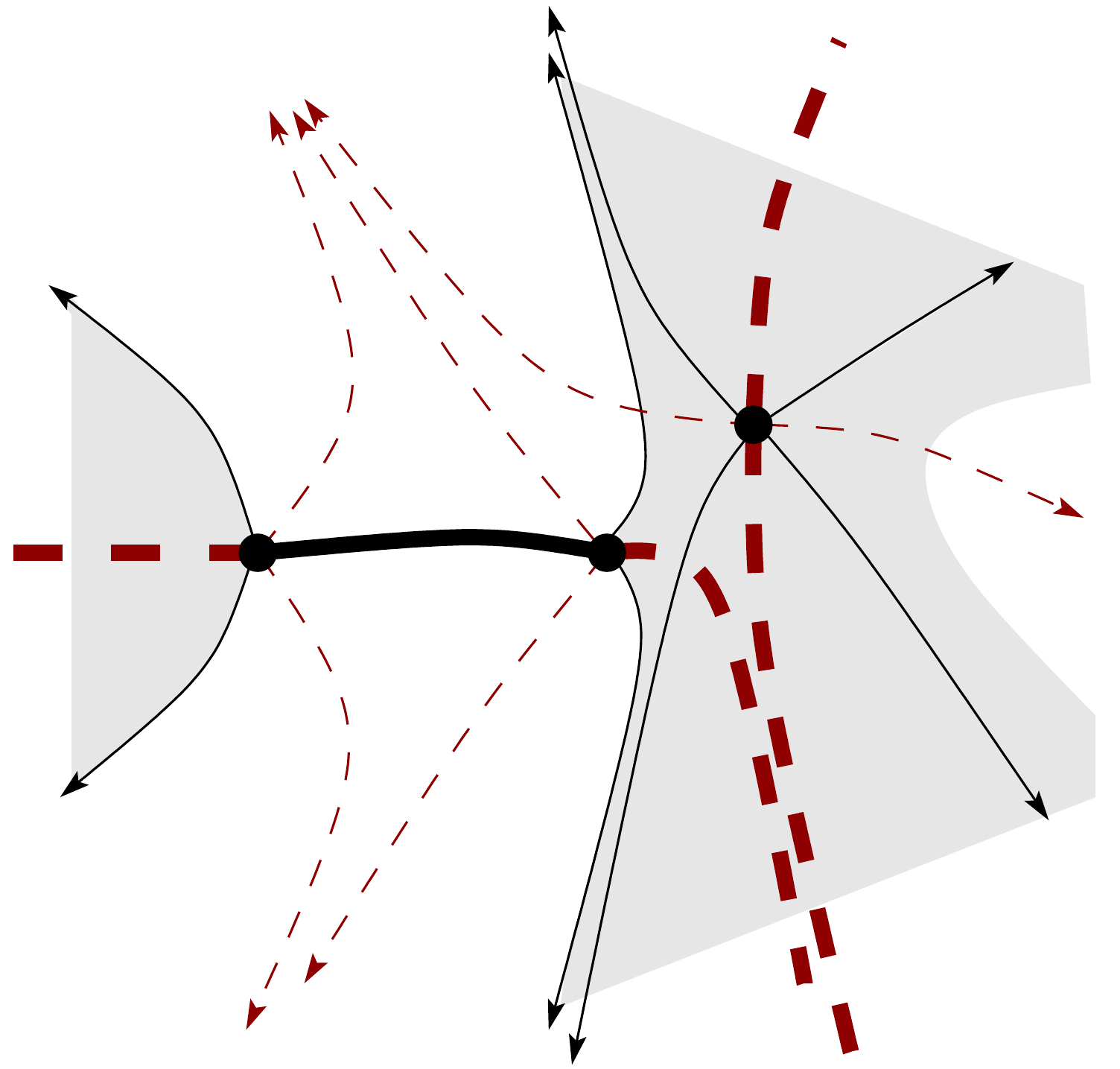}}
\caption{\small Schematic representation of the critical (solid) and critical orthogonal (dashed) graphs of $-Q(z;t)\mathrm dz^2$ when $t\in O_\mathsf{one-cut}$. The bold curves represent the preferred S-curve $\Ga_t$. Shaded region is the set where $\re\left(\int_b^zQ^{1/2}(z;t)\mathrm dz\right)<0$.}
 \label{s-curves1}
 \end{figure} 

\begin{figure}[ht!]
\subfigure[]{\includegraphics[scale=.23]{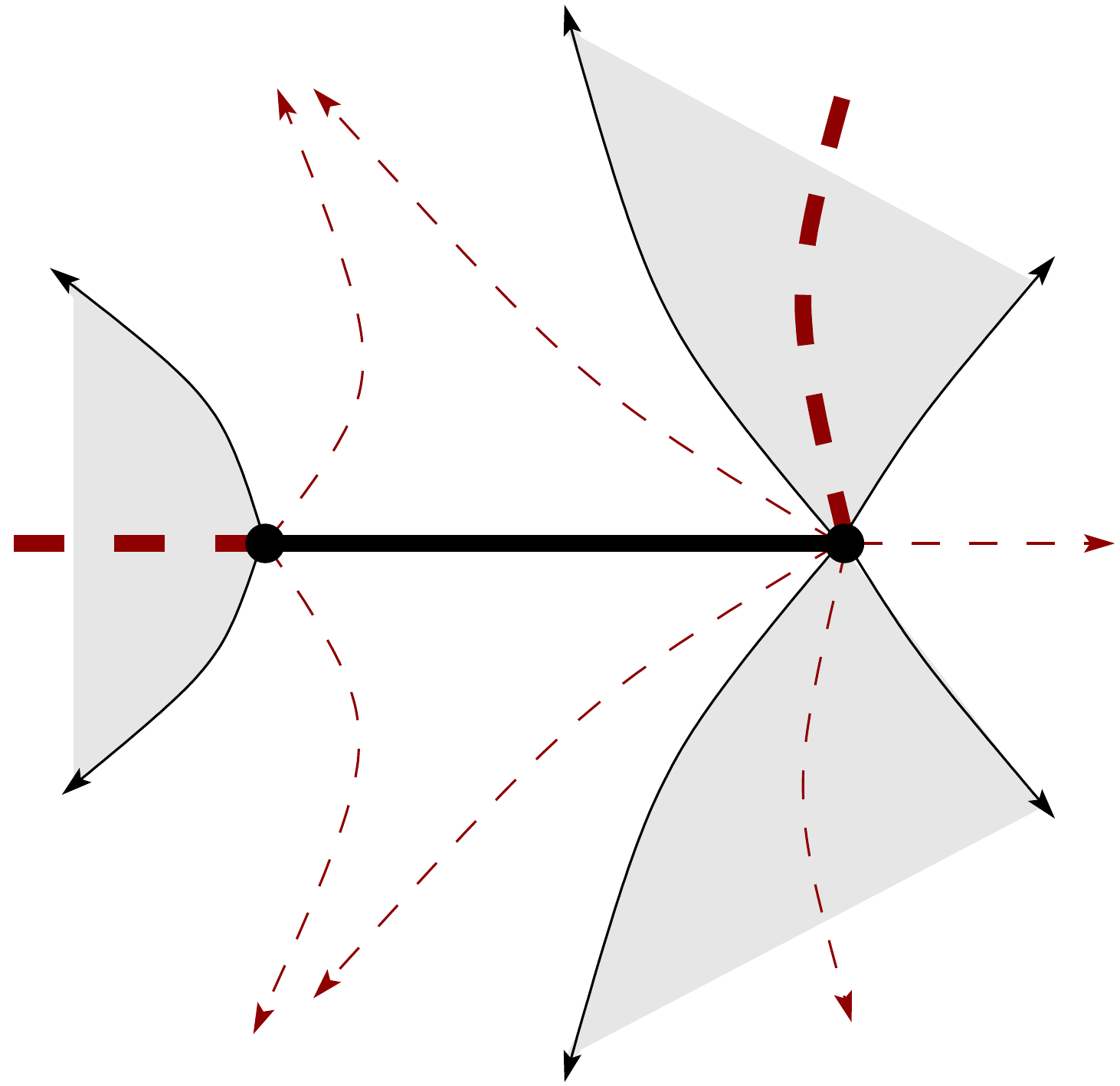}}
\subfigure[]{\includegraphics[scale=.23]{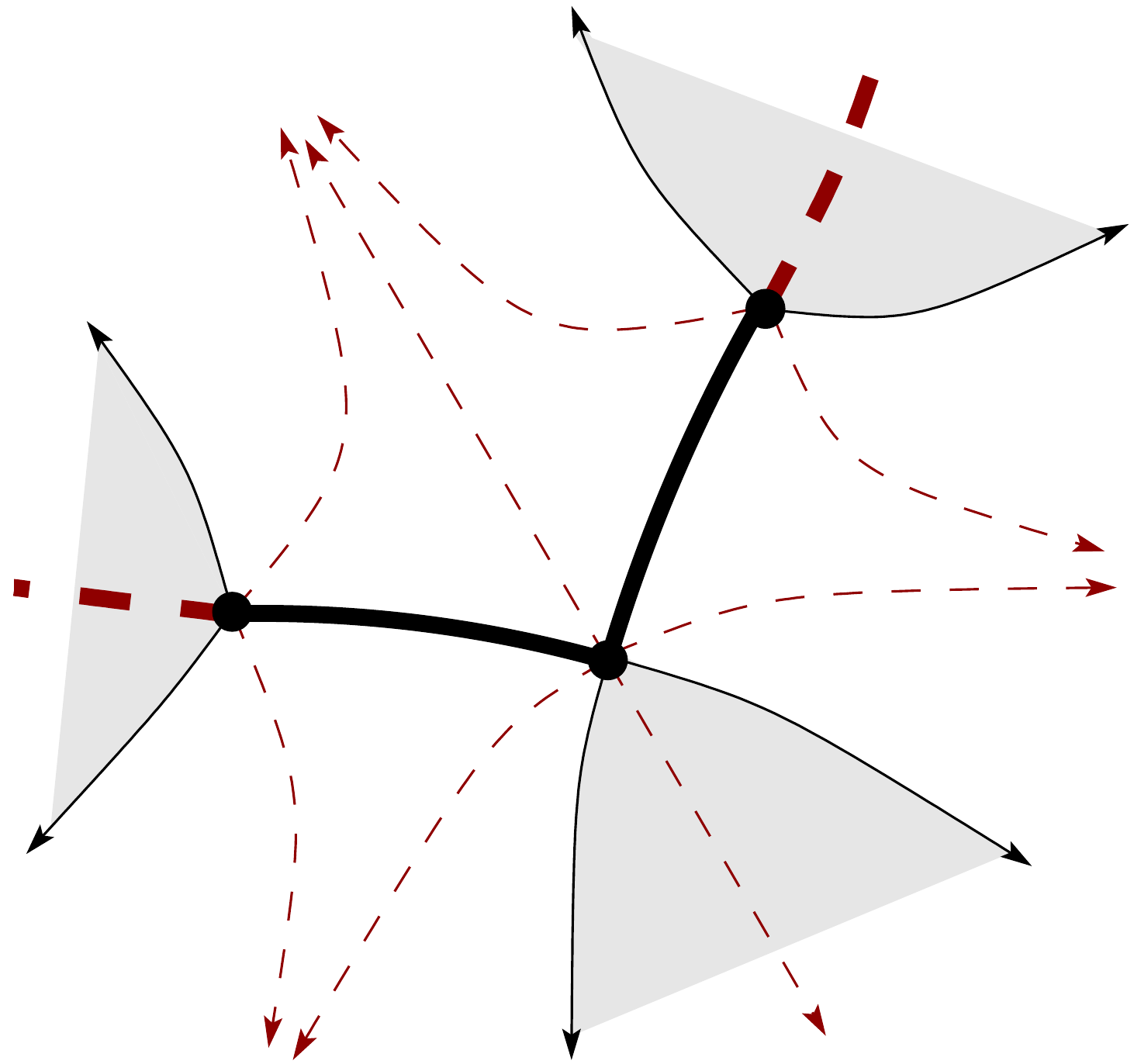}}
\subfigure[]{\includegraphics[scale=.2]{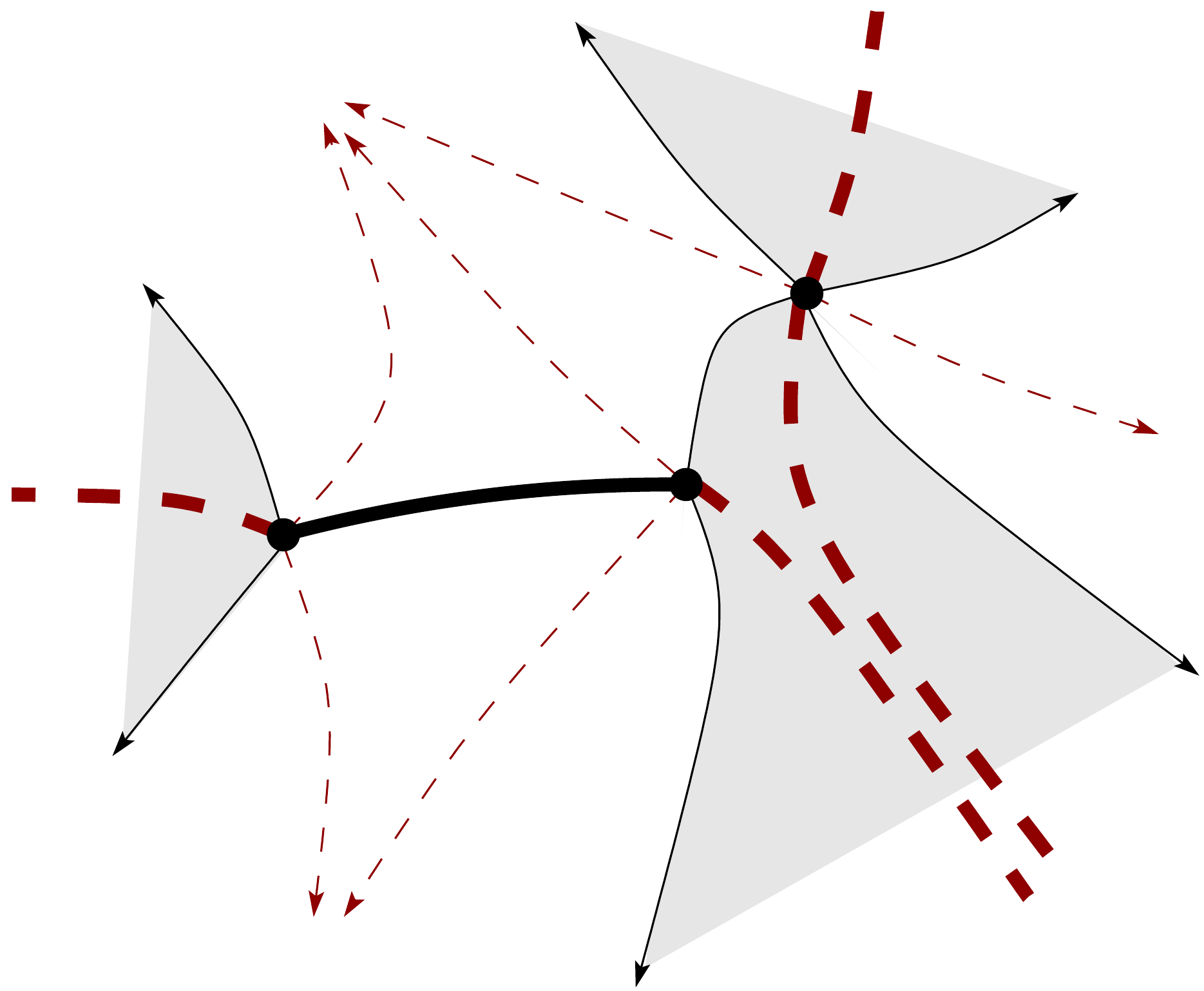}}
\caption{\small This is a continuation of Figure~\ref{s-curves1} for the case $t\in\partial O_\mathsf{one-cut}$.}
 \label{s-curves2}
 \end{figure}

\begin{theorem}
\label{geometry}
Let $\mu_t$ and $Q(z;t)$ be as in Theorem~\ref{fundamental}, $J_t=\mathsf{supp}(\mu_t)$. When $t\in\overline{O}_\mathsf{one-cut}$, the polynomial $Q(z;t)$ is of the form \eqref{ts1} with $a(t)$, $b(t)$, and $c(t)$  as in Proposition~\ref{prop:x} and the set $J_t$ consists of a single arc. Moreover,
\begin{itemize}
\item[(I)] if $t\in O_\mathsf{one-cut}$, then $J_t=\Ga[a,b]$ and an S-curve $\Ga_t\in\mathcal T$ can be chosen as
\begin{itemize}
\smallskip
\item[(a)] $\Ga\big(e^{\pi\mathrm i}\infty,a\big)\cup J_t \cup \Ga\big(b,e^{\pi\mathrm i/3}\infty\big)$ when $t$ belongs to the connected component bounded by $S\cup C_\mathsf{crit}\cup e^{2\pi\mathrm i/3}S$, see Figure~\ref{s-curves1}(a--e);
\smallskip
\item[(b)] $\Ga\big(e^{\pi\mathrm i}\infty,a\big) \cup J_t\cup \Ga(b,c) \cup \Ga\big(c,e^{\pi\mathrm i/3}\infty\big)$ when $t\in S$, see Figure~\ref{s-curves1}(f);
\smallskip
\item[(c)] $\Ga\big(e^{\pi\mathrm i}\infty,c\big) \cup \Ga(c,a) \cup J_t \cup  \Ga\big(b,e^{\pi\mathrm i/3}\infty\big)$ when $t\in e^{2\pi\mathrm i/3}S$;
\smallskip
\item[(d)] $\Ga\big(e^{\pi\mathrm i}\infty,a\big) \cup J_t\cup \Ga\big(b,e^{-\pi\mathrm i/3}\infty\big) \cup \Ga\big(e^{-\pi\mathrm i/3}\infty,c\big) \cup \Ga\big(c,e^{\pi\mathrm i/3}\infty\big)$ when $t$ belongs to the connected component bounded by $S\cup C_\mathsf{birth}^b$, see Figure~\ref{s-curves1}(g);
\smallskip
\item[(e)] $\Ga\big(e^{\pi\mathrm i}\infty,c\big) \cup \Ga\big(c,e^{-\pi\mathrm i/3}\big) \cup \Ga\big(e^{-\pi\mathrm i/3}\infty,a\big) \cup J_t \cup \Ga\big(b,e^{\pi\mathrm i/3}\infty\big)$ when $t$ belongs to the connected component bounded by $e^{2\pi\mathrm i/3}S\cup C_\mathsf{birth}^a$.
\end{itemize}
\item[(II)] if $t=t_\mathsf{cr}$ (resp. $t=e^{2\pi i/3}t_\mathsf{cr})$, then $J_t=\Ga[a,b]$, $c$ coincides with $b$ (resp. $a$), and an S-curve $\Ga_t\in\mathcal T$ can be chosen as in Case I(a), see Figure~\ref{s-curves2}(a).
\item[(III)] if $t\in C_\mathsf{split}$, then $J_t=\Ga[a,c]\cup\Ga[c,b]$ and an S-curve $\Ga_t\in\mathcal T$ can be chosen as in Case I(a), see Figure~\ref{s-curves2}(b).
\item[(IV)] if  $t\in C_\mathsf{birth}^b$ (resp. $t\in C_\mathsf{birth}^a$), then $J_t=\Ga[a,b]$ and an S-curve $\Ga_t\in\mathcal T$ can be chosen as in Case I(d) (resp. Case I(e)), see Figure~\ref{s-curves2}(c).
\end{itemize}
\end{theorem}

We prove Theorem~\ref{geometry} in Section~\ref{s:pr-geom}.

\begin{remark}
Even though we shall not dwell on this point here, the reason for the nomenclature introduced above is the following. When $t\not\in\overline O_\mathsf{one-cut}$, the double zero $c(t)$ splits into two simple zeros, say  $c_1(t)$ and $c_2(t)$. When $t$ crosses $C_\mathsf{split}$,  the trajectory connecting $a$ and $b$ will split into two, one connecting $a$ and $c_1$ and another connecting $c_2$ and $b$. When $t$ crosses  $C_\mathsf{birth}:=C_\mathsf{birth}^a\cup C_\mathsf{birth}^b$, a critical trajectory connecting $c_1$ and $c_2$ will appear while $a$ and $b$ will remain being connected by a trajectory.
\end{remark}

Let us elaborate on all the configurations appearing on Figures~\ref{s-curves1} and~\ref{s-curves2}.
\begin{figure}[ht!]
\centering
\subfigure[]{\includegraphics[scale=.4]{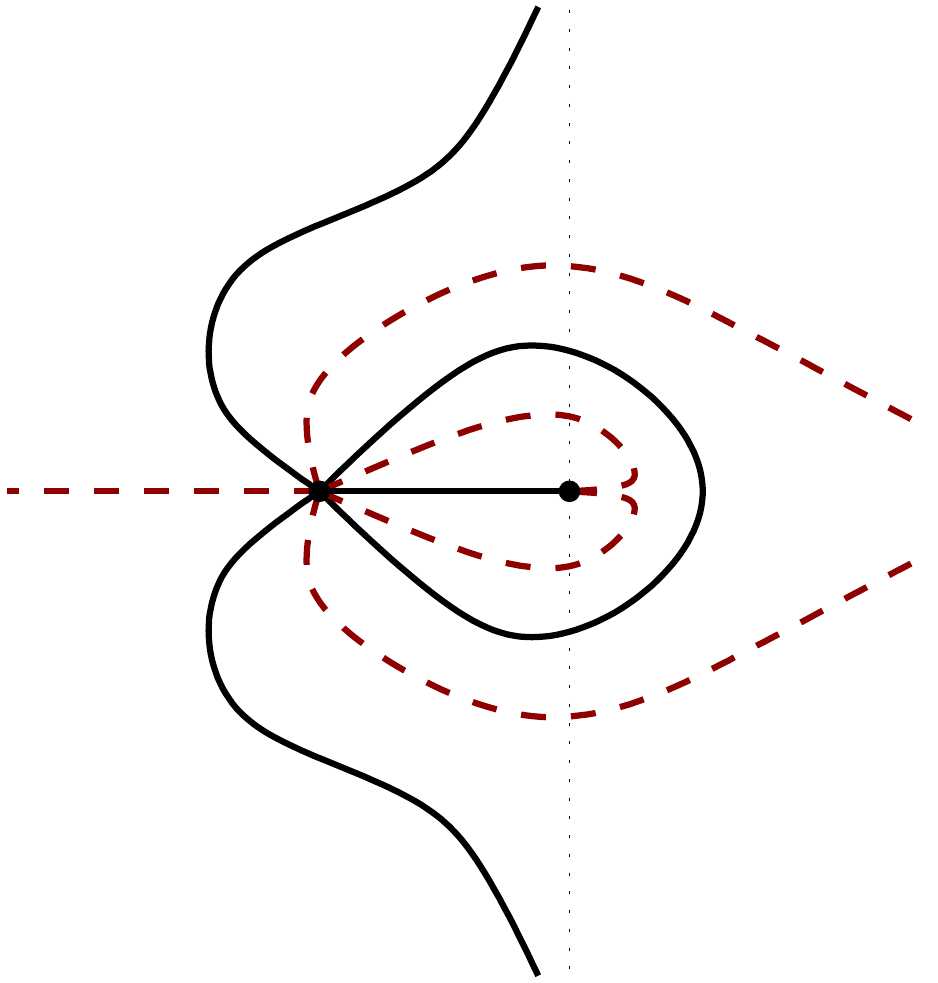}}
\begin{picture}(0,0)
\put(-20,55){$\mathcal S$}
\end{picture}
\quad
\subfigure[]{\includegraphics[scale=.65]{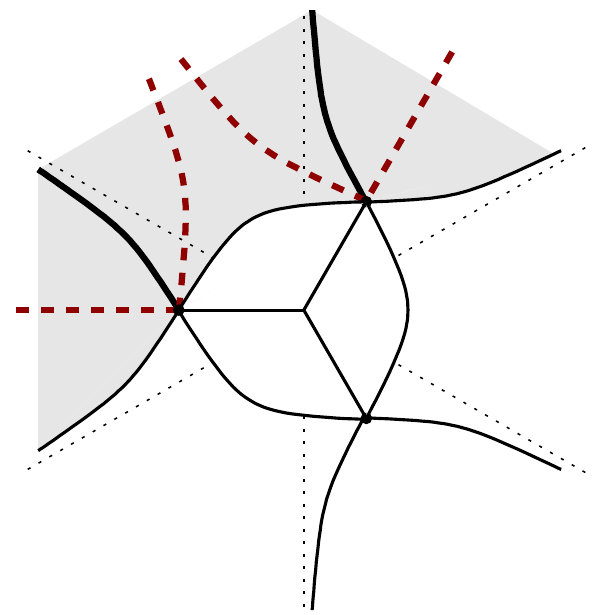}}
\begin{picture}(0,0)
\put(-117,73){$\Delta_\mathsf{crit}$}
\put(-54,100){$\Delta_\mathsf{crit}$}
\put(-80,82){$\Delta^\perp$}
\end{picture}
\quad
\subfigure[]{\includegraphics[scale=.9]{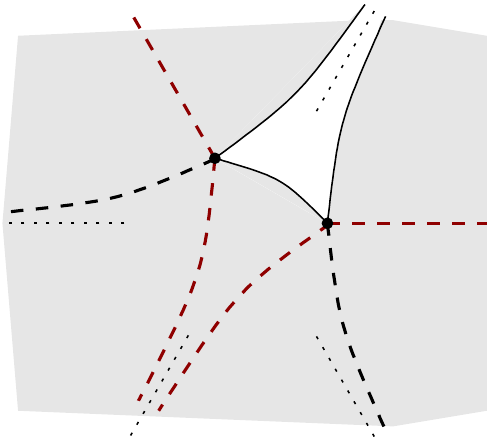}}
\begin{picture}(0,0)
\put(-41,30){$C_\mathsf{crit}^b$}
\put(-116,70){$C_\mathsf{crit}^a$}
\put(-20,60){$S$}
\put(-123,90){$e^{2\pi\mathrm i/3}S$}
\put(-70,30){$S_\mathsf{crit}^b$}
\put(-106,30){$S_\mathsf{crit}^a$}
\end{picture}
\caption{\small Schematic representations of (a) the set $\mathcal S$ (dashed lines); (b) the sets $\Delta^\perp$ (dashed lines) and $\Delta_\mathsf{crit}$ (solid curves within the shaded region); (c) the images of $\Delta^\perp$ and $\Delta_\mathsf{crit}$ under $t(x)$.}
\label{fig:loops1}
\end{figure}
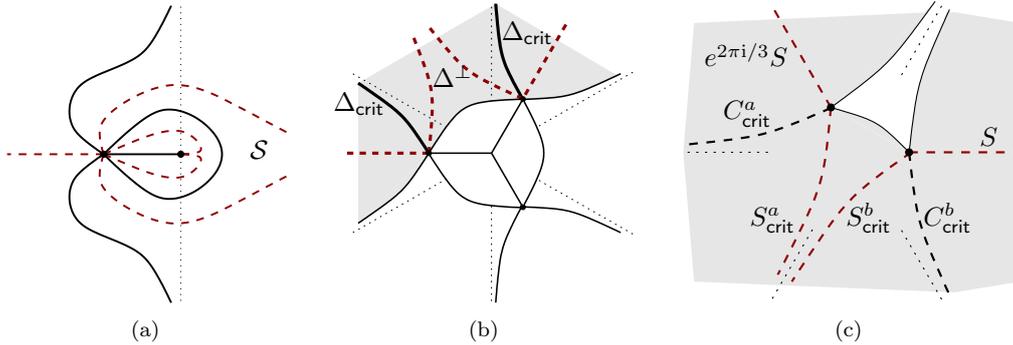
To this end, we need to introduce the totality of the orthogonal trajectories of the differential \eqref{aux-d} emanating out of $-1$, say $\mathcal S$, see Figure~\ref{fig:loops1}(a), the sets
\[
\Delta_\mathsf{crit}:= \Delta\cap\Omega_\mathsf{one-cut}, \quad \Delta^\perp := \big\{x:2x^3\in\mathcal S\big\}\cap\Omega_\mathsf{one-cut},
\]
see Figure~\ref{fig:loops1}(b), as well as the sets
\[
t(\Delta_\mathsf{crit}) =: C_\mathsf{crit}^a \cup C_\mathsf{crit}^b,  \quad t(\Delta^\perp) =: S_\mathsf{crit}^a\cup S_\mathsf{crit}^b \cup S\cup e^{2\pi\mathrm i/3}S,
\]
where $C_\mathsf{crit}^b$, $S_\mathsf{crit}^b$ and $C_\mathsf{crit}^a$, $S_\mathsf{crit}^a$ are incident with $t_\mathsf{cr}$ and $e^{2\pi\mathrm i/3}t_\mathsf{cr}$, respectively, see Figure~\ref{fig:loops1}(c).

\begin{theorem}
\label{geometry1}
The critical and critical orthogonal graphs of $-Q(z;t)\mathrm dz^2$ have the structure as on
\begin{itemize}
\item Figure~\ref{s-curves2}(a,b,c) when $t=t_\mathsf{cr}$, $t\in C_\mathsf{split}$, and $t\in C_\mathsf{birth}^b$, respectively;
\item Figure~\ref{s-curves1}(f,d,b) when $t\in S$, $t\in C_\mathsf{crit}^b$, and $t\in S_\mathsf{crit}^b$, respectively;
\item Figure~\ref{s-curves1}(g) when $t$ belongs to the component of $O_\mathsf{one-cut}$ bounded by $S\cup C_\mathsf{birth}^b$;
\item Figure~\ref{s-curves1}(e) when $t$ belongs to the component of $O_\mathsf{one-cut}$ bounded by $S\cup C_\mathsf{crit}^b$; in fact, $C_\mathsf{crit}^b$ is the reflection of $C_\mathsf{birth}^b$ across the real line and the structure is the same as on Figure~\ref{s-curves1}(g) reflected across the real line as well;
\item Figure~\ref{s-curves1}(c) when $t$ belongs to the component of $O_\mathsf{one-cut}$ bounded by $C_\mathsf{crit}^b\cup S_\mathsf{crit}^b$;
\item Figure~\ref{s-curves1}(a) when $t$ belongs to the component of $O_\mathsf{one-cut}$ bounded by $S_\mathsf{crit}^b \cup S_\mathsf{crit}^a$.
\end{itemize}
In the rest of the cases one needs to pick the reflection of $t$ across the line $L_{\pi/3}$, where
\begin{equation}
\label{line}
L_\theta := \big\{re^{\mathrm i \theta}:~r\in(-\infty,\infty)\big\},
\end{equation}
and then reflect the corresponding graph structures across the line $L_{2\pi/3}$. This symmetry is displayed on Figures~\ref{s-curves1}(a) and~\ref{s-curves2}(a).
\end{theorem}

We prove Theorem~\ref{geometry1} in Section~\ref{s:pr-geom} as well.

\section{Main Results}\label{s:main}

In what follows, we always assume that $t\in \overline O_\mathsf{one-cut}$ while $\beta_n(t,N)$ and $\ga_n(t,N)$ are recurrence coefficients \eqref{cm4} of the polynomials $P_n(z;t,N)$ satisfying orthogonality relations \eqref{cm3} with $V(z;t)$ as in \eqref{cm2} on a contour $\Ga=\Ga_t$ as in Theorem~\ref{geometry}. 

Henceforth, we use interval notation preceded by $\Ga_t$ to denote subarcs of $\Ga_t$. For example, $\Ga_t(u,v]$ stands for the subarc of $\Ga_t$ connecting $u$ and $v$, not containing $u$ and containing $v$, and $u$ precedes $v$ according to the orientation of $\Ga_t$. 

Below, we prove existence of various expansions that depend on the parameter $t$. To indicate the way of dependence, we introduce the following two notions.

\begin{definition}
\label{def:t-dep}
Given an increasing sequence $\alpha(i)\to\infty$ as $i\to\infty$, we say that an expansion
\[
A_N(t)\sim \sum_{i=0}^\infty A^{(i)}(t)N^{-\alpha(i)}
\]
holds $t$-locally uniformly if for any $T\subset \overline O_\mathsf{one-cut}$ such that $T\cap O_\mathsf{one-cut}$, $T\cap C_\mathsf{split}$, $T\cap C_\mathsf{birth}$ are compact, there exist constants $C_I(T)<\infty$ for which 
\[
|A_N(t)-\sum_{i=0}^{I-1}A^{(i)}(t)N^{-\alpha(i)}|\leq C_I(T)N^{-\alpha(I)}, \quad t\in T.
\]
Moreover, we say that an expansion is strongly $t$-locally uniform if $C_I(T)<\infty$ exists as long as $T\cap C_\mathsf{split}$ is compact, $T\cap C_\mathsf{birth}$ is closed, and there exists $\epsilon(T)>0$ for which
\begin{equation}
\label{condT}
\re\left(\int_{b(t)}^{c(t)}Q^{1/2}(z;t)\mathrm dz\right) \leq -\epsilon(T)
\end{equation}
for all $t\in T\cap O_\mathsf{one-cut}$ large with $\arg(t)\in(0,\pi/2)$. 
\end{definition}
 
To understand \eqref{condT} geometrically, notice that its left-hand side is equal to $0$ when $t\in\partial O_\mathsf{one-cut}\cup C_\mathsf{crit}^a\cup C_\mathsf{crit}^b$, see Figures~\ref{fig:C-curves} and~\ref{fig:loops1}(c), is positive when $t$ belongs to the part of $O_\mathsf{one-cut}$ bounded by $C_\mathsf{crit}^a\cup C_\mathsf{split}\cup C_\mathsf{crit}^b$, and is negative otherwise. Thus, \eqref{condT} describes the way $T$ is separated from $C_\mathsf{birth}$ when it extends to the point at infinity.

For functions that depend both on the parameter $t$ and variable $z$, we adopt the following conventions.

\begin{definition}
\label{def:zt-dep}
We say that the equality $f_N(z;t)=\mathcal O(N^{-\alpha})$ holds $(z,t)$-locally uniformly for $z\in V$ as $N\to\infty$ if for each $T$ such that $T\cap O_\mathsf{one-cut}$, $T\cap C_\mathsf{split}$, $T\cap C_\mathsf{birth}$ are compact, and any collection of compact sets $\{K_t\}_{t\in T}$ such that $K_t\subset V$, there exists $C(V;T)<\infty$ for which
\[
\big| f_N(z;t) \big| \leq C(V;T)N^{-\alpha}, \quad z\in K_t, \quad t\in T,
\]
for all $N$ large. Similarly, the notion of a property holding strongly $(z,t)$-locally uniformly for $z\in V$ should be straightforward in the view of Definition~\ref{def:t-dep}.
\end{definition}

\subsection{Asymptotic expansion of $F_N(t)$}

\begin{theorem}
\label{FreeEnergy}
Let $F_N(t)$ be the free energy \eqref{cm6}. Then it holds uniformly on subsets of $O_\mathsf{one-cut}$ satisfying \eqref{condT} that
\begin{equation}
\label{oc1}
F_N(t) \sim \sum_{k=0}^\infty F^{(2k)}(t)N^{-2k},
\end{equation}
where the functions $F^{(2k)}(t)$ are holomorphic in $O_\mathsf{one-cut}$ and extend holomorphically across each of the arcs $C_\mathsf{birth}$, $C_\mathsf{split}^a$, and $C_\mathsf{split}^b$. In particular, it holds that
\begin{equation}
\label{oc1a}
F^{(0)}(t) = 1 - \frac23 x^3(t) - \frac12\log\big(-2x(t)\big) + \int_\infty^t\int_\infty^\tau\left(-\frac1{2x(\sigma)} + \frac{7x^\prime(\sigma) + 2 \sigma x^{\prime\prime}(\sigma)}6 \right)\mathrm d\sigma\mathrm d\tau,
\end{equation}
where $x^\prime(t)$ is the derivative of $x(t)$ with respect to $t$ and the integrals can be computed along any path in $O_\mathsf{one-cut}$.
\end{theorem}

Functions $F^{(2k)}(t)$ encode information on the number of certain graphs on a Riemann surface of genus $k$.

\begin{remark}
If we relabel the functions $F^{(2g)}$ in \eqref{top2} by $\widehat F^{(2g)}$, then it holds that
\[
\left\{
\begin{array}{lll}
\widehat F^{(0)}(y) & := & \frac23 t^{3/2} - \frac14\log(4t) - F^{(0)}(t), \medskip \\
\widehat F^{(2g)}(y) & := & F^{(2g)}(t), \quad g\geq1,
\end{array}
\right. \quad y^{-1}=3(4t)^{3/4}.
\]
\end{remark}

We prove Theorem~\ref{FreeEnergy} in Section~\ref{s:ae} using Toda equations \eqref{cm6} and the asymptotic expansion of the recurrence coefficients.

\subsection{Asymptotic expansion of $\ga_n^2(t;N)$ and $\beta_n(t;N)$}

\begin{theorem}
\label{RecCoef}
Let $x(t)$ be as in Proposition~\ref{prop:x}. Assuming that $|n-N|\leq N_0$ for some absolute constant $N_0$, it holds that
\begin{equation}
\label{oc3}
\left\{
\begin{array}{llr}
\ga_n^2(t,N) &\sim& \displaystyle -\frac1{2x(t)}+\sum_{k=1}^\infty G_{\alpha_t k}(t;n-N) N^{-\alpha_t k}, \medskip \\
\be_n(t,N) &\sim& \displaystyle x(t)+\sum_{k=1}^\infty B_{\alpha_t k}(t;n-N) N^{-\alpha_t k},
\end{array}
\right.
\end{equation}
for some explicitly computable $G_{\alpha_t k}$ and $B_{\alpha_t k}$, where the expansions are $t$-locally uniform and
\begin{equation}
\label{oc2}
\alpha_t = 1, \quad \alpha_t=1/2, \quad \text{and} \quad \alpha_t=1/5
\end{equation}
for $t\in O_\mathsf{one-cut}$, $t\in C_\mathsf{birth}\cup C_\mathsf{split}$, and $t\in\big\{t_\mathsf{cr},e^{2\pi i/3}t_\mathsf{cr}\big\}$, respectively. Moreover, the expansion of $\ga_N^2(t,N)$ is strongly $t$-locally uniform. The functions $G_k(t;n-N)$ and $B_k(t;n-N)$ are holomorphic in $O_\mathsf{one-cut}$, can be holomorphically continued across each of the arcs $C_\mathsf{split}$, $C_\mathsf{birth}^a$, and $C_\mathsf{birth}^b$, and
\begin{equation}
\label{oc10}
\left\{
\begin{array}{ll}
G_{(2j-1)/5}(t;0) = 0, & t\in \big\{t_\mathsf{cr},e^{2\pi\mathrm i/3}t_\mathsf{cr}\big\}, \medskip \\
G_{2j-1}(t;0) \equiv 0,  & t\in O_\mathsf{one-cut}, 
\end{array}
\right. \quad j\in\N.
\end{equation}
\end{theorem}

Using \eqref{oc3} we can deduce certain analyticity properties of $\ga_N^2(t,N)$.

\begin{remark}
The partition function $Z_N(t)$ is an entire function of the parameter $t$. Hence, it follows from Toda equation \eqref{cm6} that $\ga_N^2(t,N)$ is a meromorphic function of $t$. Moreover,
\[
\ga_N^2(t,N) = -\frac{1}{2x(t)} + \mathcal O(N^{-1})
\]
strongly $t$-locally uniformly in $O_\mathsf{one-cut}$ by \eqref{oc3}. Hence, for any closed set $T\subset O_\mathsf{one-cut}$ satisfying \eqref{condT}, there exists an integer $N(T)$ such that $\ga_N^2(t,N)$ is holomorphic on $T$ for all $N\geq N(T)$, i.e., on some neighborhood of $T$ that belongs to $O_\mathsf{one-cut}$.
\end{remark}

Observe that the recurrence coefficients for different parameters $N$ are related. Indeed, given $N_1$ and $N_2$, it holds that
\begin{equation}
\label{oc14}
\left\{
\begin{array}{lll}
P_n(z;t_1,N_1) & = & (N_2/N_1)^{n/3} P_n(w;t_2,N_2), \medskip \\
\ga_n^2(t_1,N_1) & = & (N_2/N_1)^{2/3} \gamma_n^2(t_2,N_2), \medskip \\
\be_n(t_1,N_1) & = & (N_2/N_1)^{1/3} \beta_n(t_2,N_2), 
\end{array}
\right. 
\end{equation}
where $w:=(N_1/N_2)^{1/3}z$ and $t_2:=(N_1/N_2)^{2/3}t_1$, as follows from \eqref{cm3} and \eqref{cm4}. Hence, given $N$ and $t\in O_\mathsf{one-cut}$, asymptotic expansions \eqref{oc3} can be expanded to a larger range of indices $n$ than those covered by Theorem~\ref{RecCoef}. 

\begin{remark}
Put $\mathcal N:=\big\{(t,u):~t\in O_\mathsf{one-cut},~u\in\mathcal N_t\big\}$, where $\mathcal N_t$ is the largest open subset of $\C\setminus(-\infty,0]$ such that $ut \in O_\mathsf{one-cut}$ for all $u\in \mathcal N_t$ (clearly, $1\in\mathcal N_t$).  Define
\[
\left\{
\begin{array}{lll}
\widehat G_{2k}(t,u) & := & u^{3k-1}G_{2k}(ut;0), \medskip \\
\widehat B_k(t,u) & := & u^{(3k-1)/2} B_k(ut;0),
\end{array}
\right. \quad t\in O_\mathsf{one-cut} \text{ and } u\in \mathcal N_t,
\]
where we take the principal root of $u^{(3k-1)/2}$, which are holomorphic functions in $\mathcal N$. Then, it follows from \eqref{oc3}, \eqref{oc10}, and \eqref{oc14}, applied with $N_1=N$ and $N_2=n$, that
\begin{equation}
\label{oc15}
\left\{
\begin{array}{llr}
\ga_n^2(t,N) & \sim & \displaystyle \frac{-1}{2ux(ut)}+\sum_{k=1}^\infty \widehat G_{2k}(t,u) N^{-2k}, \medskip \\
\be_n(t,N) & \sim & \displaystyle \frac{x(ut)}{\sqrt u} + \sum_{k=1}^\infty \widehat B_k(t,u) N^{-k},
\end{array}
\right. \quad u=\left(\frac nN\right)^{-2/3},
\end{equation}
whenever $(n/N)^{-2/3}\in \mathcal N_t$, where the expansions are locally uniform in $t$ and $u$. 
\end{remark}

In fact, following \cite[Section~5]{BI}, we can improve on the expansion of $\beta_n(t,N)$.

\begin{theorem}
\label{RecCoef1}
There exist holomorphic in $\mathcal N$ functions $\widetilde B_{2k}(t,v)$ such that
\begin{equation}
\label{oc16}
\be_n(t,N) \sim \sum_{k=0}^\infty \widetilde B_{2k}(t,v) N^{-2k}, \quad v=\left(\frac {n+1/2}N\right)^{-2/3},
\end{equation}
whenever $((n+1/2)/N)^{-2/3}\in\mathcal N_t$, where the expansion is locally uniform in $t$ and $v$.
\end{theorem}

Theorems~\ref{RecCoef} and~\ref{RecCoef1} are proven in Section~\ref{s:ae}.

\subsection{Strong asymptotics of $P_n(z;t,N)$}

To describe the asymptotics of the orthogonal polynomials themselves, we need to introduce complexified equilibrium potential
\begin{equation}
\label{oc4}
g(z;t) := \int\log(z-s)\mathrm d\mu_t(s), \quad z\in\C\setminus\Ga_t\big(e^{\pi\mathrm i}\infty,b\big],
\end{equation}
where we take the principal branch of $\log(\cdot-s)$ holomorphic outside of $\Ga_t\big(e^{\pi\mathrm i}\infty,s\big]$ and $\mu_t$ is the equilibrium measure defined in \eqref{em6}. Since $\mu_t$ is a probability measure, it holds that
\begin{equation}
\label{oc5}
e^{g(z;t)} = z+ \mathcal{O}(1) \quad \text{as} \quad z\to\infty
\end{equation}
and this function is holomorphic in $\C\setminus J_t$. In fact, the function $e^g$ can be written explicitly.

\begin{proposition}
\label{prop:eg}
Let, as before, $a(t)$ and $b(t)$ be the endpoints of $J_t$, see \eqref{ts6}. In what follows, we set
\[
\sqrt{(z-a(t))(z-b(t))}\sim z \quad \text{as} \quad z\to\infty
\]
to be the square root with the branch cut along $J_t$. Then the function
\begin{equation}
\label{oc9}
D(z;t) := \exp\left\{\left(3V(z;t) - 2x^3(t) + \big(z^2+zx(t)-2t\big)\sqrt{(z-a(t))(z-b(t))}\right)/6\right\}
\end{equation}
is holomorphic in $\overline\C\setminus J_t$. Moreover, it is non-vanishing there, $D(\infty;t)=1$, and it has continuous traces on $J_t$ that satisfy
\begin{equation}
\label{oc11}
D_+(s;t)D_-(s;t) = \exp\left\{V(s;t)-2x^3(t)/3\right\}, \quad s\in J_t.
\end{equation}
That is, $D(z;t)$ is the Szeg\H{o} function of $V(\cdot;t)_{|J_t}$ normalized to have value $1$ at infinity. Furthermore, it holds that
\begin{equation}
\label{oc12}
e^{g(z;t)} = \frac{D(z;t)}{\sqrt{2x(t)}}\frac{A(z;t)}{B(z;t)},
\end{equation}
where the functions $A$ and $B$ are defined by
\begin{equation}
\label{oc6}
\left\{
\begin{array}{lll}
A(z;t) &:=& \displaystyle  \frac12\left(\left(\frac{z-b(t)}{z-a(t)}\right)^{1/4} + \left(\frac{z-a(t)}{z-b(t)}\right)^{1/4}\right), \medskip \\
B(z;t) &:=& \displaystyle  \frac{\mathrm i}2\left(\left(\frac{z-b(t)}{z-a(t)}\right)^{1/4} - \left(\frac{z-a(t)}{z-b(t)}\right)^{1/4}\right),
\end{array}
\right.
\end{equation}
and the branches of the $1/4$-roots are principal and have the branch cuts along $J_t$; in particular, $A(\infty;t)=1$ and $B(\infty;t)=0$. The function $F(z;t):=-\mathrm iA(z;t)/B(z;t)$ can be equivalently written as
\begin{equation}
\label{oc13}
F(z;t) = \frac2{b(t)-a(t)}\left(z-\frac{b(t)+a(t)}2 + \sqrt{(z-a(t))(z-b(t))}\right)
\end{equation}
and is holomorphic and non-vanishing in $\C\setminus J_t$, has a simple pole at infinity, and its traces on $J_t$ multiply to $1$.
\end{proposition}

We prove Proposition~\ref{prop:eg} in Section~\ref{s:g}. 

\begin{theorem}
\label{global}
Let  $\alpha_t$ be as in Theorem~\ref{FreeEnergy} and $|N-n|\leq N_0$ for some fixed constant $N_0$. Then
\begin{equation}
\label{oc7}
P_n(z;t,N) = \left(1+\mathcal{O}\big(N^{-\alpha_t}\big)\right) A(z;t) D^{N-n}(z;t)e^{ng(z;t)},
\end{equation}
$(z,t)$-locally uniformly for $z\in\C\setminus J_t$ (or $z\in\C\setminus\big(J_t\cup\{c\}\big)$ when $t\in C_\mathsf{birth}$), where $\mathcal{O}$-term vanishes at $z=\infty$. In particular, $\deg(P_n(\cdot;t,N))=n$ for all $N$ large. Moreover,
\begin{multline}
\label{oc8}
P_n(s;t,N) = \left(1+\mathcal{O}\big(N^{-\alpha_t}\big)\right) A_+(s;t)D_+^{N-n}(s;t)e^{ng_+(s;t)} + \\ + \left(1+\mathcal{O}\big(N^{-\alpha_t}\big)\right) A_-(s;t)D_-^{N-n}(s;t)e^{ng_-(s;t)},
\end{multline}
$(s,t)$-locally uniformly for $s\in \Ga_t(a,b)$ (or $s\in\Ga_t(a,c)\cup\Ga_t(c,b)$ when $t\in C_\mathsf{split}$). When $n=N$, $\mathcal O$-terms in \eqref{oc7} and \eqref{oc8} are strongly $(z,t)$-locally uniform.
\end{theorem}

Theorem~\ref{global} is proven in Section~\ref{s:aa}. Combining Theorem~\ref{global} with observation \eqref{oc14}, we obtain the following corollary.

\begin{corollary}
Given $t\in O_\mathsf{one-cut}$, assume that $u:=\lim_{N\to\infty}(n/N)^{-2/3}$ exists and $ut\in O_\mathsf{one-cut}$. Then it holds locally uniformly in $\C\setminus J_{ut}$ that
\[
u^{n/2}P_n\big(z/\sqrt u;t,N\big) = \big(1+o(1)\big) A(z;ut) e^{ng(z;ut)}.
\]
\end{corollary}

\section{S-curves}
\label{s:pr-geom}

For brevity, we set $\varpi_t:=-Q(z;t)\mathrm dz^2$. 

\subsection{Critical graphs: local structure}
\label{ss:ls}

 The differential $\varpi_t$ has two critical points of order $1$, namely $a,b$, a critical point of order $2$, namely  $c$, (unless $c$ coincides with either $a$ or $b$ in which case $\varpi_t$ has critical points of orders $1$ and $3$), and a critical point of order $-8$ at infinity. All other points are regular with respect to $\varpi_t$ (order $0$).

Through each regular point of $\varpi_t$ passes exactly one trajectory and one orthogonal trajectory, which are orthogonal to each other at the point. Two distinct (orthogonal) trajectories meet only at critical points \cite[Theorem~5.5]{Strebel}. 

As $Q(z;t)$ is a polynomial, no finite union of (orthogonal) trajectories can form a closed Jordan curve while a trajectory and an orthogonal trajectory can intersect at most once \cite[Lemmas~8.3]{Pommerenke}. Furthermore, (orthogonal) trajectories of $\varpi_t$ cannot be recurrent (dense in two-dimensional regions) \cite[Theorem~3.6]{Jenkins}. 

If $z_0\in\{a,b,c\}$ has order $m$, there are $m+2$ critical trajectories emanating from $z_0$ at angles
\begin{equation*}
\label{traj-at-crit}
\big((2k+1)\pi-\arg Q^{(m)}(z_0;t)\big)/(m+2), \quad k\in\{0,\ldots,m+1\},
\end{equation*}
see \cite[Theorem~7.1]{Strebel}. Thus, there are 3 critical trajectories of $\varpi_t$ emanating from $a$, 3 critical trajectories emanating from $b$, and 4 emanating from $c$ (under the condition $c\neq a,b$). Since the point at infinity is a pole of order $8$, there are 6 distinguished directions, namely,
\begin{equation*}
\label{traj-at-infty}
\pi/6+k\pi/3, \quad k\in\{0,\ldots,5\},
\end{equation*}
in which the trajectories can approach it. Moreover, there is a neighborhood of infinity such that every trajectory entering this neighborhood necessarily tends to infinity, \cite[Theorem~7.4]{Strebel}. The above discussion applies to the orthogonal trajectories as well. In particular, they can approach infinity only at the angles $k\pi/3$, $k\in\{0,\ldots,5\}$.

A geodesic polygon with respect to $\varpi_t$ is a Jordan curve in $\overline\C$ that consists of a finite number of trajectories and orthogonal trajectories of $\varpi_t$. According to Teichm\"uller's lemma \cite[Theorem~14.1]{Strebel}, it holds that
\begin{equation}
\label{teichmuller}
\sum_{z\in P}\left(1-\theta(z)\frac{\mathsf{ord}(z)+2}{2\pi}\right) = 2 + \sum_{z\in \mathsf{int}(P)} \mathsf{ord}(z),
\end{equation}
where $P$ is a geodesic polygon, $\mathsf{ord}(z)$ is order of $z$ with respect to $\varpi_t$, and $\theta(z)\in[0,2\pi]$, $z\in P$, is the interior angle of $P$ at $z$. Clearly, both sums in \eqref{teichmuller} are finite as only critical points of $\varpi_t$ and vertices of the polygon have a non-zero contribution.

To simplify the forthcoming discussion, let us observe that the differential $\varpi_t$ possesses several symmetries. Firstly, notice that when $t$ belongs to the subregion of $O_\mathsf{one-cut}$ bounded by $C_\mathsf{birth}^b$ and $C_\mathsf{crit}^b$, see Figures~\ref{fig:C-curves} and~\ref{fig:loops1}(c),we have
\begin{equation}
\label{diff-sym1}
x(\overline t)=\overline{x(t)} \quad \Rightarrow \quad \varpi_t(z) = \overline{\varpi_{\overline t}(\overline z)}.
\end{equation}
That is, for such $t$, the critical (orthogonal) graph coincides with the reflection across the real axis of the critical (orthogonal) graph for $\overline t$. Secondly, it holds that
\begin{equation}
\label{diff-sym2}
x\big(\overline te^{2\pi\mathrm i/3}\big)=\overline{x(t)}e^{4\pi\mathrm i/3} \quad \Rightarrow \quad \varpi_t(z) = \overline{\varpi_{\overline te^{2\pi\mathrm i/3}}\big(\overline ze^{4\pi\mathrm i/3}\big)}.
\end{equation}
That is, the critical (orthogonal) graph for $t$ coincides with the reflection across the line $L_{2\pi/3}$, see \eqref{line}, of the critical (orthogonal) graph for $\overline te^{2\pi\mathrm i/3}$ (which is the reflection of $t$ across the line $L_{\pi/3}$). Symmetries \eqref{diff-sym1} and \eqref{diff-sym2} yield that we need to concern ourselves only with the case
\begin{equation}
\label{restrict}
x\in \overline\Omega_\mathsf{one-cut} \quad \text{and} \quad 2\pi/3\leq\arg(x)\leq\pi.
\end{equation}
Notice also that \eqref{diff-sym1} and \eqref{diff-sym2} are precisely the symmetries described in Theorem~\ref{geometry1}.

\subsection{Critical graphs via level lines}

To continue, it will be convenient to observe the following. Let $R(z):=\sqrt{(z-a)(z-b)}$ be the branch holomorphic outside of some arc, say $\gamma_{ab}$, joining $a$ and $b$ and such that $R(z)=z+\mathcal{O}(1)$ as $z\to\infty$. Expressing $a$, $b$, and $c$ through $x$ via \eqref{ts6}, we have
\begin{equation}
\label{ge0}
(z+x)R(z) = 2Q^{1/2}(z;t) = z^2 + \frac1x - x^2 + \frac2z + \mathcal O\big(z^{-2}\big)
\end{equation}
as $z\to\infty$. Therefore, the function
\begin{eqnarray}
I_x(z) &: =& \int_b^z(s+x)R(s)\mathrm ds = 2\int_b^zQ^{1/2}(s;t)\mathrm ds \nonumber \\
\label{ge1}
& = & \frac13R^3(z) +x(z-x)R(z) + \log\left(\frac{z - x + R(z)}{z - x - R(z)}\right)
\end{eqnarray}
is defined up to an addition of an integer multiple of $4\pi\mathrm i$ (depending on the path of integration) and is analytic (multi-valued) in $\C\setminus\gamma_{ab}$. From the previous subsection we know that there are $3$ trajectories emanating from $a$ and the three from $b$. As there are only three finite critical points, there always exists at least one trajectory out of $a$ and at least one trajectory out of $b$ that extends to infinity. Pick one such trajectory for $a$, say $\ga_a$. Then $I_x(z)$ is a well-defined holomorphic function in $\C\setminus(\ga_a\cup\ga_{ab})$. Write,
\[
U_x(z) := \re(I_x(z)) \quad \mbox{and} \quad V_x(z) := \im(I_x(z)).
\]
Then we can see from \eqref{ge1} that $U_x(z)$ is a harmonic function in $\C\setminus\gamma_{ab}$ while $V_x(z)$ can be defined harmonically in $\C\setminus(\ga_a\cup\ga_{ab})$. Since $U_x(a)=U_x(b)=0$, the zero level set of $U_x$ contains the trajectories emanating from both $a$ and $b$ and is independent of the choice of $\ga_{ab}$ (the analytic continuation of $I_x(z)$ across $\ga_{ab}$ is given by $-I_x(z)$ that preserves the zero level set of $U_x$). Similarly, the orthogonal trajectories out of $b$ are part of the zero level set of the selected branch of $V_x$ while the orthogonal trajectories out of $a$ are part of $2\pi$ and $-2\pi$-level sets of $V_x$.

It is obvious from their definition that the harmonic functions $U_x$ continuously depend on the parameter $x$. Hence, their corresponding level sets converge to each other in Hausdorff metric on any compact subset of $\C$ (to see this around $\gamma_{ab}$, recall that $U_x$ can always be harmonically continued across $\gamma_{ab}$). Moreover, if we subtract from $U_x$ the real part of the polynomial part of the first two terms in \eqref{ge1} and $\log |z|$, the obtained function will be harmonic at infinity and will continuously depend on $x$. Therefore, we can control the behavior of the level sets of $U_x$ not only on compact subsets of $\C$ but around the point at infinity as well. Thus, if for some fixed $x_0$ all four critical trajectories out of $c=-x_0$ approach infinity, then the critical trajectories out of $c=-x$ will approach infinity in the same directions for all $x$ in a small neighborhood of $x_0$. Hence, if $x$  belongs to an open connected set on which $U_x(-x)\neq0$ (this necessarily implies that trajectories out of $c$ cannot end at $a$ or $b$), then the trajectories out $c$ approach infinity in the same directions for each $x$ on this set. Similar considerations hold for $V_x$ as well.

\subsection{Critical graphs: transitions}\label{Sec_critgraphs}

It follows from the previous subsection the structure of the critical (orthogonal) graph can change only when $U_x(c)=0$ ($V_x(c)=0,\pm2\pi$). Let us identify for which $x$ these harmonic functions vanish at $c$. From the choice of the branch of the square root we have that $R(-x)=-2x\sqrt{1+1/2x^3}$, where the root is equal to 1 when $x=\infty$. Moreover, we see from \eqref{restrict} that we are interested only in the values $\im(x^3)\geq0$. Hence,
\begin{eqnarray}
I_x(-x) & = &  -\frac{8x^3}3\left(1+\frac1{2x^3}\right)^{3/2} + 4x^3\left(1+\frac1{2x^3}\right)^{1/2} +  \log\left(\frac{1 + \sqrt{1+1/2x^3}}{1 -\sqrt{1+1/2x^3}}\right) \nonumber \\
& = & \frac23\int_{-1}^{2x^3}\left(1+\frac1s\right)^{3/2}\mathrm ds, \label{pheec}
\end{eqnarray}
where the path of integration lies in the upper half-plane. That is, we need to understand the integral \eqref{pheec} of the quadratic differential \eqref{aux-d} in the upper half plane. From the general principles, we see that $U_x(-x)=0$ ($V_x(-x)=0$) if and only if $2x^3$ belongs to a (orthogonal) trajectory emanating from $-1$. 

Differential  \eqref{aux-d} has a zero of order $3$ at $-1$, a pole of order $3$ at the origin, and a pole of order $4$ at infinity. Thus, there are $5$ trajectories emanating from $-1$, one of which is clearly $(-1,0)$. There is one distinguished approach of the origin, which is necessarily along the negative real axis since $(-1,0)$ is a trajectory. There are two distinguished directions at infinity, which are along the imaginary axis. Moreover, according to the three pole theorem \cite[Theorem~3.6]{Jenkins}, this differential does not have any recurrent trajectories. The last fact implies that the four trajectories out of $-1$ (excluding $(-1,0)$) either approach infinity or form loops. Going through the possible cases and using Teichm\"uller's lemma, we see that the trajectories emanating from $-1$ at the angles $\pm2\pi/5$ form a loop\footnote{The loop crosses the real line approximately at $0.6349131623$.}, and the other two approach infinity (it is a simple calculus exercise to see that they cannot touch the real line). Hence, Figure~\ref{fig:loops}(a) is indeed correct.

On the other hand, the local structure of the critical orthogonal trajectories near critical points must be the same. It is obvious that $(-\infty,-1)$ and $(0,\infty)$ are critical orthogonal trajectories. Thus, by repeating the same analysis, we get that the critical orthogonal graph is as on Figure~\ref{fig:loops1}(a) with the orthogonal trajectory $(0,\infty)$ not displayed. 

\begin{figure}[ht!]
\centering
\includegraphics[scale=.9]{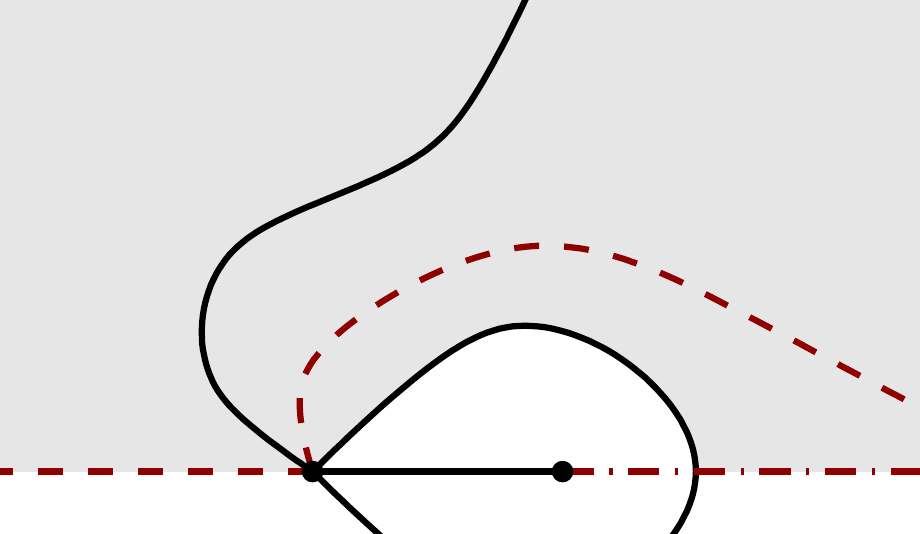}
\begin{picture}(0,0)
\put(-230,20){$V_x(c)=0$}
\put(-230,7){$U_x(c)<0$}
\put(-160,120){$V_x(c)>0$}
\put(-110,120){$U_x(c)=0$}
\put(-65,70){$V_x(c)=0$}
\put(-47,59){$U_x(c)>0$}
\put(-55,20){$V_x(c)=-\pi$}
\put(-55,7){$U_x(c)>0$}
\put(-125,58){$V_x(c)<0$}
\put(-125,39){$U_x(c)=0$}
\end{picture}
\caption{\small The values $U_x(c)$ and $V_x(c)$ ($c=-x$) when $2x^3$ belongs to the displayed trajectories and orthogonal trajectories of \eqref{aux-d}.}
\label{fig:UxVx}
\end{figure}
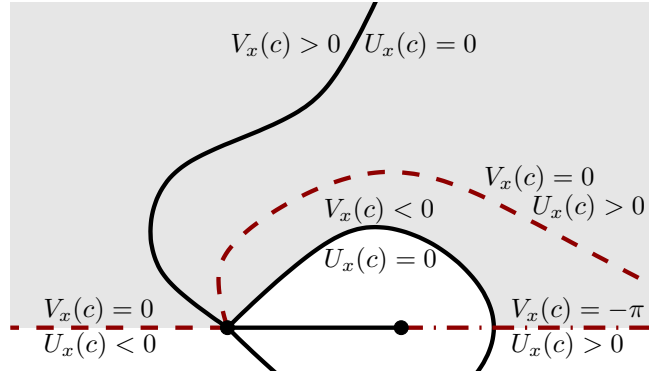

Combining the above analysis with explicit computations and \eqref{pheec}, we see that the values of $U_x(c)$ and $V_x(c)$ are as displayed on Figure~\ref{fig:UxVx}.
 
 \subsection{Critical graph: global structure}
 
\begin{figure}[ht!]
\centering
\includegraphics[scale=.9]{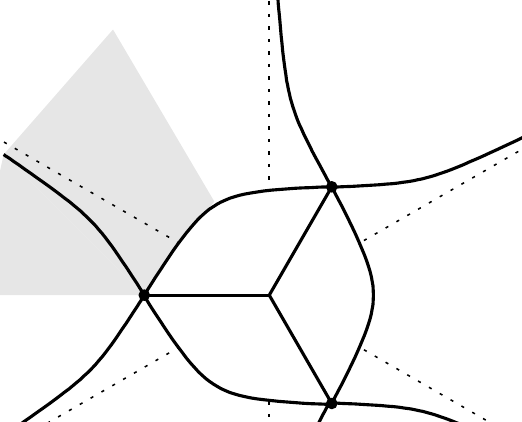}
\begin{picture}(0,0)
\put(-100,85){$\Omega_{ab}$}
\put(-40,86){$\Omega_a$}
\put(-140,30){$\Omega_b$}
\put(-82,45){$\Omega_*$}
\put(-84,65){$\Delta_\mathsf{split}$}
\put(-42,70){$\Delta_\mathsf{birth}^a$}
\put(-64,100){$\Delta_\mathsf{crit}^a$}
\put(-147,10){$\Delta_\mathsf{birth}^b$}
\put(-130,45){$\Delta_\mathsf{crit}^b$}
\end{picture}
\caption{\small Domains $\Omega_*$, $\Omega_a$, $\Omega_b$, $\Omega_{ab}$ and their boundaries; the shaded region is the one from \eqref{restrict}.}
\label{fig:Delta}
\end{figure}
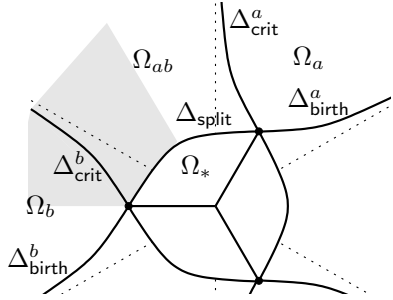

Let $\Delta_\mathsf{split}$ and $\Delta_\mathsf{birth}^e$ be as defined before \eqref{ts5}, while $\Delta_\mathsf{crit}$ as defined before Theorem~\ref{geometry1}. Write $\Delta_\mathsf{crit}=\Delta_\mathsf{crit}^a\cup\Delta_\mathsf{crit}^b$, where $\Delta_\mathsf{crit}^e$ and $\Delta_\mathsf{birth}^e$ are incident with the same point, see Figure~\ref{fig:Delta}. Then it follows from the preceding subsection that
\[
U_x(c)=0 \quad \Leftrightarrow \quad x \in \Delta_\mathsf{birth}^a \cup \Delta_\mathsf{birth}^b\cup \Delta_\mathsf{crit}^a\cup\Delta_\mathsf{crit}^b\cup\big\{-2^{1/3}\big\}\cup\big\{2^{1/3}e^{\pi\mathrm i/3}\big\}.
\]
Denote by $\Omega_e$ the subdomain of $\Omega_\mathsf{one-cut}$ bounded by $\Delta_\mathsf{birth}^e$ and $\Delta_\mathsf{crit}^e$, $e\in\{a,b\}$, and by $\Omega_{ab}$ the subdomain bounded by $\Delta_\mathsf{crit}^a$, $\Delta_\mathsf{crit}^b$, and $\Delta_\mathsf{split}$, see Figure~\ref{fig:Delta}. Recall that we only need to study the cases when $x$ satisfies \eqref{restrict}. 

Let $x\in\Omega_{ab}$. From what precedes we know that the trajectories out of $c$ approach infinity in the same four directions. The case $x\in L_{2\pi/3}$, see \eqref{line}, has been worked out in  \cite{HKL} (to obtain the setting of \cite{HKL} one needs to perform the transformations $z\mapsto e^{\pi\mathrm i/6}z$ and $t\mapsto e^{\pi\mathrm i/3}K$ in \eqref{cm2}). It was shown that there exists $r^*$ such that for $x=re^{2\pi\mathrm /3}$, $r<r^*$, there are no critical trajectory of $\varpi_t$ connecting $a$ and $b$ and for $r>r^*$ such a trajectory exists and the critical (orthogonal) graph is as on Figure~\ref{s-curves1}(a). Thus, the trajectories out of $c$ approach infinity at the angles
\[
7\pi/6+k\pi/3, \quad k\in\{0,1,2,3\},
\]
for each $x\in\Omega_{ab}$.  Now, if there always exists a trajectory connecting $a$ and $b$, the other two trajectories out of $b$ must approach infinity at the angles $\pi/6$ and $\pi/2$ and the trajectories out of $a$ must approach infinity at the angles $5\pi/6$ and $7\pi/6$ by Teichm\"uller's lemma \eqref{teichmuller}, which would finish the description of the critical graph in this case.

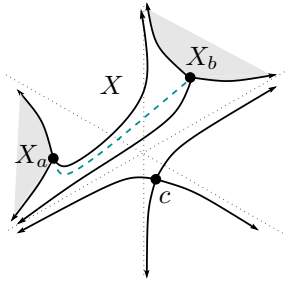
\begin{figure}[ht!]
\includegraphics[scale=.25]{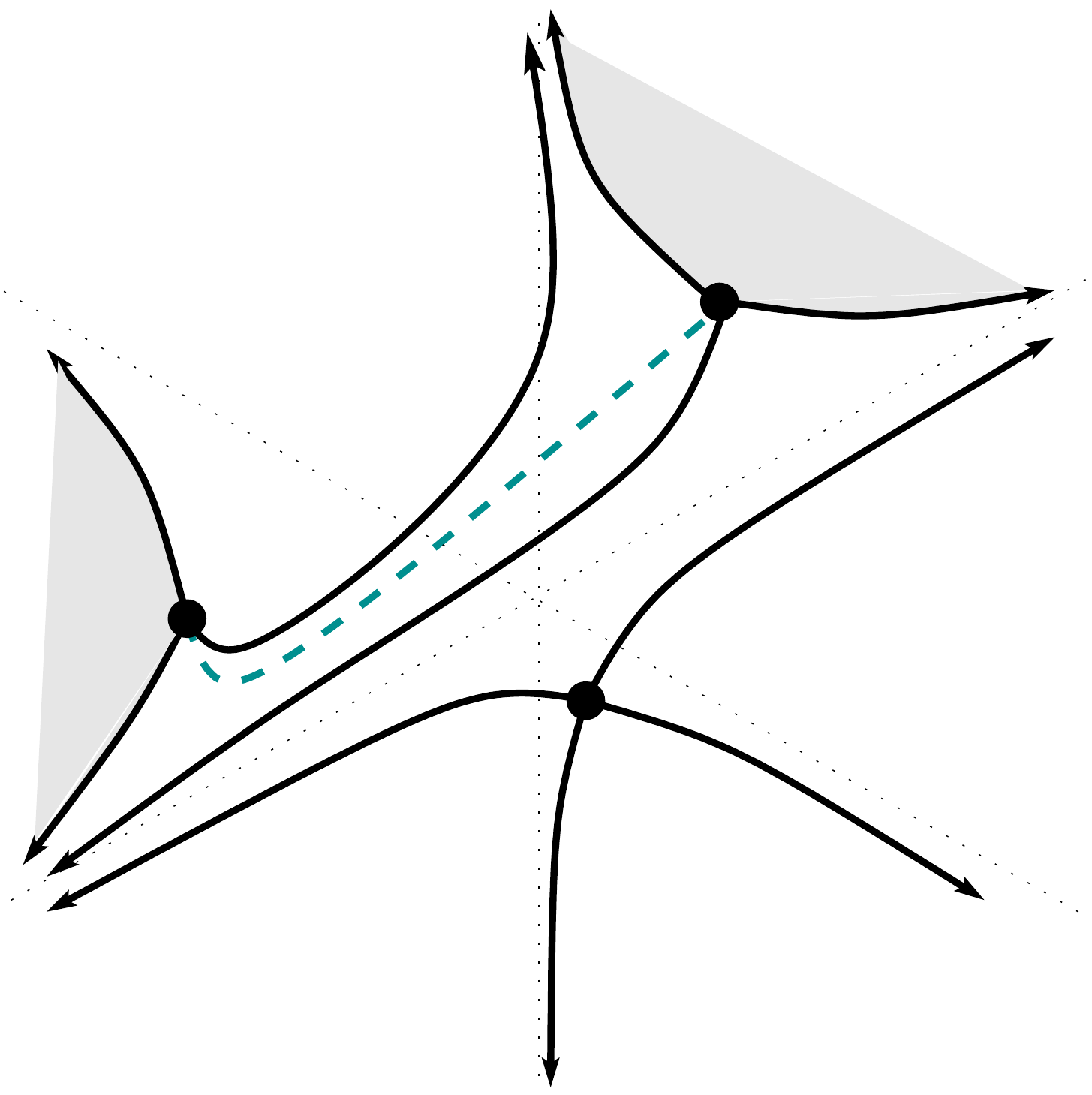}
\begin{picture}(0,0)
\put(-105,44){$X_a$}
\put(-41,81){$X_b$}
\put(-51,29){$c$}
\put(-73,70){$X$}
\end{picture}
 \caption{\small The dashed arc is $\gamma_{ab}$, the shaded regions are part of the open set $\{z:~U_x(z)<0\}$.}
 \label{fig:xOmab}
 \end{figure}

Assume to the contrary that such a trajectory does not exist. It follows from Teichm\"uller's lemma that both $a$ and $b$ must belong to the sector of opening $\pi$ at infinity delimited by the trajectories out of $c$.  In  this case two trajectories out one of the points $a,b$ will approach infinity in the directions $5\pi/6$ and $7\pi/6$, forming a sector say $X_a$, and two trajectories of the other point will approach infinity in the directions $\pi/6$ and $\pi/2$, forming  a sector say $X_b$, see Figure~\ref{fig:xOmab}. We further can choose the arc $\gamma_{ab}$ outside of $X_a\cup X_b$, that is, belonging to geodesic polygon with four corners $a,b,\infty,\infty$ and respective angles $2\pi/3,2\pi/3,0,0$, see the dashed arc on Figure~\ref{fig:xOmab}. Denote by $X$ the region that does not contain $c$ and is bounded by $\ga_{ab}$ and a part of $\partial X_a\cup\partial X_b$. As trajectories cannot intersect, $X$ contains one trajectory arc, say $T$. Recall that around each simple zero of $\varpi_t$, the differential can be written as $(3/2)^2\zeta\mathrm d\zeta^2$ for some local parameter $\zeta$, \cite[Theorem~6.1]{Strebel}. This means that $U_x(z)$ has constant sign locally in $X_a$ and $X_b$, but it also is continuous and cannot vanish there. Hence, it has a constant sign in each of these sectors. As we can always choose a branch of the logarithm in \eqref{ge1} so that $I_x(z)$ is holomorphic in the closure $X_a\cup X\cup X_b$, we have $I_x(z)=z^3(1/3+o(1))$ as $z\to\infty$ uniformly in $X_a\cup X\cup X_b$. Thus, $U_x(z)$ has the same sign in $X_a$ and $X_b$. The same local structure and continuity yield that $U_x(z)$ has the opposite sign in $X\setminus T$ and is zero on $T$ by construction. As $U_x(z)$ is harmonic in $X$, the latter contradicts the maximum principle and therefore our assumption is false. Therefore, when $x\in\Omega_{ab}$ the critical graph of $\varpi_t$ has indeed the structure as on Figure~\ref{s-curves1}(a,b,c).

Let $x\in\Omega_b$, which is also connected.  The case $x\in L_\pi\cap\overline\Omega_\mathsf{one-cut}$ has been investigated in \cite{BD1}. It was shown that the critical and critical orthogonal graphs are as on Figure~\ref{s-curves1}(f) when $x\in L_\pi \cap\Omega_b$ and Figure~\ref{s-curves2}(a) when $x=-2^{1/3}$. This fixes the behavior of the trajectories out of $c$. Arguing as in the previous paragraph, we get that $a$ and $b$ must be connected by a trajectory and therefore the behavior of the whole critical graph is fixed, see Figure~\ref{s-curves1}(e,f,g).

Let now $x\in\Delta_\mathsf{crit}^b$. Continuity with respect to parameter implies that the structure of the critical graph should be obtained through the limiting process from within both $\Omega_{ab}$ and $\Omega_b$, which necessarily yields that it must be as on Figure~\ref{s-curves1}(d).
 
\begin{figure}[ht!]
\includegraphics[scale=.25]{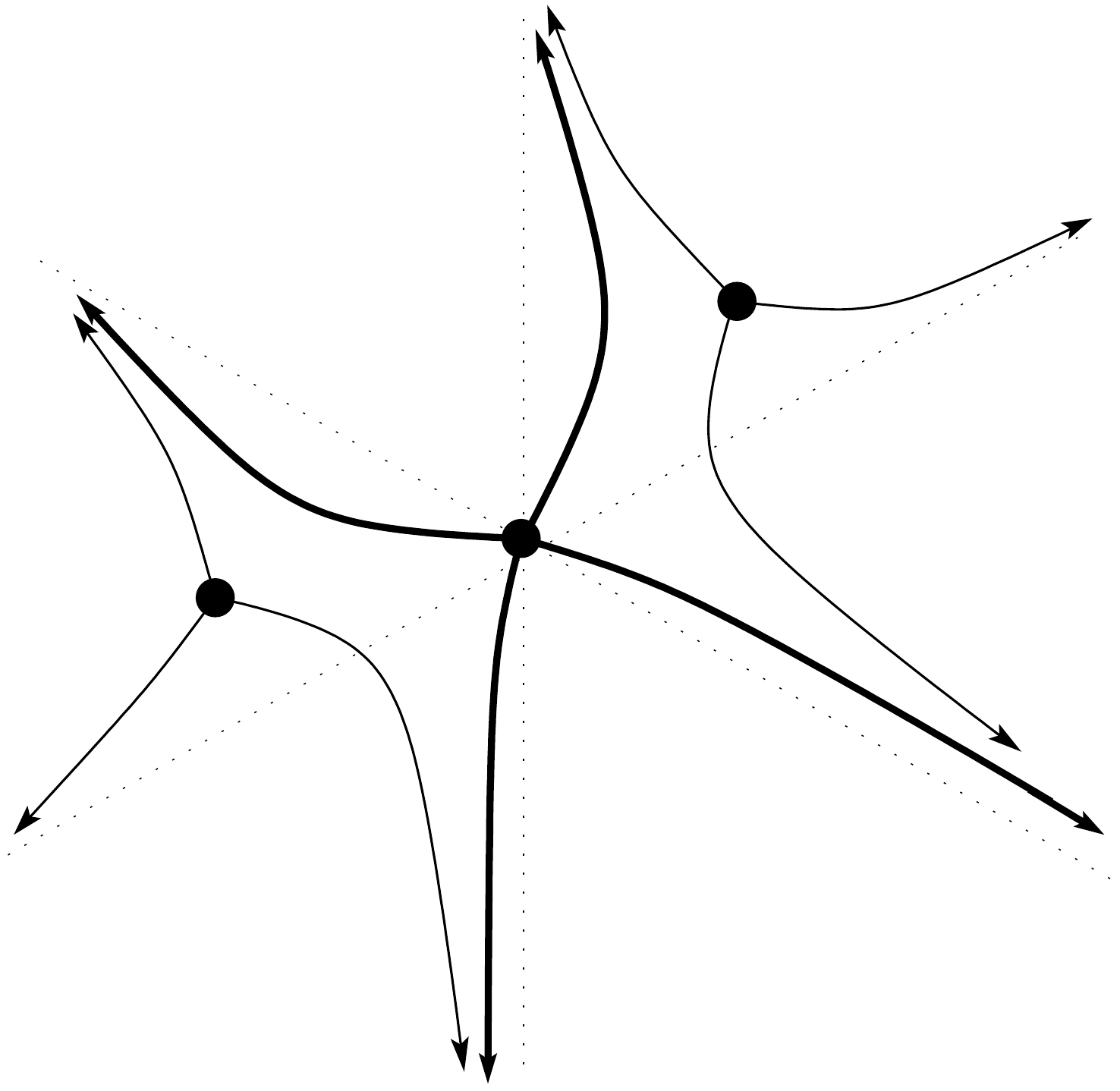}
\caption{\small The critical graph of $\varpi_t$ when $t$ crosses $\Delta_\mathsf{split}$.}
\label{fig:nc}
\end{figure}

Finally, let $x\in\Delta_\mathsf{split}$. Denote by $\Omega_*$ a domain whose boundary contains $\Delta_\mathsf{split}$ that has empty intersection with $\Omega_\mathsf{one-cut}$ and $\Delta$ (defined after \eqref{ts4}), see Figure~\ref{fig:Delta}. It was shown in \cite{HKL} that the critical graph and critical orthogonal graphs for $x\in L_{2\pi/3}\cap\Omega_*$ are as in Figure~\ref{fig:nc}.
Hence, the trajectories out of $c$ approach infinity in the directions $-\pi/6$, $\pi/2$, $5\pi/6$, and $3\pi/2$. It follows from Teichm\"uller's lemma \eqref{teichmuller} that the points $a$ and $b$ must be separated by the trajectories out of $c$. Hence, there are no trajectory joining $a$ and $b$. Continuity with respect to the parameter immediately yields that the critical graph of $\varpi_t$ is as on Figure~\ref{s-curves2}(b) when $x\in\Delta_\mathsf{split}$.
 
 \subsection{Critical orthogonal graph: global structure}

In what follows we shall refer to the \emph{key observation}:  given an unbounded domain whose boundary consists of critical trajectories that are consecutive at each point of intersection (such intersections have zero contribution to the left-hand side of \eqref{teichmuller} and the two trajectories extending to infinity necessarily form an angle of magnitude $\pi/3$ there), \eqref{teichmuller} implies that any orthogonal trajectory entering this domain will remain inside. Recall further, that orthogonal trajectories cannot intersect.

\begin{figure}[ht!]
\centering
\includegraphics[scale=1]{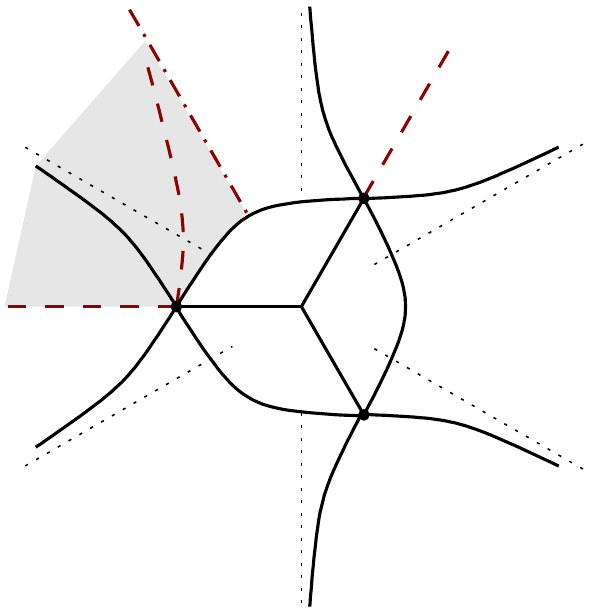}
\begin{picture}(0,0)
\put(-90,55){$V_x(c)=0$}
\put(-110,25){$V_x(c)>0$}
\put(-42,70){$V_x(c)=-\pi$}
\put(-95,7){$V_x(c)=0$}
\end{picture}
\caption{\small The dashed lines are those where $V_x(c)=0$, solid lines are those where $U_x(c)=0$, and $V_x(c)=-\pi$ on the dashed-dotted line. The shaded region is the one from \eqref{restrict}.}
\label{fig:COG}
\end{figure}
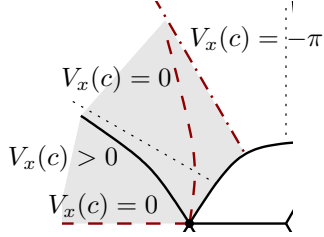

Let $x\in L_{2\pi/3}\cap\overline\Omega_\mathsf{one-cut}$. In this case the graphs must be symmetric with respect to the line $L_{2\pi/3}$ by \eqref{diff-sym2}. This symmetry, the global structure of the critical graph, and the key observation yield that the orthogonal critical graph is as on Figure~\ref{s-curves1}(a) or Figure~\ref{s-curves2}(b). 

When $x\in\Delta_\mathsf{crit}^b$, the global structure of the critical graph and key observation along fix the critical orthogonal graph to be as on Figure~\ref{s-curves1}(d), see also Figure~\ref{s-curves1}(c,e). 

Consider $x$ in the region bounded by $L_{2\pi/3}$ and $\Delta_\mathsf{crit}^b$. The critical graph is always the same, see Figure~\ref{s-curves1}(a,b,c). The key observation fixes three orthogonal trajectories out of $c$ except for the one, say $T_x$, that becomes $\Ga(c,e^{2\pi\mathrm i/3}\infty)$ when $x\in L_{2\pi/3}\cap\overline\Omega_\mathsf{one-cut}$, see Figure~\ref{s-curves1}(a), and becomes $\Ga(c,e^{\pi\mathrm i/3}\infty)$ when $x\in\Delta_\mathsf{crit}^b$, see Figure~\ref{s-curves1}(d). Notice also that fixing $T_x$ fixes the entire critical orthogonal graph as orthogonal trajectories cannot intersect. Observe also that besides becoming a short orthogonal trajectory $\Ga(c,b)$, those are the only options for $T_x$. Indeed, it could have happened that $T_x=\Ga(c,e^{\pi\mathrm i}\infty)$, but then it would necessarily hold that $T_x=\Ga(c,a)$ for some $x$ by continuity. In this case we would have $V_x(c)=-2\pi$, which is impossible as $V_x(c)\geq -\pi$ in the considered region, see Figures~\ref{fig:UxVx} and~\ref{fig:COG}.  Thus, the critical graph is as on Figure~\ref{s-curves1}(a,b,c) in the considered region. Since $V_x(c)<0$ when $T_x=\Ga(c,e^{2\pi\mathrm i/3}\infty)$, $V_x(c)=0$ when $T_x=\Ga(c,b)$, and $V_x(c)>0$ when $T_x=\Ga(c,e^{\pi\mathrm i/3}\infty)$, the corresponding claims of Theorem~\ref{geometry1} follow.

When $x\in L_\pi\cap\overline\Omega_\mathsf{one-cut}$, we know that the critical orthogonal graph must be symmetric with respect to the real axis by \eqref{diff-sym1}. This symmetry, the global structure of the critical graph, and the key observation imply that the critical orthogonal graph must be as on Figure~\ref{s-curves1}(f) or Figure~\ref{s-curves2}(a).

Finally, let $x$ belong to the region bounded by $L_\pi\cap\overline\Omega_\mathsf{one-cut}$ and $\Delta_\mathsf{crit}^b$. Since the critical graph is always the same, the critical orthogonal graph can be only as on Figure~\ref{s-curves1}(e,f,g). Continuity considerations similar to the ones above imply that it is as on Figure~\ref{s-curves1}(e) for the considered $x$. This finishes the proof of Theorem~\ref{geometry1} and therefore of Theorem~\ref{geometry}.

\section{$g$-Function}
\label{s:g}

In this section we discuss properties of $g(z;t)$ defined in \eqref{oc4}. We consider the parameter $t\in \overline O_\mathsf{one-cut}$ to be fixed and stop indicating the dependence on $t$ of the various quantities appearing below whenever this does not introduce ambiguity.

\subsection{Global properties} 

It follows directly from definition \eqref{oc4} that
\[
\partial_z g(z;t) = \int\frac{\mathrm d\mu_t(s)}{z-s},
\]
where $\partial_z:=(\partial_x-\mathrm i\partial_y)/2$. Therefore, we can deduce from \eqref{em2} and \eqref{em5} that
\begin{equation}
\label{g1}
g(z;t) = \frac{V(z;t)-\ell_t^*}2 + \int_b^zQ^{1/2}(s;t)\mathrm ds,
\end{equation}
where, as usual, we take the branch $Q^{1/2}(z;t)=\frac12z^2+\mathcal{O}(z)$, and $\ell_t^*$ is a constant such that $\re(\ell_t^*)=\ell_t$ (the explicit expression for $\ell_t^*$ can be obtained from \eqref{ge1} and the fact that $g(z;t)=\log z+\mathcal O(z^{-1})$ as $z\to\infty$). In the view of \eqref{g1}, define
\begin{equation}
\label{g2}
\phi_e(z) := 2\int_e^zQ^{1/2}(s;t)\mathrm ds, \quad e\in\{a,b\},
\end{equation}
holomorphically in $\C\setminus\Ga_t\big[b,e^{\pi\mathrm i/3}\infty\big)$ when $e=b$, and in $\C\setminus\Ga_t\big(e^{\pi\mathrm i}\infty,a\big]$ when $e=a$ (observe that \eqref{ge1} provides an explicit formula for these functions). It follows from \eqref{em6} and \eqref{g2} that
\begin{equation}
\label{g3}
\left\{
\begin{array}{ll}
\displaystyle \phi_b(z) = \phi_a(z) \pm 2\pi \mathrm i, & z\in\C\setminus\Ga_t, \medskip \\
\displaystyle \phi_{b\pm}(s) = \pm2\pi \mathrm i\mu_t\big(\Ga_t[s,b]\big), & s\in\Ga_t(a,b), \medskip \\
\end{array}
\right.
\end{equation}
where, in the first relation, the plus sign is used if $z$ lies to the left of $\Ga_t$ and the minus sign if $z$ lies to the right of $\Ga_t$. By combining \eqref{g1} and \eqref{g3} we get that
\begin{equation}
\label{g4}
g_+(s;t) - g_-(s;t) = \left\{
\begin{array}{rl}
0, & s\in\Ga_t\big(b,e^{\pi\mathrm i/3}\infty\big), \medskip\\
\pm\phi_{b\pm}(s), & s\in \Ga_t(a,b), \medskip \\
2\pi \mathrm i, & s\in\Ga_t\big(e^{\pi\mathrm i}\infty,a\big),
\end{array}
\right.
\end{equation}
and that
\begin{equation}
\label{g5}
g_+(s;t) + g_-(s;t) - V(s;t) + \ell_t^* = \left\{
\begin{array}{rl}
\phi_b(s), & s\in\Ga_t\big(b,e^{\pi\mathrm i/3}\infty\big), \medskip \\
0, & s\in \Ga_t(a,b), \medskip \\
\phi_a(s), & s\in\Ga_t\big(e^{\pi\mathrm i}\infty,a\big).
\end{array}
\right.
\end{equation}

To control the error terms in Theorem~\ref{global}, we need to have precise information  on the behavior of $\phi_e$ around $a$, $b$, and $c$, when the latter belongs to $\Ga_t$. This is exactly the goal of the following two subsections.

\subsection{Local analysis at $e\in\{a,b\}$, $e\neq c$}
\label{ss:LA}

Given $e\in\{a,b\}$, $e\neq c$, set
\begin{equation}
\label{Ue1}
U_e := \big\{z:~|z-e|< \delta_e\rho(t)|a-b|\big\},
\end{equation}
where $\delta_e\in(0,1]$ to be adjusted later and we shall specify the function $\rho(t)$ further below in Section~\ref{ssec:Ue}. Define
\begin{equation}
\label{g6}
J_e:=U_e\cap J_t \quad \text{and} \quad I_e:=U_e\cap(\Ga_t\setminus J_t),
\end{equation}
where the arcs $J_e$ and $I_e$ inherit their orientation from $\Ga_t$. Since $I_e$ is a subarc of the orthogonal trajectory of $\varpi_t=-Q(z;t)dz^2$, it holds that
\begin{equation}
\label{g7} 
\phi_e(s)<0, \quad s\in I_e.
\end{equation}
Moreover, we get from \eqref{g3} that
\begin{equation}
\label{g8}
\phi_{e\pm}(s) = \pm2\pi \mathrm i\varepsilon_e\mu_t(J_{s,e})  = 2\pi e^{\pm\varepsilon_e\frac{3\pi}2\mathrm i}\mu_t(J_{s,e}),
\end{equation}
where $J_{s,e}$ is the subarc of $J_e$ with endpoints $e$ and $s$,
\begin{equation}
\label{g9}
\varepsilon_e :=
\left\{
\begin{array}{rl}
1, & e=b, \smallskip \\
-1, & e=a,
\end{array}
\right.
\end{equation}
and the second equality in \eqref{g8} follows from \eqref{g7} and the fact $|\phi_e(z)|\sim|z-e|^{3/2}$.  Thus, we can define an analytic branch of $(-\phi_e)^{2/3}(z)$ that is positive on $I_e$. Then \eqref{g8} yields that
\[
(-\phi_e)^{2/3}_\pm(s) = -\big(2\pi \mu_t(J_{s,e})\big)^{2/3}, \quad s\in J_e,
\]
that is, $(-\phi_e)^{2/3}(z)$ is holomorphic across $J_e$. Since $(-\phi_e)^{2/3}(z)$ has a simple zero at $e$, it is conformal in some region around $e$. It will be clear from the choice of $\rho(t)$ in Section~\ref{ssec:Ue} that this region contains $U_e$. Thus, we get that $(-\phi_e)^{2/3}(z)$ maps $U_e$ conformally onto some neighborhood of the origin and satisfies
\begin{equation}
\label{g10}
\left\{
\begin{array}{lcl}
(-\phi_e)^{2/3}(J_e) &\subset& (-\infty,0), \medskip \\
(-\phi_e)^{2/3}(I_e) &\subset& (0,\infty).
\end{array}
\right.
\end{equation}
Furthermore, if we define $(-\phi_e)^{1/6}(z)$ to be holomorphic in $U_e\setminus J_e$ and positive on $I_e$, then
\begin{equation}
\label{g11}
(-\phi_e)^{1/6}_+(s) = \varepsilon_e \mathrm i(-\phi_e)^{1/6}_-(s), \quad s\in J_e.
\end{equation}

\subsection{Local analysis at $c$} 
\label{ss:LASC}

Assume that $t\in\big\{t_\mathsf{crit},e^{2\pi\mathrm i/3}t_\mathsf{crit}\big\}$. Then either $c=b$ or $c=a$. In any case $c$ is a triple zero of $Q$. Define $\varepsilon_c$ by \eqref{g9}. Define $U_c$ by \eqref{Ue1} with
\begin{equation}
\label{Ue2}
\rho(t_\mathsf{cr}):=\rho(e^{2\pi\mathrm i/3}t_\mathsf{cr}):=1/3,
\end{equation}
where again $\delta_c\in(0,1]$. Let $I_c$ and $J_c$ be as in \eqref{g6}. It follows from  \eqref{ts1} that $\phi_c(z)\sim|z-c|^{5/2}$ as $z\to c$. Since $\phi_c<0$ on $I_c$ and the angle between $I_c$ and $J_c$ is $3\pi/5$, we can define a branch of $\phi_c^{2/5}$ that is conformal around $c$, in fact, in $U_c$ (see the analysis in the next section) and is negative on $J_c$. That is, $\phi_c^{2/5}$ maps $U_c$ conformally onto some neighborhood of the origin and satisfies
\begin{equation}
\label{g12}
\left\{
\begin{array}{lcl}
\phi_c^{2/5}(J_c) &\subset& \big\{z:~\arg(z)=\varepsilon_c\pi\big\}, \medskip \\
\phi_c^{2/5}(I_c) &\subset& \big\{z:~\arg(z)=\varepsilon_c2\pi/5\big\}.
\end{array}
\right.
\end{equation}
Moreover, \eqref{g11} is replaced in this case by
\begin{equation}
\label{g13}
\phi_{c+}^{1/10}(s) = \varepsilon_c \mathrm i\phi_{c-}^{1/10}(s), \quad s\in J_c.
\end{equation}

Let now $t\in C_\mathsf{split}$. Determining the left and right sides of $\Ga_t $ by its orientation, set
\begin{equation}
\label{g14}
\varphi(z) := \left\{
\begin{array}{rl}
-\phi_b(z), & z\text{ is to the left of }\Ga_t, \medskip \\
\phi_b(z), & z\text{ is to the right of }\Ga_t.
\end{array}
\right.
\end{equation}
It follows from \eqref{g3} that $\varphi-\varphi(c)$ is holomorphic across $J_t$, vanishes at $c$, is negative purely imaginary on $\Ga_t(a,c)$, and positive purely imaginary on $\Ga_t(c,b)$. Moreover, \eqref{ts1} yields that $|\varphi(z)-\varphi(c)|\sim|z-c|^2$ as $z\to c$. Therefore, we can define a branch of $(\varphi-\varphi(c))^{1/2}$ that is conformal around $c$ and satisfies
\begin{equation}
\label{g15}
\left\{
\begin{array}{lcl}
(\varphi-\varphi(c))^{1/2}\left(\Ga_t(c,b)\right) & \subset & \big\{z:~\arg(z)=\pi/4\big\}, \medskip \\
(\varphi-\varphi(c))^{1/2}\left(\Ga_t(a,c)\right) & \subset & \big\{z:~\arg(z)=3\pi/4\big\}.
\end{array}
\right.
\end{equation}
As before, we attach a circular neighborhood to $c$ of the form
\begin{equation}
\label{Ue3}
U_c := \big\{z:~|z-c|<\delta_c\rho(t)|a-b|\big\},
\end{equation}
where, as in \eqref{Ue1}, $\rho(t)$ is a function that will be specified in the next section (in particular, it will ensure conformality of $(\varphi-\varphi(c))^{1/2}$ in $U_c$), and $\delta_c\in(0,1]$.

Finally, let us consider the case $t\in C_\mathsf{birth}^e$, $e\in\{a,b\}$. Define $\phi_c:=\phi_e-\phi_e(c)$. Notice that $\phi_c$ has a double zero at $c$ and it is real negative on $\Ga_t$ around $c$. Hence, we can select a branch of $(-\phi_c)^{1/2}$ that is conformal in $U_c$ of the form \eqref{Ue3}, satisfies
\begin{equation}
\label{g16}
(-\phi_c)^{1/2}\big(\Ga_t\cap U_c\big) \subset \R,
\end{equation}
and preserves the orientation (positive direction on $\Ga_t$ is mapped into the positive direction on $\R$).  As it will be important latter, let us also observe that $\phi_e(c)$ is purely imaginary.

\subsection{Neighborhoods $U_e$}
\label{ssec:Ue}

The goal of this section is to specify the function $\rho(t)$ appearing in the definition of the neighborhoods of $U_e$, $e\in\{a,b,c\}$, in \eqref{Ue1} and \eqref{Ue3}. We would like to show that this function can be chosen in such a fashion that the corresponding map is conformal in $U_e$ and the image of $U_e$ under this map contains a disk
\begin{equation}
\label{imageUe}
\big\{z:~|z|<\delta_e\tilde\rho(t)/32\big\},
\end{equation}
where $\tilde\rho(t)$ is a continuous positive function in $\overline O_\mathsf{one-cut}\setminus\big\{t_\mathsf{cr},e^{2\pi\mathrm i/3}t_\mathsf{cr}\big\}$ that is separated from zero when $t\to\infty$, and the constant $1/32$ is introduced for convenience only.

The main tool in showing that the above requirement can be met is the Basic Structure Theorem, see \cite[Theorem~3.5]{Jenkins}. It states in particular that the function $\phi_e(z)$, defined in \eqref{g2}, is conformal in each connected component of the complement of the joint critical graph (critical and critical orthogonal) of $\varpi_t$ (see Figures~\ref{s-curves1} and~\ref{s-curves2} for the possible configurations of this graph). Recall that $\re\phi_e(z)$ is constant on the critical trajectories and $\im\phi_e(z)$ is constant on the critical orthogonal trajectories. 
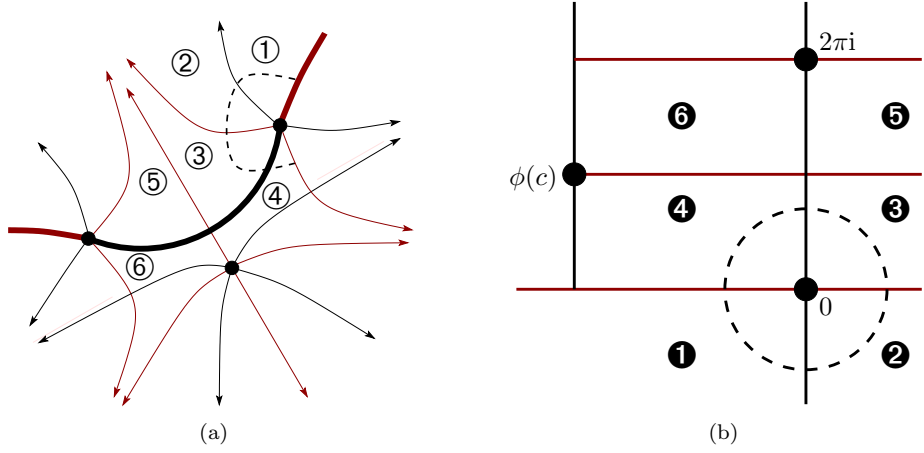
\begin{figure}[ht!]
\subfigure[]{\includegraphics[scale=.35]{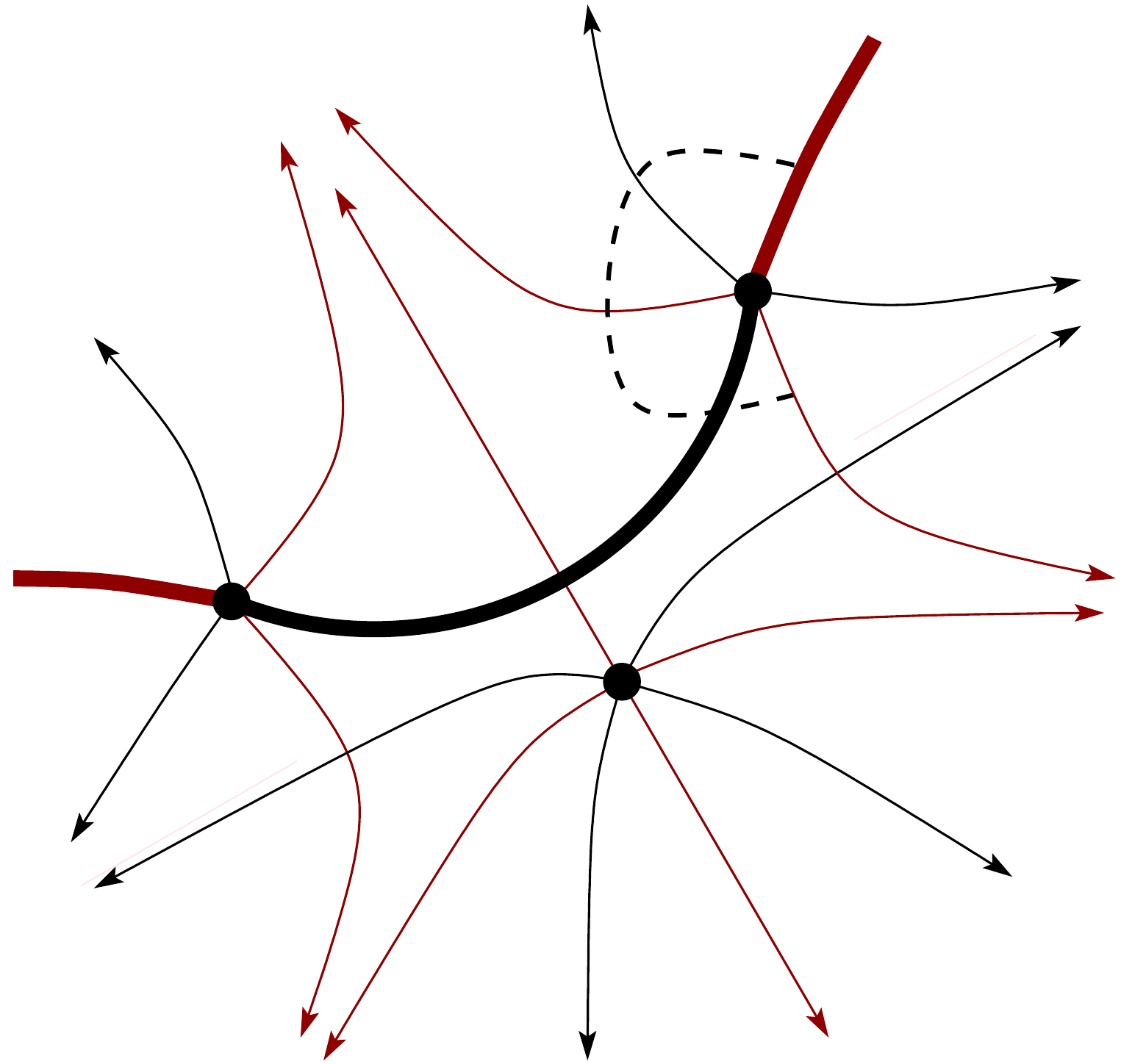}
\begin{picture}(0,0)
\put(-65,130){\LARGE\ding{192}}
\put(-95,125){\LARGE\ding{193}}
\put(-90,90){\LARGE\ding{194}}
\put(-107,80){\LARGE\ding{196}}
\put(-61,75){\LARGE\ding{195}}
\put(-112,48){\LARGE\ding{197}}
\end{picture}
}\quad\quad\quad
\subfigure[]{\includegraphics[scale=.6]{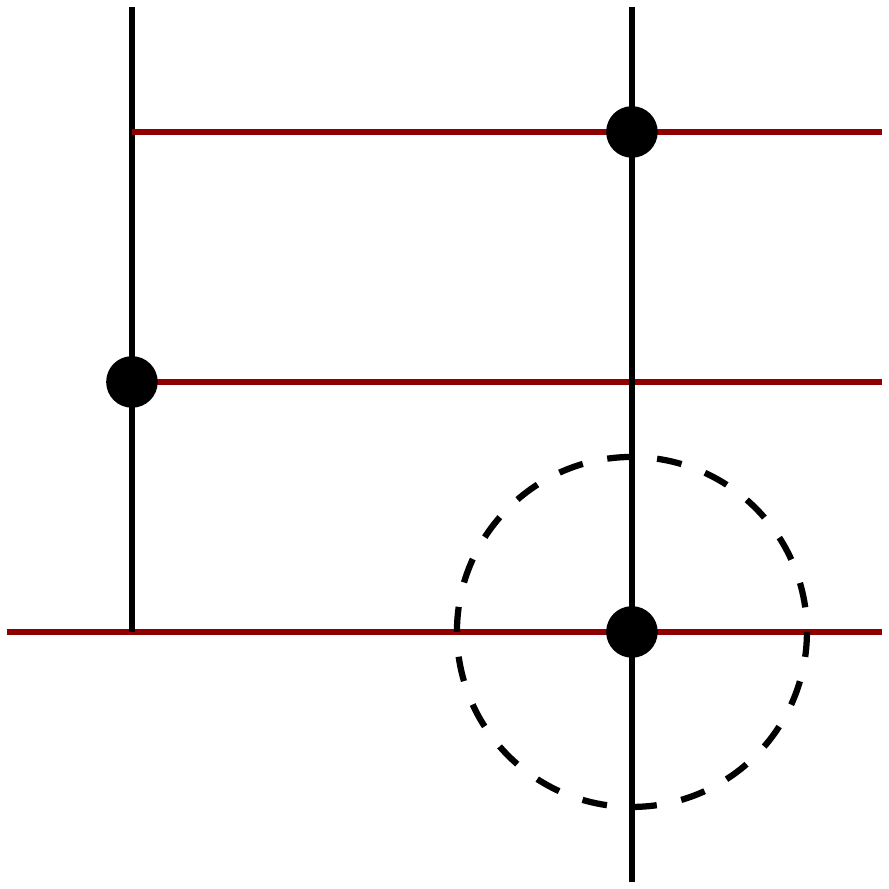}
\begin{picture}(0,0)
\put(-100,15){\LARGE\ding{202}}
\put(-20,15){\LARGE\ding{203}}
\put(-20,70){\LARGE\ding{204}}
\put(-20,105){\LARGE\ding{206}}
\put(-100,70){\LARGE\ding{205}}
\put(-100,105){\LARGE\ding{207}}
\put(-43,35){$0$}
\put(-43,135){$2\pi\mathrm i$}
\put(-159,84){$\phi(c)$}
\end{picture}
}
\caption{\small (a) The critical graph of $\varpi_t$ with some of the connected components of its complement numbered; (b) The images of the numbered components from part (a) under a certain branch of the map $\phi_b(z)$. The interior of the dashed circle from panel (b) is mapped conformally by $\phi_b^{-1}(z)$ onto the corresponding set bounded by the dashed curve and joint critical graph on panel (a).}
\label{conf-maps}
\end{figure} 
Therefore, each connected component of the complement of the joint critical graph is mapped by $\phi_e(z)$ into a quadrant, semi-infinite strip, or a rectangle, see Figure~\ref{conf-maps}. Moreover, if two such regions share a side that is the image of the same part of the joint critical graph, then $\phi_e(z)$ extends conformally through this side. As $|\phi_a(b)|=|\phi_b(a)|=2\pi$ for any $t$, the inverse of the restriction of $\phi_e$ to any connected component is conformal in the intersection of the image of the component under $\phi_e(z)$ and any disk of radius at most $\min\big\{2\pi,|\phi_e(c)|\big\}$. In particular, the inverse of $(-\phi_e)^{2/3}$ is conformal in a disk
\[
\big\{z:~|z|<\tilde\rho_e(t)/2\big\}, \quad \tilde\rho_e(t) := \left(\min\big\{2\pi,|\phi_e(c)|\big\}\right)^{2/3}.
\]
It follows from a direct computation and \eqref{ts6} that
\begin{equation}
\label{derivative}
\left|\left((-\phi_e)^{2/3}\right)^\prime(e)\right| = \frac{\sqrt[3]{96}}{|a-b|}\left|1+\frac{\mathrm i\varepsilon_e}{\sqrt 2x^{3/2}}\right|^{2/3}
\end{equation}
for any $t\neq t_\mathsf{ct},e^{2\pi\mathrm i/3}t_\mathsf{cr}$. Koebe's Quarter Theorem then yields that $(-\phi_e)^{2/3}$ is conformal in any disk \eqref{Ue1} with
\[
\rho(t) := \min\left\{\frac13,\frac{\tilde\rho(t)}{8\sqrt[3]{96}}\left|1+\frac{\mathrm i\varepsilon_e}{\sqrt 2x^{3/2}}\right|^{-2/3}\right\}, \quad \tilde\rho(t) := \min\big\{\tilde\rho_a(t),\tilde\rho_b(t)\big\},
\]
which, together with Koebe's Quarter Theorem used once more, implies that \eqref{imageUe} indeed takes place (we bound $\rho(t)$ by $1/3$ for convenience only). Notice that the rate of decay to zero of $\tilde\rho(t)$ as $t\to\big\{t_\mathsf{ct},e^{2\pi\mathrm i/3}t_\mathsf{cr}\big\}$ can be deduced from \eqref{pheec} as $|\phi_b(c)|=|I_x(-x)|$ and the fact that an analogous formula holds for $|\phi_a(c)|$.

It is not hard to see that the above argument can be applied to the conformal map around $c$ when $t\in C_\mathsf{split}\cup C_\mathsf{birth}$ to show that the conclusion \eqref{imageUe} still holds for $U_c$ as in \eqref{Ue3} with $\tilde\rho(t)$ now defined as $\min\big\{\tilde\rho_a(t),\tilde\rho_b(t),\tilde\rho_c(t)\big\}$, where
\[
\tilde\rho_c^2(t) = \left\{
\begin{array}{ll}
\min\big\{|\phi_c(a)|,|\phi_c(b)|\big\}, & t\in C_\mathsf{birth}, \medskip \\
\min\big\{|\varphi(a)-\varphi(c)|,|\varphi(b)-\varphi(c)|\big\}, &  t\in C_\mathsf{split}.
\end{array}
\right.
\]

\subsection{Functions $D(z;t)$, $A(z;t)$, and $B(z;t)$}

Let $D(z;t)$ be given by \eqref{oc9}. As in \eqref{ge0} we can compute that
\[
\sqrt{(z-a)(z-b)} = z - x + \frac1{xz} + \frac1{z^2} + \frac1{z^3}\left(x-\frac1{2x^2}\right)  + \frac1{z^4}\left(x^2 - \frac3{2x}\right) + \mathcal O\left(\frac1{z^5}\right)
\]
and therefore
\begin{equation}
\label{Dt-series}
\big(z^2+zx -2t\big) \sqrt{(z-a)(z-b)} = z^3 - 3zt + 2x^3 + \frac3{2x^2z} + \mathcal O\left(\frac1{z^3}\right),
\end{equation}
from which the analyticity and normalization at infinity follow. The equality in \eqref{oc11} is a trivial consequence of the behavior of the  square root along the branch cut.

Define $F(z)$ as in \eqref{oc13}. Since
$\left(\frac{s-b}{s-a}\right)^{1/4}_+ = \mathrm i \left(\frac{s-b}{s-a}\right)^{1/4}_-$ for $s\in J_t$, it holds that
\begin{equation}
\label{rh9}
A_\pm(s;t) = \pm B_\mp(s;t) \quad \Rightarrow \quad F_+(s)F_-(s) \equiv 1, \quad s\in J_t.
\end{equation}
Notice also that neither $A$ nor $B$ is equal to zero in $\C$. Indeed, denote by $\mathcal R$ the Riemann surface realized as two copies of $\overline\C\setminus\ J_t$ glued to each other crosswise along $J_t$. Lift $A^2$ to one of the sheets of $\mathcal R$ (a copy of $\overline\C\setminus\ J_t$) and $B^2$ to another. It follows from \eqref{rh9} that thus defined function is rational on $\mathcal R$. As it only has two poles (at the places that project to $a$ and $b$), it has exactly two zeros. Since $B^2$ has a double zero at infinity, the claim follows.

We further deduce from the previous paragraph that $F(z)$ is non-vanishing and finite in $\C\setminus J_t$ and has a simple pole at infinity. Hence, it follows from \eqref{rh9} that by lifting $F$ to one of the sheets of $\mathcal R$ and $F^{-1}$ to another, we construct a rational function with a single pole and a single zero, both projecting to the point at infinity. Clearly, a similar lift of the right-hand side of \eqref{oc13} and its reciprocal to $\mathcal R$ produces a rational function with the same properties. The normalization at infinity then gives \eqref{oc13}. Observe also that the above argument applied to the ratio of the right- and left-hand sides of \eqref{oc12} together with \eqref{oc11}, \eqref{g5},  \eqref{rh9}, and the normalization at infinity implies the validity of \eqref{oc12} as well.

For further use, let us also record several estimates. By the very definition of $U_a$ in \eqref{Ue1} and \eqref{Ue2}, we have that
\begin{equation}
\label{est-AB1}
\frac 2{3\delta_a\rho(t)}\leq\frac1{\delta_a\rho(t)}-1 \leq \left|\frac{s-b}{s-a}\right| \leq 1 + \frac1{\delta_a\rho(t)}\leq \frac 4{3\delta_a\rho(t)}, \quad s\in\partial U_a,
\end{equation}
where we used the estimate $\delta_a\rho(t)\leq 1/3$ and our convention \eqref{Ue2}. Since an analogous bound holds on $\partial U_b$, we get that
\begin{equation}
\label{est-AB2}
|A(z;t)|,|B(z;t)| \leq \big(\min\{\delta_a,\delta_b\}\rho(t)\big)^{-1/4}, \quad z\in \overline \C\setminus \big( U_a \cup U_b \big),
\end{equation}
where the bound extends outside of $U_a\cup U_b$ by the maximum modulus principle applied on the lift $\partial U_a\cup \partial U_b$ to $\mathcal R$ to the rational function on $\mathcal R$ comprised of the lifts of $A^2$ and $B^2$. Similarly, we deduce from \eqref{est-AB1} that
\begin{equation}
\label{est-AB3}
|A(s;t)|^{-1},|B(s;t)|^{-1} \leq 8\big(\max\{\delta_a,\delta_b\}\rho(t)\big)^{1/4}, \quad s\in \partial U_a \cup \partial U_b,
\end{equation}
and therefore it follows from the very definition of $F(z)$ as the left-hand side of \eqref{oc13} that
\begin{equation}
\label{est-AB4}
|F^{\pm1}(z)| \leq 8, \quad z\in \overline U_a\cup \overline U_b,
\end{equation}
independently of $t$, where we apply the argument with $\delta_a=\delta_b=1$ on the corresponding $\partial U_a\cup \partial U_b$ and then extend the bound inside by the maximum modulus principle applied on $\mathcal R$. In fact, it also holds that
\begin{equation}
\label{F-estimate}
|F^{-1}(z)| = \mathcal O(1)
\end{equation}
uniformly for $z\in\overline\C\setminus J_t$ and $t\in \overline O_\mathsf{one-cut}$. Indeed, by the maximum modulus principle and the analyticity of $F^{-1}(z)$ in $\overline\C\setminus J_t$, we only need to prove \eqref{F-estimate} for the traces $F^{-1}_\pm(s)$, $s\in J_t$. Moreover, the compactness argument shows that it is sufficient to consider only $|t|$ large. As explained in Section~\ref{ssec:Ue}, in such situations the inverse of $(-\phi_e)^{2/3}$, $e\in\{a,b\}$, is conformal in the disk of radius $(2\pi)^{2/3}$. Moreover, $(-\phi_a)^{2/3}(b)$ or $(-\phi_b)^{2/3}(a)$, depending on whether $e=a$ or $e=b$, belongs to the boundary of this disk. Hence,
\[
J_t\subset \left((-\phi_a)^{2/3}\right)^{-1}(U)\cup \left((-\phi_b)^{2/3}\right)^{-1}(U), \quad U:= \left\{z:~|z|<2(2\pi)^{2/3}/3\right\}.
\] 
Then it follows from Koebe's distortion theorem and \eqref{derivative} that
\[
|s-e| \leq \mathsf{const}|a-b|, \quad s\in J_t\cap\left((-\phi_e)^{2/3}\right)^{-1}(U),
\]
for some absolute constant. Therefore, $|s-e| \leq \mathsf{const}|a-b|$ for all $s\in J_t$ and $e\in\{a,b\}$. This estimate and the explicit expression
\[
F^{-1}(z) = \frac2{b-a}\left(z-\frac{b+a}2-\sqrt{(z-a)(z-b)}\right)
\]
immediately imply the desired bound on $J_t$.

\section{Asymptotic Analysis}
\label{s:aa}

\subsection{Initial Riemann-Hilbert problem}

In what follows, it will be convenient to set
\[
\boldsymbol I:=\left(\begin{matrix} 1 & 0 \\ 0 & 1 \end{matrix}\right), \quad \sigma_1:=\left(\begin{matrix} 0 & 1 \\ 1 & 0 \end{matrix}\right), \quad \text{and} \quad \sigma_3:=\left(\begin{matrix} 1 & 0 \\ 0 & -1 \end{matrix}\right).
\]
We are seeking solutions of the following sequence of Riemann-Hilbert problems for $2\times2$ matrix-valued functions (\rhy):
\begin{itemize}
\label{rhy}
\item[(a)] $\boldsymbol Y$ is analytic in $\C\setminus\Ga_t$ and $\lim_{\C\setminus\Ga_t\ni z\to\infty}\boldsymbol Y(z)z^{-n\sigma_3}=\boldsymbol I$;
\item[(b)] $\boldsymbol Y$ has continuous traces on $\Ga_t\setminus\{a,b,c\}$ that satisfy
\[
\boldsymbol Y_+(s) = \boldsymbol Y_-(s) \left(\begin{matrix} 1 & e^{-NV(s;t)} \\ 0 & 1 \end{matrix}\right).
\]
\end{itemize}

The connection of \hyperref[rhy]{\rhy} to orthogonal polynomials was first demonstrated by Fokas, Its, and Kitaev in \cite{FIK} and lies in the following. If the solution of \hyperref[rhy]{\rhy} exists, then it is necessarily of the form
\begin{equation}
\label{rh1}
\boldsymbol Y(z) = \left(\begin{matrix}
P_n(z) & \big(\mathcal{C}P_ne^{-NV}\big)(z) \medskip \\
-\frac{2\pi \mathrm i}{h_{n-1}}P_{n-1}(z) & -\frac{2\pi \mathrm i}{h_{n-1}}\big(\mathcal{C}P_{n-1}e^{-NV}\big)(z)
\end{matrix}\right),
\end{equation}
where $P_n$ is the polynomial satisfying orthogonality relations \eqref{cm3}, $h_k$ are constants defined by \eqref{cm5}, and $\mathcal{C}f$ is the Cauchy transform of a function $f$ given on $\Ga_t$, i.e.,
\[
(\mathcal{C}f)(z) = \frac1{2\pi \mathrm i}\int_{\Ga_t}\frac{f(s)}{s-z}ds.
\]

Below, we show the solvability of \hyperref[rhy]{\rhy} for all $|n-N|\leq N_0$ and $N$ large enough following the framework of the steepest descent analysis introduced by Dieft and Zhou \cite{DZ}. The latter lies in a series of transformations which reduce the initial problem to a problem with jumps asymptotically close to the identity.

\subsection{Renormalized Riemann-Hilbert problem}

Suppose that $\boldsymbol Y$ is a solution of \hyperref[rhy]{\rhy}. Put
\begin{equation}
\label{rh2}
\boldsymbol T := e^{N\ell_t^*\sigma_3/2}\left(\frac4{b-a}\right)^{(n-N)\sigma_3}\boldsymbol Y(z) e^{-N\big(g+\ell_t^*/2\big)\sigma_3}F^{(N-n)\sigma_3},
\end{equation}
where the function $g$ is defined by \eqref{oc4}, $\ell_t^*$ is introduced in \eqref{g1}, and $F=-\mathrm iA/B$ is a function from \eqref{oc13}. Then
\[
\boldsymbol T_+ = \boldsymbol T_- \left(\begin{matrix} (F_+/F_-)^{N-n}e^{-N(g_+-g_-)} & (F_+F_-)^{n-N}e^{N(g_+ + g_- -V+\ell_t^*)} \\ 0 & (F_-/F_+)^{N-n}e^{-N(g_- - g_+)} \end{matrix}\right),
\]
on $\Ga_t$, and therefore we deduce from \eqref{oc5}, \eqref{oc13}, \eqref{g4}, \eqref{g5}, and \eqref{rh9} that $\boldsymbol T$ solves \rht:
\begin{itemize}
\label{rht}
\item[(a)] $\boldsymbol T$ is analytic in $\C\setminus\Ga_t$ and $\lim_{\C\setminus\Ga_t\ni z\to\infty}\boldsymbol T(z)=\boldsymbol I$;
\item[(b)] $\boldsymbol T$ has continuous traces on $\Ga_t\setminus\{a,b,c\}$ that satisfy
\[
\boldsymbol T_+ = \boldsymbol T_- \left\{
\begin{array}{rll}
\left(\begin{matrix} 1 & F^{2(n-N)}e^{N\phi_b} \\ 0 & 1 \end{matrix}\right), & \text{on} & \Ga_t\big(b,e^{\pi\mathrm i/3}\infty\big), \medskip \\
\left(\begin{matrix} F_+^{2(N-n)}e^{-N\phi_{b+}} & 1 \\ 0 & F_-^{2(n-N)}e^{-N\phi_{b-}} \end{matrix}\right), & \text{on} & \Ga_t(a,b), \medskip \\
\left(\begin{matrix} 1 & F^{2(n-N)}e^{N\phi_a} \\ 0 & 1 \end{matrix}\right), & \text{on} &\Ga_t\big(e^{\pi\mathrm i}\infty,a\big).
\end{array}
\right.
\]
\end{itemize}

Clearly, if \hyperref[rht]{\rht} is solvable and $\boldsymbol T$ is the solution, then by inverting \eqref{rh2} one obtains a matrix $\boldsymbol Y$ that solves \hyperref[rhy]{\rhy}.

\subsection{Lens opening} 

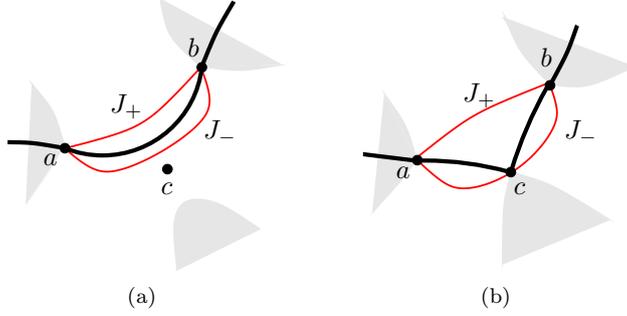
\begin{figure}[!h]
\subfigure[]{\includegraphics[scale=.25]{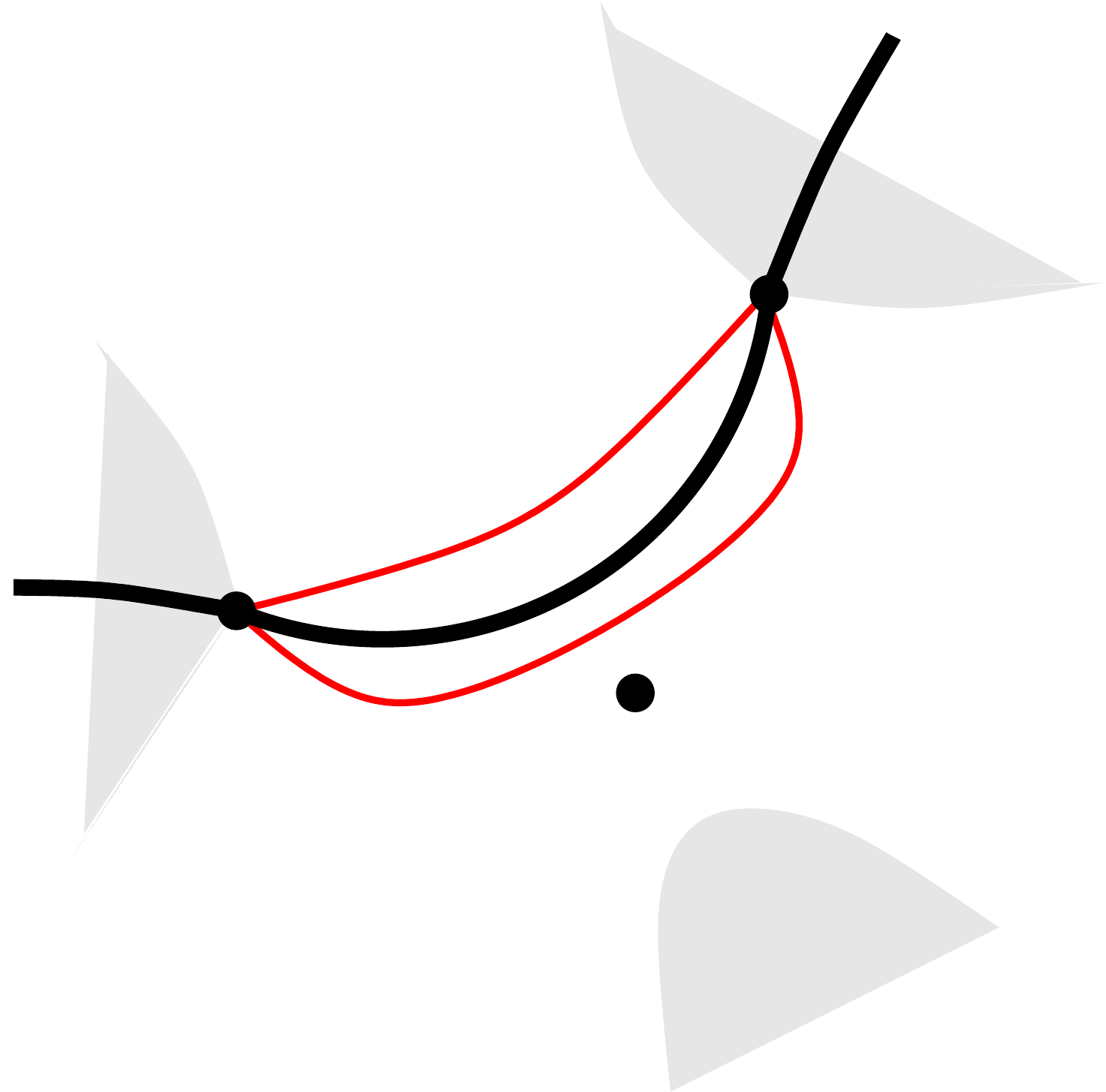}}
\begin{picture}(0,0)
\put(-95,40){$a$}
\put(-41,81){$b$}
\put(-51,29){$c$}
\put(-70,60){$J_+$}
\put(-35,50){$J_-$}
\end{picture}
\qquad
\subfigure[]{\includegraphics[scale=.25]{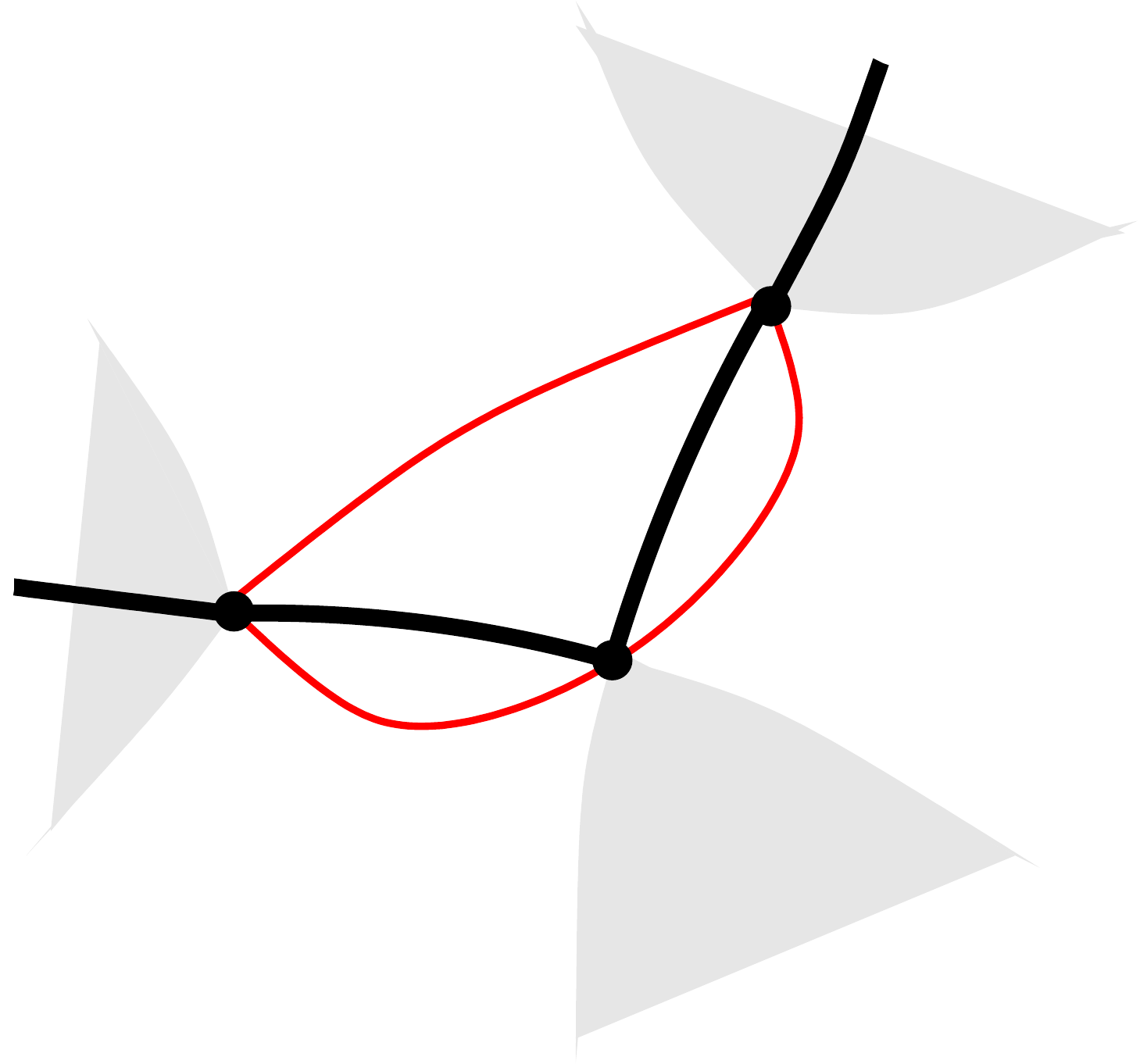}}
\begin{picture}(0,0)
\put(-95,35){$a$}
\put(-41,77){$b$}
\put(-51,29){$c$}
\put(-70,64){$J_+$}
\put(-32,50){$J_-$}
\end{picture}
\caption{\small The thick arcs represent $\Ga_t$ and thiner arcs represent $J_\pm$.  Shaded region is the set where $\re\big(\phi_b(z)\big)<0$.}
\label{lens}
 \end{figure}

As usual in the steepest descent analysis of matrix Riemann-Hilbert problems for orthogonal polynomials, the next step is based on the identity
\[
\left(\begin{matrix} a_+ & 1 \\ 0 & a_- \end{matrix}\right) =  \left(\begin{matrix} 1 & 0 \\ a_- & 1 \end{matrix}\right) \left(\begin{matrix} 0 & 1 \\ -1 & 0 \end{matrix}\right)\left(\begin{matrix} 1 & 0 \\ a_+ & 1 \end{matrix}\right), \quad a_+a_-\equiv1,
\]
which is applicable by \eqref{g3} and \eqref{rh9}. To carry it out, we shall introduce two additional arcs. Denote by $J_\pm$ smooth homotopic deformations of $J_t$ within the region $\re(\phi_b(z))>0$ such that $J_+$ lies to the left and $J_-$ to the right of $J_t$, see Figure~\ref{lens}.
 Moreover, we shall fix the way these arcs emanate from $e\in\{a,b\}$. Namely, let $U_e$ be a disk centered at $e$ as described in Sections~\ref{ss:LA}--\ref{ssec:Ue}. Assume first that we are in a generic situation when $e\neq c$. Then we require that
\begin{equation}
\label{rh3}
\arg\big((-\phi_e)^{2/3}(z)\big) = \pm\varepsilon_e(2\pi/3), \quad z\in U_e\cap J_\pm,
\end{equation}
where $\varepsilon_e$ is defined by \eqref{g9}. The latter is always possible due to \eqref{g10}. Suppose now that $e=c\in\{a,b\}$. Then we require that 
\begin{equation}
\label{rh4}
\arg\big(\phi_c^{2/5}(z)\big) = \pm\varepsilon_c(4\pi/5), \quad z\in U_c\cap J_\pm,
\end{equation}
where $\varepsilon_c$ is still defined by \eqref{g9} and such a choice is possible according to \eqref{g12}. In addition, when $c\in J_t\setminus\{a,b\}$, it necessarily holds that $J_-$ touches $J_t$ at $c$. We shall choose $J_-$ around $c$ so that
\begin{equation}
\label{rh5}
(\varphi-\varphi(c))^{1/2}(U_c\cap J_-) \subset \R,
\end{equation}
where $\varphi$ is defined by \eqref{g14} and such a choice is possible due to \eqref{g15}. 

Denote by $O_\pm$ the open sets delimited by $J_\pm$ and $J_t$. Set
\begin{equation}
\label{rh6}
\boldsymbol S(z) := \boldsymbol T(z) \left\{
\begin{array}{ll}
\left(\begin{matrix} 1 & 0 \\ \mp F^{2(N-n)}(z)e^{-N\phi_b(z)} & 1 \end{matrix}\right), & z\in O_\pm, \medskip \\
\boldsymbol I, & \text{otherwise}.
\end{array}
\right.
\end{equation}
Then, if $\boldsymbol T$ solves \hyperref[rht]{\rht}, $\boldsymbol S$ solves \rhs:
\begin{itemize}
\label{rhs}
\item[(a)] $\boldsymbol S$ is analytic in $\C\setminus(\Ga_t\cup J_+\cup J_-)$ and $\lim_{\C\setminus\Ga_t\ni z\to\infty}\boldsymbol S(z)=\boldsymbol I$;
\item[(b)] $\boldsymbol S$ has continuous traces on $\Ga_t\setminus\{a,b,c\}$ that satisfy \hyperref[rht]{\rht}(b) on $\Ga_t\big(e^{\pi\mathrm i}\infty,a\big)$ and $\Ga_t\big(b,e^{\pi\mathrm i/3}\infty\big)$, as well as
\[
\boldsymbol S_+(s) = \boldsymbol S_-(s) \left\{
\begin{array}{rl}
\left(\begin{matrix} 0 & 1 \\ -1 & 0 \end{matrix}\right), & s\in J_t, \medskip \\
\left(\begin{matrix} 1 & 0 \\ F^{2(N-n)}(s)e^{-N\phi_b(s)} &1\end{matrix}\right), & s\in J_\pm.
\end{array}
\right.
\]
\end{itemize}
As before, since transformation \eqref{rh6} is invertible, a solution of \hyperref[rhs]{\rhs} yields a solution of \hyperref[rht]{\rht}.

\subsection{Model solution} 

The model Riemann-Hilbert problem \rhm~ is obtained from \hyperref[rhs]{\rhs} by removing from the jump matrices in \hyperref[rhs]{\rhs}(b) the quantities that are asymptotically zero. Thus, we are seeking the solution of \rhm:
\begin{itemize}
\label{rhm}
\item[(a)] $\boldsymbol M$ is analytic in $\overline\C\setminus J_t$ and $\boldsymbol M(\infty)=\boldsymbol I$;
\item[(b)] $\boldsymbol M$ has continuous traces on $\Ga_t(a,b)$ that satisfy
\[
\boldsymbol M_+(s) = \boldsymbol M_-(s) \left(\begin{matrix} 0 & 1 \\ -1 & 0 \end{matrix}\right), \quad s\in \Ga_t(a,b).
\]
\end{itemize}

Using \eqref{rh9}, one can easily verify that \hyperref[rhm]{\rhm} is solved by
\begin{equation}
\label{rh10}
\boldsymbol M(z) := \left(\begin{matrix} A(z;t) & -B(z;t) \\ B(z;t) & A(z;t) \end{matrix}\right),
\end{equation}
where  $A(z;t)$ and $B(z;t)$ are given by \eqref{oc6}. Observe also that $\det(\boldsymbol M)\equiv1$ in $\C$.

\subsection{Local parametrices} 

The jumps discarded in \hyperref[rhm]{\rhm} are not uniformly close to the identity around $e\in\{a,b,c\}$ (the point $e=c$ is included when $t\in\partial O_\mathsf{one-cut}$). The goal of this section is to solve \hyperref[rhs]{\rhs} within the disks $U_e$ introduced in Sections~\ref{ss:LA}--\ref{ssec:Ue} with a certain matching condition on $\partial U_e$. More precisely, given $e\in\{a,b,c\}$, we are looking for a matrix-valued function $\boldsymbol P_e$ that solves \rhp:
\begin{itemize}
\label{rhp}
\item[(a)] $\boldsymbol P_e$ has the same analyticity properties as $\boldsymbol S$ restricted to $U_e$, see \hyperref[rhs]{\rhs}(a);
\item[(b)] $\boldsymbol P_e$ satisfies the same jump relations as $\boldsymbol S$ restricted to $U_e$, see \hyperref[rhs]{\rhs}(b);
\item[(c)] $\boldsymbol P_e=\boldsymbol M\big(\boldsymbol I+\boldsymbol{\mathcal{O}}(N^{-\alpha_e})\big)$ holds uniformly on $\partial U_e$ as $N\to\infty$ for some $\alpha_e>0$.
\end{itemize}

\subsubsection{Parametrix $\boldsymbol P_e$ around $e\in\{a,b\}$, $e\neq c$}

Let $U_e$, $J_e$, and $I_e$ be as in \eqref{g6}. In this section we are looking for a matrix function $\boldsymbol P_e$ that is holomorphic in $U_e\setminus(\Ga_t\cup J_+\cup J_-)$, fulfills \hyperref[rhp]{\rhp}(c), and whose traces satisfy
\begin{equation}
\label{rh11}
\boldsymbol P_{e+}(s) = \boldsymbol P_{e-}(s) \left\{
\begin{array}{rl}
\left(\begin{matrix} 0 & 1 \\ -1 & 0 \end{matrix}\right), & s\in J_e, \medskip \\
\left(\begin{matrix} 1 & 0 \\ F^{2(N-n)}(s)e^{-N\phi_e(s)} &1\end{matrix}\right), & s\in J_\pm\cap U_e, \medskip \\
\left(\begin{matrix} 1 & F^{2(n-N)}(s)e^{N\phi_e(s)} \\ 0 & 1 \end{matrix}\right), & s\in I_e.
\end{array}
\right.
\end{equation}
Notice that we replaced $\phi_b$ by $\phi_e$ as compared to \hyperref[rhs]{\rhs}(b). Such a substitution is possible due to the first relation in \eqref{g3}.

Let $\boldsymbol A$ be the Airy matrix \cite{DKMVZ,DKMVZ2}. That is, it is analytic in $\C\setminus\big((-\infty,\infty)\cup L_-\cup L_+\big)$,  $L_\pm:=\big\{z:~\arg(z)=\pm2\pi/3\big\}$, and satisfies
\[
\boldsymbol A_+(s) = \boldsymbol A_-(s) \left\{
\begin{array}{rl}
\left(\begin{matrix} 0 & 1 \\ -1 & 0 \end{matrix}\right), & s\in (-\infty,0), \medskip \\
\left(\begin{matrix} 1 & 0 \\ 1 & 1 \end{matrix}\right), & s\in L_\pm, \medskip \\
\left(\begin{matrix} 1 & 1 \\ 0 & 1 \end{matrix}\right), & s\in (0,\infty),
\end{array}
\right.
\]
where the real line is oriented from $-\infty$ to $\infty$ and the rays $L_\pm$ are oriented towards the origin. It is known that $\boldsymbol A$ has the following asymptotic expansion at infinity:
\begin{equation}
\label{rh12}
\boldsymbol A(\zeta)e^{\frac23\zeta^{3/2}\sigma_3} \sim \frac{\zeta^{-\sigma_3/4}}{\sqrt2}\sum_{k=0}^\infty \left(\begin{matrix} s_k & 0 \\ 0 & t_k \end{matrix}\right)\left(\begin{matrix} (-1)^k & \mathrm i \\ (-1)^k \mathrm i & 1 \end{matrix}\right)  \left(\frac23\zeta^{3/2}\right)^{-k},
\end{equation}
where the expansion holds uniformly in $\C\setminus\big((-\infty,\infty)\cup L_-\cup L_+\big)$, and
\[
s_0=t_0=1, \quad s_k=\frac{\Gamma(3k+1/2)}{54^kk!\Gamma(k+1/2)}, \quad t_k=-\frac{6k+1}{6k-1}s_k, \quad k\geq1.
\]

Set $\boldsymbol A_b:=\boldsymbol A$ and $\boldsymbol A_a:=\sigma_3\boldsymbol A\sigma_3$. It can be easily checked that $\sigma_3\boldsymbol A\sigma_3$ has the same jumps as $\boldsymbol A$ only with the reversed orientation of the real line and the rays $L_\pm$. Moreover, one needs to replace each occurrence of  $\mathrm i$ by $-\mathrm i$ in \eqref{rh12} when describing the asymptotic expansion $\sigma_3\boldsymbol A\sigma_3$. Then \eqref{g3}, \eqref{g10}, and \eqref{rh3}  yield that the matrix function
\begin{equation}
\label{rh13}
\boldsymbol P_e(z) = \boldsymbol E_e(z)\boldsymbol A_e\left(\big[-(3/4)N\phi_e(z)\big]^{2/3}\right)e^{-N\phi_e(z)\sigma_3/2}F^{(N-n)\sigma_3}(z)
\end{equation}
satisfies  \hyperref[rhp]{\rhp}(a) and \hyperref[rhp]{\rhp}(b) for any matrix function $\boldsymbol E_e$ holomorphic in $U_e$. Thus, we only need to choose $\boldsymbol E_e$ so that \hyperref[rhp]{\rhp}(c) is fulfilled. Choose
\[
\boldsymbol E_e(z) := \boldsymbol M(z)F^{(n-N)\sigma_3}(z) \left(\begin{matrix} 1 & -\varepsilon_e \mathrm i \\ -\varepsilon_e \mathrm i & 1 \end{matrix}\right) \frac{\big[-(3/4)N\phi_e(z)\big]^{\sigma_3/6}}{\sqrt2}
\]
whose analyticity in $U_e$ follows \hyperref[rhm]{\rhm}(b), \eqref{rh9}, and \eqref{g11} with $\varepsilon_e$ given by \eqref{g9}. Then we deduce from \eqref{rh12} and \eqref{rh13} that
\begin{equation}
\label{rh14}
\boldsymbol P_e(z) \sim \boldsymbol M(z)F^{(n-N)\sigma_3}(z) \left(\boldsymbol I + N^{-\alpha_e}\sum_{k=0}^\infty \boldsymbol P_{e,k}(z) N^{-k} \right)F^{(N-n)\sigma_3}(z),
\end{equation}
where the expansion inside the parenthesis holds uniformly on $\partial U_e$ and locally uniformly for $t\in\overline O_\mathsf{one-cut}\setminus\big\{t_\mathsf{cr},e^{2\pi\mathrm i/3}t_\mathsf{cr}\big\}$ by \eqref{imageUe}, the parameter $\alpha_e=1$, and
\begin{equation}
\label{rh15}
\boldsymbol P_{e,k-1}(z) = \left(\begin{matrix} 1 & -\varepsilon_e \mathrm i \\ -\varepsilon_e \mathrm i & 1 \end{matrix}\right) \left(\begin{matrix} s_k & 0 \\ 0 & t_k \end{matrix}\right)\left(\begin{matrix} (-1)^k & \varepsilon_e \mathrm i \\ \varepsilon_e(-1)^k \mathrm i & 1 \end{matrix}\right) \left(-\frac{\phi_e(z)}2\right)^{-k}, \quad k\geq1.
\end{equation}

\subsubsection{Parametrix $\boldsymbol P_c$ around $c\in\{a,b\}$}

The local problem at $e=c$ is formulated exactly as before with the jumps as in \eqref{rh11}. However, the above solution does not apply because $(-\phi_c)^{2/3}$ is no longer conformal (we shall replace it by $\phi_c^{2/5}$) and the arcs $I_c$ and $J_c$ no longer form an angle $\pi$ at $c$ (it is $3\pi/5$). 

As in the previous subsection, we shall need an auxiliary matrix-valued function. This time it depends on two parameters: $\alpha,\lambda\in\C$, and solves the following Riemann-Hilbert problem (\rhpsi): 
\begin{itemize}
\label{rhpsi}
\item[(a)] $\boldsymbol\Psi_\alpha$ is a holomorphic matrix function in $\C\setminus\big((-\infty,0]\cup L_{1-}\cup L_{1+}\cup L_{2-}\cup L_{2+}\big)$,  where $L_{k\pm}:=\big\{z:~\arg(z)=\pm2k\pi/5\big\}$, the rays $(-\infty,0)$ and $L_{2\pm}$ are oriented towards the origin while $L_{1\pm}$ are oriented away from the origin;
\item[(b)] $\boldsymbol\Psi_\alpha$ has continuous traces on $(-\infty,0)\cup L_{1-}\cup L_{1+}\cup L_{2-}\cup L_{2+}$ that satisfy
\[
\boldsymbol\Psi_{\alpha+}(s;\lambda) = \boldsymbol\Psi_{\alpha-}(s;\lambda) \left\{
\begin{array}{rl}
\left(\begin{matrix} 0 & 1 \\ -1 & 0 \end{matrix}\right), & s\in (-\infty,0), \medskip \\
\left(\begin{matrix} 1 & 0 \\ 1 & 1 \end{matrix}\right), & s\in L_{2\pm}, \medskip \\
\left(\begin{matrix} 1 & 1-\alpha \\ 0 & 1 \end{matrix}\right), & s\in L_{1+}, \medskip \\
\left(\begin{matrix} 1 & \alpha \\ 0 & 1 \end{matrix}\right), & s\in L_{1-};
\end{array}
\right.
\]
\item[(c)] $\boldsymbol\Psi_\alpha$ satisfies
\[
\boldsymbol\Psi_\alpha(\zeta;\lambda) = \frac{\zeta^{-\sigma_3/4}}{\sqrt2} \left(\begin{matrix} 1 & -\mathrm i \\ 1 & \mathrm i \end{matrix}\right) \left(\boldsymbol I+ \boldsymbol{\mathcal{O}}\left(\zeta^{-1/2}\right)\right)e^{(\frac45\zeta^{5/2}+\lambda\zeta^{1/2})\sigma_3} 
\]
uniformly in $\C\setminus\big((-\infty,0]\cup L_{1-}\cup L_{1+}\cup L_{2-}\cup L_{2+}\big)$.
\end{itemize}

\hyperref[rhpsi]{\rhpsi} characterizes \emph{tronqu\'ee} solutions of Painlev\'e I equation \cite{Kap}. That is, $\boldsymbol\Psi_\alpha(\zeta;\lambda)$ satisfies the following system of linear ODEs:
\begin{equation}
\label{ode-p1}
\left\{
\begin{array}{lll}
\partial_\zeta(\boldsymbol\Psi_\alpha(\zeta;\lambda))\boldsymbol\Psi_\alpha(\zeta;\lambda)^{-1} &=& \left(\begin{matrix}  -\partial_\lambda y_\alpha & 2\zeta^2+2\zeta y_\alpha + \lambda+2y_\alpha^2 \medskip \\  2\zeta-2y_\alpha  & \partial_\lambda y_\alpha \end{matrix}\right) \medskip \\
\partial_\lambda(\boldsymbol\Psi_\alpha(\zeta;\lambda))\boldsymbol\Psi_\alpha(\zeta;\lambda)^{-1} &=& \left(\begin{matrix}  0 & \zeta+2y_\alpha \smallskip\\  1  & 0 \end{matrix}\right)
\end{array}
\right.
\end{equation}
with the functions $y_\alpha(\lambda)$ forming a one parameter family of solutions to $y^{\prime\prime}(\lambda)=6y^2(\lambda)+\lambda$ and satisfying $y_\alpha(\lambda)=\sqrt{-\lambda/6}\big(1+\mathcal O\big((-\lambda)^{-5/2}\big)\big)$ as $|\lambda|\to\infty$, $|\arg(\lambda)-\pi|<2\pi/5$, where the parameter $\alpha$ appears when describing the more detailed asymptotics of $y_\alpha$. In particular, each such solution is pole free in the sector $|\arg(\lambda)-\pi|<2\pi/5$ for $|\lambda|$ large, and therefore is a tronqu\'ee solution as designated by Boutroux \cite{Bout}. Moreover, the functions $y_0(\lambda)$ and $y_1(\lambda)$ are known to be \emph{tritronqu\'ee} solutions as they are asymptotically pole free in sectors $|\arg(\lambda)-7\pi/5|<4\pi/5$ and $|\arg(\lambda)-3\pi/5|<4\pi/5$, respectively.

It is known that \hyperref[rhpsi]{\rhpsi} is solvable if and only if $\lambda$ is not a pole of the corresponding solution $y_\alpha$ \cite[Section~4.6]{DK}. It is also known that tritronqu\'ee solutions are pole free in a disk around the origin \cite[Theorem~1]{CHT}. Hence, the matrices $\boldsymbol\Psi_0(\cdot;0)$ and $\boldsymbol\Psi_1(\cdot;0)$ exist and have the properties described by \hyperref[rhpsi]{\rhpsi}.

Set $\boldsymbol B_b:=\boldsymbol \Psi_0(\cdot;0)$ and $\boldsymbol B_a:=\sigma_3\boldsymbol \Psi_1(\cdot;0)\sigma_3$. As before, one can check that $\sigma_3\boldsymbol \Psi_\alpha\sigma_3$ has the same jumps as $\boldsymbol \Psi_\alpha$ only with the reversed orientation of the rays. Moreover, one needs to replace the anti-diagonal elements in \hyperref[rhpsi]{\rhpsi}(c) by their negatives when describing the behavior of $\sigma_3\boldsymbol \Psi_\alpha\sigma_3$ at infinity. Then \eqref{g3}, \eqref{g12}, and \eqref{rh4} yield that the matrix function
\begin{equation}
\label{rh16}
\boldsymbol P_c(z) = \boldsymbol E_c(z)\boldsymbol B_c\left(\big[(5/8)N\phi_c(z)\big]^{2/5}\right)e^{-N\phi_c(z)\sigma_3/2}F^{(N-n)\sigma_3}(z)
\end{equation}
satisfies  \hyperref[rhp]{\rhpc}(a) and \hyperref[rhp]{\rhpc}(b) for any matrix function $\boldsymbol E_c$ holomorphic in $U_c$. Thus, we only need to choose $\boldsymbol E_c$ so that \hyperref[rhp]{\rhpc}(c) is fulfilled. Choose
\[
\boldsymbol E_c(z) := \boldsymbol M(z) F^{(n-N)\sigma_3}(z) \left(\begin{matrix} 1 & \varepsilon_c  \\ \varepsilon_c \mathrm i & -\mathrm i \end{matrix}\right)\frac{ \big[(5/8)N\phi_c(z)\big]^{\sigma_3/10} } {\sqrt 2},
\]
whose analyticity in $U_c$ follows \eqref{g13}, \eqref{rh9}, and \hyperref[rhm]{\rhm}(b). It can be readily verified that \eqref{rh16} satisfies \eqref{rh14} uniformly on $\partial U_c$ with $\alpha_c=1/5$ and
\begin{equation}
\label{rh17}
\boldsymbol P_{c,k}(z) = \left(\begin{matrix} 1 & 0  \\ 0 & \varepsilon_c \end{matrix}\right) \boldsymbol \Psi_k \left(\begin{matrix} 1 & 0  \\ 0 & \varepsilon_c \end{matrix}\right)\left[\frac58\phi_c(z)\right]^{\sigma_3/5}
\end{equation}
where the $\boldsymbol O\big(\zeta^{-1/2}\big)\sim\sum_{k=1}^\infty \boldsymbol\Psi_k\zeta^{-k/2}$ is the error term from \hyperref[rhpsi]{\rhpsi}(c) and the matrices $\boldsymbol\Psi_k$ can be found recursively using \eqref{ode-p1}.

\subsubsection{Parametrix $\boldsymbol P_c$ around $c\in\Ga_t(a,b)$}

Recall that $U_c$ is given by \eqref{Ue3}. We always can adjust the constant $\delta_c$ so that $J_+\cap U_c=\varnothing$. In this case $\boldsymbol P_c$ is a holomorphic matrix in $U_c\setminus(J_\Gamma\cup J_-)$ that fulfills \hyperref[rhp]{\rhp}(c) and whose traces satisfy
\begin{equation}
\label{rh18}
\boldsymbol P_{c+}(s) = \boldsymbol P_{c-}(s) \left\{
\begin{array}{rl}
\left(\begin{matrix} 0 & 1 \\ -1 & 0 \end{matrix}\right), & s\in (U_c\cap J_t)\setminus\{c\}, \medskip \\
\left(\begin{matrix} 1 & 0 \\ F^{2(N-n)}(s)e^{-N\phi_b(s)} &1\end{matrix}\right), & s\in (U_c\cap J_-)\setminus\{c\}.
\end{array}
\right.
\end{equation}
Let $\boldsymbol C$ be the following matrix-valued function:
\begin{equation}
\label{rh19}
\boldsymbol C(\zeta) := \left(
\begin{matrix}
e^{\zeta^2} & 0 \smallskip \\
b(\zeta) & e^{-\zeta^2}
\end{matrix}
\right), \quad
b(\zeta) := \frac12e^{-\zeta^2}\left\{
\begin{array}{rl}
\mathrm{erfc}\big(-\mathrm i\sqrt 2\zeta\big), & \im(\zeta)>0, \medskip \\
-\mathrm{erfc}\big(\mathrm i\sqrt 2\zeta\big), & \im(\zeta)<0.
\end{array}
\right.
\end{equation}
Equivalently, we could have defined $b(\zeta)$ as $\pm(2\pi)^{-1/2}U(1/2;\mp2\mathrm i\zeta)$, $\pm\im(\zeta)>0$, see \cite[Eq. (12.7.5)]{DLMF}, where $U(a;z)$ is a parabolic cylinder function solving \cite[Eq. (12.2.2)]{DLMF}. Observe that
\[
b_+(x) - b_-(x) = \frac12 e^{-x^2}\left(\mathrm{erfc}\big(-\mathrm i\sqrt 2x\big) + \mathrm{erfc}\big(\mathrm i\sqrt 2x\big)\right) = e^{-x^2}, 
\] 
for $x\in\R$ and therefore
\begin{equation}
\label{rh20}
\boldsymbol C_+ = \boldsymbol C_-\left(\begin{matrix} 1 & 0 \\ 1 & 1 \end{matrix}\right) \quad \text{on} \quad \R.
\end{equation}
Moreover, since $\re(-\mathrm i\zeta)>0$ when $\im(\zeta)>0$ and $\re(\mathrm i\zeta)>0$ when $\im(\zeta)<0$, it holds that
\[
b(\zeta) \sim \frac{\mathrm ie^{\zeta^2}}{\sqrt{2\pi}}\sum_{k=0}^\infty\frac{\Gamma(k+1/2)}{2^{k+1}\Gamma(1/2)}\zeta^{-(2k+1)} =: e^{\zeta^2}\sum_{k=0}^\infty b_k\zeta^{-(2k+1)}
\]
uniformly in the upper and lower half-planes by \cite[Eq. (7.12.1) or Eq. (12.9.1)]{DLMF}. Thus, we deduce that
\begin{equation}
\label{rh21}
\boldsymbol C(\zeta) = \left(
\begin{matrix}
1 & 0 \smallskip \\
e^{-\zeta^2} b(\zeta) & 1
\end{matrix}
\right)e^{\zeta^2\sigma_3} \sim \left(\boldsymbol I + \sum_{k=0}^\infty\left(\begin{matrix} 0 & 0 \\ b_k & 0 \end{matrix} \right) \zeta^{-(2k+1)} \right)e^{\zeta^2\sigma_3} \quad \text{as} \quad \zeta\to\infty,
\end{equation}
where the expansion is uniform in the lower and upper half-planes. Set
\[
\boldsymbol J(z) := \left\{
\begin{array}{rr}
\left(\begin{matrix} 0 & -1 \\ 1 & 0 \end{matrix}\right), & z\in U_c^+, \smallskip \\
\boldsymbol I, & z\in U_c^-,
\end{array}
\right.
\]
where $U_c^+$ is the part of $U_c$ that lies to the left of $J_t$ and $U_c^-$ that lies to the right of it. We claim that \hyperref[rhp]{\rhpc} is solved by
\begin{multline}
\label{rh22}
\boldsymbol P_c(z) = \boldsymbol M(z)F^{(n-N)\sigma_3}(z)\boldsymbol J(z) e^{N\varphi(c)\sigma_3/2}\boldsymbol C\left(\sqrt{N/2}\big[\varphi-\varphi(c)\big]^{1/2}(z)\right)\times \\ \times\boldsymbol J^{-1}(z)e^{-N\phi_b(z)\sigma_3/2}F^{(N-n)\sigma_3}(z).
\end{multline}
Indeed, \hyperref[rhp]{\rhpc}(a) is  satisfied due to  the choice of the branch of $(\varphi-\varphi(c))^{1/2}$, see \eqref{g15}, and the choice of $J_-$, see \eqref{rh5}. Further, since $\boldsymbol J=\boldsymbol I$ in $U_c^-$, analyticity of $\boldsymbol M$, $\phi_b$, and $F$ across $J_-\setminus\{c\}$ as well as \eqref{rh5} and \eqref{rh20} imply that $\boldsymbol P_c$ has the jump there as in \eqref{rh18}. Moreover, observe that
\[
\boldsymbol M(z)F^{(n-N)\sigma_3}(z)\boldsymbol J(z) \quad \text{and} \quad e^{N\varphi(c)\sigma_3/2}\boldsymbol C\left(\sqrt{N/2}\big[\varphi-\varphi(c)\big]^{1/2}(z)\right)
\]
are analytic across $J_t$ by \hyperref[rhm]{\rhm}(b), \eqref{rh9}, and the choice of $\boldsymbol J$. Hence, since $\phi_{b+}+\phi_{b-}\equiv0$ on $J_t$ by \eqref{g3}, it follows from the definition of $\boldsymbol J$ that $\boldsymbol P_c$ has the jumps across $J_t\setminus\{c\}$ as in \eqref{rh18}. That is, \hyperref[rhp]{\rhpc}(b) is fulfilled as well. Finally, we get from \eqref{g14}, \eqref{imageUe}, and \eqref{rh21} that \eqref{rh22} satisfies \eqref{rh14} uniformly on $\partial U_c$ and locally uniformly on $C_\mathsf{split}$ with $\alpha_c=1/2$ and $\boldsymbol P_{c,k}$ given by
\begin{equation}
\label{rh23}
\frac{2^{k+1/2}e^{-N\varphi(c)}}{(\varphi(z)-\varphi(c))^{k+1/2}} \left( \begin{matrix} 0 & -b_k \\ 0 & 0 \end{matrix} \right) \quad \text{and} \quad \frac{2^{k+1/2}e^{-N\varphi(c)}}{(\varphi(z)-\varphi(c))^{k+1/2}} \left( \begin{matrix} 0 & 0 \\ b_k & 0 \end{matrix} \right)
\end{equation}
in $U_c^+$ and $U_c^-$, respectively (recall also that $|e^{\varphi(c)}|=1$).

\subsubsection{Parametrix $\boldsymbol P_c$ around $c\in\Ga_t\big(e^{\pi\mathrm i}\infty,a\big)\cup\Ga_t\big(b,e^{\pi\mathrm i/3}\infty\big)$}

Put $e=a$ if $c\in\Ga_t\big(e^{\pi\mathrm i}\infty,a\big)$ and $e=b$ if $c\in\Ga_t\big(b,e^{\pi\mathrm i/3}\infty\big)$. We are seeking a matrix function holomorphic in $U_c\setminus\Ga_t$ that fulfills \hyperref[rhp]{\rhpc}(c) and whose traces satisfy
\[
\boldsymbol P_{c+}(s) = \boldsymbol P_{c-}(s)\left(\begin{matrix} 1 & F^{2(n-N)}(s)e^{N\phi_e(s)} \\ 0 & 1 \end{matrix}\right), \quad s\in\Ga_t\cap U_c.
\]
The Riemann-Hilbert problem \hyperref[rhp]{\rhpc} is solved by
\begin{multline}
\label{rh24}
\boldsymbol P_c(z) = \boldsymbol M(z)F^{(n-N)\sigma_3}(z)e^{N\phi_e(c)\sigma_3/2}\sigma_1\boldsymbol C\left(\big[-(N/2)\phi_c(z)\big]^{1/2}\right)\sigma_1\times \\ \times e^{-N\phi_e(z)\sigma_3/2}F^{(N-n)\sigma_3}(z),
\end{multline}
where $\boldsymbol C$ is defined by \eqref{rh19} and $(-\phi_c)^{1/2}$ is the branch chosen in \eqref{g16}. Indeed, it can be readily verified using \eqref{rh20} that
\begin{equation}
\label{rh25}
\left(\sigma_1\boldsymbol C\sigma_1\right)_+ = \left(\sigma_1\boldsymbol C\sigma_1\right)_- \left(\begin{matrix} 1 & 1 \\ 0 & 1 \end{matrix}\right) \quad \text{on} \quad \R.
\end{equation}
As $\boldsymbol M$ and $F$ are holomorphic in $U_c$, $(-\phi_c)^{1/2}$ is conformal there, satisfies \eqref{g16}, and preserves the orientation, we see that \hyperref[rhp]{\rhpc}(a) is fulfilled. The above properties and \eqref{rh25} yield that \hyperref[rhp]{\rhpc}(b) is fulfilled as well. Finally, we get from \eqref{rh21} that
\[
\sigma_1\boldsymbol C(\zeta)\sigma_1 \sim \left(\boldsymbol I + \sum_{k=0}^\infty\left(\begin{matrix} 0 & b_k \\ 0 & 0 \end{matrix} \right) \zeta^{-(2k+1)} \right)e^{-\zeta^2\sigma_3} \quad \text{as} \quad \zeta\to\infty
\]
uniformly in the lower and upper half-planes. Therefore, by \eqref{imageUe} and since $\phi_c(z)=\phi_e(z)-\phi_e(c)$, \eqref{rh24} satisfies \eqref{rh14} with the expansion in parenthesis being uniform on $\partial U_c$ and closed subsets of $C_\mathsf{birth}^a\cup C_\mathsf{birth}^b$, $\alpha_c=1/2$, and
\begin{equation}
\label{rh26}
\boldsymbol P_{c,k}(z) = \frac{2^{k+1/2}e^{N\phi_e(c)}}{(-\phi_c(z))^{k+1/2}} \left( \begin{matrix} 0 & b_k \\ 0 & 0 \end{matrix} \right)
\end{equation}
(again, notice that $|e^{\phi_e(c)}|=1$).

\subsection{Riemann-Hilbert problem with small jumps}

\begin{figure}[!h]
\includegraphics[scale=.4]{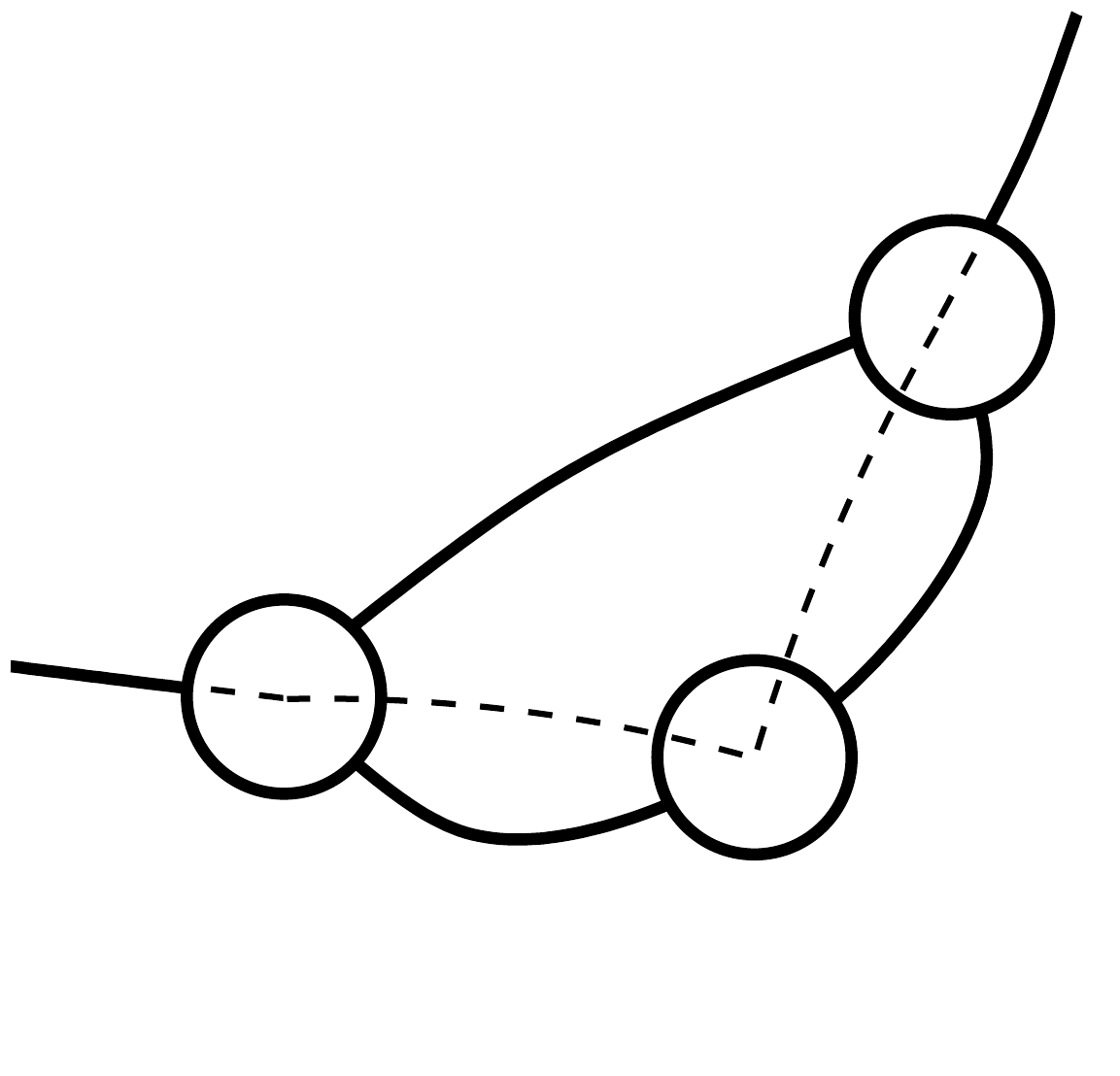}
\begin{picture}(0,0)
\put(-131,35){$\partial U_a$}
\put(-10,77){$\partial U_b$}
\put(-33,31){$\partial U_c$}
\put(-70,80){$J_+$}
\put(-23,52){$J_-$}
\put(-90,22){$J_-$}
\put(-8,110){$\Ga_t\big(b,e^{\pi\mathrm i/3}\infty\big)$}
\put(-163,53){$\Ga_t\big(e^{\pi\mathrm i}\infty,a\big)$}
\end{picture}
\caption{\small The contour $\Sigma_{\boldsymbol R}$ (continuous lines). The dashed lines represent the part of $\Ga_t$ that does not belong to $\Sigma_{\boldsymbol R}$.}
\label{SigmaR}
 \end{figure}

Set $\Sigma_{\boldsymbol R} := \big[\big(\Ga_t\big(e^{\pi\mathrm i}\infty,a\big)\cup J_+\cup J_-\cup\Ga_t\big(b,e^{\pi\mathrm i/3}\infty\big)\big)\cap D\big]\cup \big[\cup_e \partial U_e\big]$, $D:= \C\setminus\cup_e \overline U_e$, where $e$ runs over $a$ and $b$ as well as $c$ when $t\in\partial O_\mathsf{one-cut}$ (in what follows, we shall always understand the symbol $\cup_e$ this way), see Figure~\ref{SigmaR}. Consider \rhr:
\begin{itemize}
\label{rhr}
\item[(a)] $\boldsymbol R$ is holomorphic in $\C\setminus\Sigma_{\boldsymbol R}$ and $\lim_{\C\setminus\Ga_t\ni z\to\infty}\boldsymbol R(z)=\boldsymbol I$;
\item[(b)] $\boldsymbol R$ has continuous traces on $\Sigma_{\boldsymbol R}^\circ$ that satisfy
\[
\boldsymbol R_+(s) =  \boldsymbol R_-(s) \left\{
\begin{array}{ll}
\boldsymbol P_e(s)\boldsymbol M^{-1}(s), & s\in\partial U_e, \medskip \\
\boldsymbol M(s)  \left(\begin{matrix} 1 & 0 \\  F^{2(N-n)}(s)e^{-N\phi_b(s)} & 1 \end{matrix}\right) \boldsymbol M^{-1}(s), & s\in J_{\pm}\cap D,
\end{array}
\right.
\]
where $\partial U_e$ is oriented clockwise, $\boldsymbol M$ is given by \eqref{rh10}, and $\boldsymbol P_e$ is given by \eqref{rh13}, \eqref{rh16}, \eqref{rh22}, or \eqref{rh24} depending on $e$; as well as
\[
\boldsymbol R_+(s) = \boldsymbol R_-(s)  \boldsymbol M(s) \left(\begin{matrix} 1 & F^{2(n-N)}(s)e^{N\phi_e(s)} \\ 0 & 1 \end{matrix}\right) \boldsymbol M^{-1}(s), 
\]
for $s\in\Ga_t\big(b,e^{\pi\mathrm i/3}\infty\big) \cap D$ with $e=b$ and for $s\in\Ga_t\big(e^{\pi\mathrm i}\infty,a\big)\cap D$ with $e=a$ (observe that $\boldsymbol M^{-1}$ is well defined since $\det(\boldsymbol M)\equiv 1$).
\end{itemize}

Let us show that the jump matrices in \hyperref[rhr]{\rhr}(b) are uniformly close to $\boldsymbol I$. To this end, set
\begin{equation}
\label{Delta}
\boldsymbol \Delta := \boldsymbol R_-^{-1}\boldsymbol R_+ - \boldsymbol I
\end{equation}
to be the deviation of the jumps of $\boldsymbol R$ from the identity matrix. 

Firstly, it follows from \hyperref[rhr]{\rhr}(b) and \eqref{rh14} with $\boldsymbol P_{e,k}$ given by \eqref{rh15}, \eqref{rh17}, \eqref{rh23}, or \eqref{rh26} that 
\begin{equation}
\label{rh27}
\boldsymbol \Delta \sim N^{-\alpha_e}\sum_{k=0}^\infty \left(\boldsymbol MF^{(n-N)\sigma_3}\boldsymbol P_{e,k}F^{(N-n)\sigma_3}\boldsymbol M^{-1}\right) N^{-k} \quad \text{on} \quad \partial U_e.
\end{equation}
The above expansions of $\boldsymbol\Delta(s)$ are uniform in $s$ on each $\partial U_e$. Moreover, the expansions on $\partial U_a$ and $\partial U_b$ are also locally uniform in $\overline O_\mathsf{one-cut}\setminus\big\{t_\mathsf{cr},e^{2\pi\mathrm i/3} t_\mathsf{cr}\big\}$ by \eqref{est-AB2} and \eqref{est-AB4}. Furthermore, the expansion on $\partial U_c$ is uniform on compact subsets of $C_\mathsf{split}$, $C_\mathsf{birth}^a$, and $C_\mathsf{birth}^b$ by \eqref{est-AB2} and a compactness argument applied to $\max_{s\in\partial U_c} |F^{\pm1}(s)|$. In addition, the expansion on $\partial U_c$ is uniform on closed subsets of $C_\mathsf{birth}^a$ and $C_\mathsf{birth}^b$ when $n=N$ because the term $F^{(n-N)\sigma_3}$ is no longer present. Altogether, we get by looking at the first term in expansion \eqref{rh27} that
\begin{equation}
\label{rh28}
\|\boldsymbol\Delta\|_{L^\infty(\cup_e\partial U_e)} \leq C_0(t,\delta)N^{-\alpha_t}, \quad \alpha_t = \min_{e}\alpha_e, \quad \delta:=\min_e\delta_e,
\end{equation}
where the constants $\delta_e$ were introduced in \eqref{Ue1} and \eqref{Ue3}, $C_0(t,\delta)$ can be chosen to depend continuously on $t$ and $\delta$ with additional property of being bounded as $t\to\infty$ for each fixed $\delta>0$ when $n=N$.

Secondly, since $\Ga_t\setminus J_t$ consists of orthogonal trajectories of $-Q(z;t)\mathrm dz^2$, it holds by \eqref{g2} that $\phi_e(s)<0$ on the corresponding part of $(\Ga_t\setminus J_t)\cap D$. More precisely, there exists a constant $0<C_1(t,\delta)<1$ such that
\begin{equation}
\label{rh31}
\big |F^{2(n-N)}(s)e^{N\phi_e(s)}\big|<C_1^N(t,\delta), \quad s\in \big[\Ga_t\setminus J_t \big] \cap D,
\end{equation}
for all $N$ large. Since the quantities on the left-hand side of \eqref{rh31} depend on $t$ continuously, one can clearly choose $C_1(t,N)$ to be a continuous function of $t$ and $\delta$. Hence, a simple compactness argument shows that the estimate \eqref{rh31} is $(s,t)$-locally uniform. In addition, notice that $\phi_e$ is monotone on each connected piece of $\Ga_t\setminus(J_t\cup\{c\})$. Therefore, $|e^{N\phi_e(s)}|$ achieves its largest value on $(\Ga_t\setminus J_t)\cap D$ either at $c$ (when $c$ belongs to $\Ga_t$) or on $\partial D$. Then, assuming $n=N$, it follows from the properties of the function $\tilde\rho(t)$ in \eqref{imageUe} and the way we construct the conformal maps in \eqref{g10} and \eqref{g16} that $C_1(t,N)$ is bounded away from $1$ on closed subsets of $C_\mathsf{birth}^a$ and $C_\mathsf{birth}^b$, and subsets $T\subset O_\mathsf{one-cut}$ satisfying \eqref{condT}. That is, \eqref{rh31} is strongly $(s,t)$-locally uniform in the notation of Definition~\ref{def:zt-dep}.

Lastly, recall also that the arcs $J_\pm$ were chosen so that $\re(\phi_b(s))>0$ on $J_\pm\cap D$. In fact, we can choose them to be level lines of $\re(\phi_b)$. Clearly, the maximal level will depend on the maximal value of $\re(\phi_b)$ on $\partial U_a\cup \partial U_b$ (also on $\partial U_c$ when $t\in C_\mathsf{split}$) as well as $\re(\phi_b(c))$.  Therefore, there exists a constant $0<C_2(t,\delta)<1$ such that
\begin{equation}
\label{rh34}
\big |F^{2(N-n)}(s)e^{-N\phi_b(s)}\big|<C_2^N(t,\delta), \quad s\in J_\pm \cap D.
\end{equation}
Again, it holds that $C_2(t,\delta)$ depends on its parameters continuously and is bounded away from $1$ when $t$ belongs to closed subsets of $\overline O_\mathsf{one-cut}\setminus\big\{t_\mathsf{ct},e^{2\pi\mathrm i/3}t_\mathsf{cr}\big\}$ and $n=N$.

Summarizing, we get from \eqref{rh27}, \eqref{est-AB2}, \eqref{rh31}, and \eqref{rh34} that
\begin{equation}
\label{rh29}
\boldsymbol \Delta = \boldsymbol{\mathcal{O}}\left(\max\left\{C_1^N(t,\delta),C_2^N(t,\delta)\right\}\right) \quad \text{on} \quad \Sigma_{\boldsymbol R}\setminus\bigcup \partial U_e.
\end{equation}
Estimates \eqref{rh28} and \eqref{rh29} show that $\boldsymbol \Delta$ is uniformly close to zero on $\Sigma_{\boldsymbol R}$. Since the entries of $\boldsymbol \Delta$ are geometrically small as $\Ga_t\ni s\to\infty$, $\boldsymbol \Delta$ is close to zero in $L^2$-norm as well. Then it follows from the same analysis as in \cite[Corollary~7.108]{Deift} that $\boldsymbol R$ exists for all $N$ large and
\begin{equation}
\label{rh30}
\big|\boldsymbol R - \boldsymbol I \big| \leq c_0(t,\delta)N^{-\alpha_t}
\end{equation}
in any matrix norm, where $c_0(t,\delta)$ continuously depends on $t$ and $\delta$, blows up as $\delta\to0$ or $t\to\infty$, but is bounded as $t\to\infty$ along either $C_\mathsf{birth}^a$, $C_\mathsf{birth}^b$, or subsets $T\subset O_\mathsf{one-cut}$ satisfying \eqref{condT} when $n=N$.

\subsection{Solution of \hyperref[rhy]{\rhy}}
\label{ss:solution}

Given $\boldsymbol R$, the solution of \hyperref[rhr]{\rhr}, it is straightforward to verify that \hyperref[rhs]{\rhs} is solved by
\begin{equation}
\label{rh32}
\boldsymbol S = \left\{
\begin{array}{ll}
\boldsymbol R\boldsymbol M & \text{in} \quad D\setminus\big[(\Ga_t\setminus J_t)\cup J_+\cup J_-\big], \medskip \\
\boldsymbol R\boldsymbol P_e & \text{in} \quad U_e.
\end{array}
\right.
\end{equation}

Let $K_t$ be a compact set in $\C\setminus\Ga_t$. We can always adjust quantities $\delta_e$ in \eqref{Ue1} and \eqref{Ue3} as well as the arcs $J_\pm$ so that $K_t$ lies entirely within one of the unbounded components of the complement of $\Sigma_{\boldsymbol R}$. Then it follows from \eqref{rh2}, \eqref{rh6}, and \eqref{rh32} that
\begin{equation}
\label{rh33}
\boldsymbol Y(z) = \big((b-a)/4\big)^{(n-N)\sigma_3}e^{-N\ell_t^*\sigma_3/2}\boldsymbol R(z)\boldsymbol M(z)e^{N(g(z;t)+\ell_t^*/2)\sigma_3}F^{(n-N)\sigma_3}(z)
\end{equation}
on $K_t$. Subsequently, by using \eqref{oc12}, \eqref{rh10}, and the definition of $F$, we see that 
\begin{eqnarray*}
[\boldsymbol Y]_{11}(z) &=& \big([\boldsymbol R]_{11}(z)[\boldsymbol M]_{11}(z) + [\boldsymbol R]_{12}(z) [\boldsymbol M]_{21}(z)\big) D^{N-n}(z;t)e^{ng(z;t)} \\
&=& \left([\boldsymbol R]_{11}(z) + [\boldsymbol R]_{12}(z)F^{-1}(z)\right)A(z;t)D^{N-n}(z;t)e^{ng(z;t)}.
\end{eqnarray*}
Equation \eqref{oc7} now follows from \eqref{F-estimate} and \eqref{rh30}. To handle compact sets $K_t$ in $\C\setminus (J_t\cup\{c\})$, it is enough to consider only the sets belonging to sufficiently small Hausdorff neighborhood of $\Ga_t$. In this case the curve $\Ga_t$ can be deformed locally around $K$ in such a fashion that \eqref{rh31} still holds on a deformed curve, perhaps with a different constant. As the rest of the analysis is the same, the full claim \eqref{oc7} follows. The (strongly) $(z,t)$-locally uniform character of \eqref{oc7} follows from the continuity properties of $c_0(t,\delta)$ in \eqref{rh30}.

Take now $K_t\subset \Ga_t(a,b)\setminus\{c\}$. Then it follows from \eqref{rh2}, \eqref{rh6}, \eqref{rh32}, and \eqref{oc12} that
\begin{multline*}
[\boldsymbol Y]_{11} = [\boldsymbol R]_{11}\left([\boldsymbol M]_{11\pm} \pm [\boldsymbol M]_{12\pm}F_\pm^{2(N-n)}e^{-N\phi_{b\pm}}\right)e^{ng_\pm}D_\pm^{N-n} + \\
+ [\boldsymbol R]_{12}\left( [\boldsymbol M]_{21\pm} \pm  [\boldsymbol M]_{22\pm}F_\pm^{2(N-n)}e^{-N\phi_{b\pm}} \right)e^{ng_\pm}D_\pm^{N-n}.
\end{multline*}
Hence, we get from \eqref{rh10}, \eqref{rh9}, \eqref{g4}, \eqref{oc12}, and some algebra that
\begin{multline*}
[\boldsymbol Y]_{11} = [\boldsymbol R]_{11}\left(A_+D_+^{N-n}e^{ng_+} + A_-D_-^{N-n}e^{ng_-}\right) + \\
+ [\boldsymbol R]_{12} \left(B_+D_+^{N-n}e^{ng_+} + B_-D_-^{N-n}e^{ng_-}\right).
\end{multline*}
Therefore, we deduce from the definition of $F$ that
\[
[\boldsymbol Y]_{11} = \left([\boldsymbol R]_{11}+[\boldsymbol R]_{12}F_+^{-1}\right) A_+D_+^{N-n}e^{ng_+} +  \left([\boldsymbol R]_{11}+[\boldsymbol R]_{12}F_-^{-1}\right) A_-D_-^{N-n}e^{ng_-}.
\]
Formula \eqref{oc8} now follows from \eqref{F-estimate} and \eqref{rh30}.

\section{Asymptotic Expansions} 
\label{s:ae}

In this section, we first improve \eqref{rh30} to a full asymptotic expansion following the technique of \cite[Theorem~7.8 and Theorem~7.10]{DKMVZ2}. Then we show how recurrence coefficients appear within the matrix $\boldsymbol Y$ and use \eqref{rh33} and the expansion of $\boldsymbol R$ to prove Theorem~\ref{RecCoef}. Next, we discuss the so-called \emph{string equations} and prove Theorem~\ref{RecCoef1} using them. Finally, we employ Theorem~\ref{RecCoef} and Toda equation \eqref{cm6} to prove Theorem~\ref{FreeEnergy}.

\subsection{Error matrix $\boldsymbol R$}

Let us show that
\begin{equation}
\label{ae1}
\boldsymbol R(z) \sim \boldsymbol I + \sum_{i=1}^\infty \boldsymbol r_i(z;n-N)N^{-\alpha_t i},
\end{equation}
(strongly when $n=N$) $(z,t)$-locally uniformly in $\overline\C$, see Definitions~\ref{def:t-dep} and~\ref{def:zt-dep}. Indeed, as shown in \cite[Theorem~7.8]{DKMVZ2}, it holds that
\begin{equation}
\label{ae2}
\boldsymbol R = \boldsymbol I +  \sum_{k=0}^\infty\mathcal{C}\left(\mathcal{C}^k_{\boldsymbol\Delta}(\boldsymbol I)\boldsymbol\Delta\right),
\end{equation}
where, as before, $\mathcal C$ is the Cauchy transform, $\boldsymbol\Delta$ was defined in \eqref{Delta}, and
\begin{equation}
\label{ae3}
\mathcal C_{\boldsymbol\Delta}(\boldsymbol F):=\mathcal C_-(\boldsymbol{\Delta F}), \quad \boldsymbol F\in L^2(\Sigma_{\boldsymbol R}).
\end{equation}
It follows from \eqref{rh28} and \eqref{rh29} that
\begin{equation}
\label{ae4}
\|\mathcal C_{\boldsymbol \Delta}\| = \mathcal{O}\left(N^{-\alpha_t}\right)
\end{equation}
(strongly when $n=N$) $t$-locally uniformly. Define
\begin{equation}
\label{ae5}
\boldsymbol\Delta_l(s) := N^{-\alpha_e}\sum_{k=0}^{l-1} \left(\boldsymbol M F^{(n-N)\sigma_3} \boldsymbol P_{e,k} F^{(N-n)\sigma_3} \boldsymbol M^{-1}\right)(s)N^{-k}, \quad s\in\partial U_e,
\end{equation}
and set $\boldsymbol\Delta_l\equiv0$ on $\Sigma_{\boldsymbol R}\setminus\bigcup\partial U_e$. Observe that the coefficient next to $N^{-k}$ depends neither on $n$ nor $N$ individually, but does depend on the difference $n-N$. Clearly, it follows from \eqref{rh27} and \eqref{rh29} that $\boldsymbol\Delta_l$ approximate $\boldsymbol\Delta$, that is,
\begin{equation}
\label{ae6}
\|\boldsymbol\Delta - \boldsymbol\Delta_l\|_{L^1(\Sigma_{\boldsymbol R})} + \|\boldsymbol\Delta - \boldsymbol\Delta_l\|_{L^2(\Sigma_{\boldsymbol R})} + \|\boldsymbol\Delta - \boldsymbol\Delta_l\|_{L^\infty(\Sigma_{\boldsymbol R})} = \mathcal{O}\left(N^{-(\alpha_t+l)}\right)
\end{equation}
(strongly when $n=N$) $t$-locally uniformly. Define $\mathcal C_{\boldsymbol \Delta_l}$ as in \eqref{ae3} with $\boldsymbol \Delta$ replaced by $\boldsymbol \Delta_l$. Clearly, $\mathcal C_{\boldsymbol \Delta_l}$ satisfies \eqref{ae4}. Moreover, it holds that
\begin{equation}
\label{ae7}
\boldsymbol R - \boldsymbol I - \sum_{k=0}^{l-1}\mathcal{C}\left(\mathcal{C}^k_{\boldsymbol\Delta_l}(\boldsymbol I)\boldsymbol\Delta_l\right) = \boldsymbol{\mathcal O}\left(N^{-\alpha_t(l+1)}\right)
\end{equation}
(strongly when $n=N$) $(z,t)$-locally uniformly in $\overline \C$. Indeed, similarly to \eqref{rh30}, equations \eqref{rh28}, \eqref{ae4}, and deformation of $\Sigma_{\boldsymbol R}$ technique yield that
\begin{equation}
\label{ae8}
\sum_{k=l}^\infty\mathcal{C}\left(\mathcal{C}^k_{\boldsymbol\Delta}(\boldsymbol I)\boldsymbol\Delta\right) =\boldsymbol{\mathcal O}\left(N^{-\alpha_t(l+1)}\right)
\end{equation}
(strongly when $n=N$) $t$-locally uniformly. Furthermore, it holds by \eqref{ae6} and an analogous argument that
\begin{equation}
\label{ae9}
\sum_{k=0}^{l-1}\mathcal{C}\left(\mathcal{C}^k_{\boldsymbol\Delta}(\boldsymbol I)\left(\boldsymbol\Delta-\boldsymbol\Delta_l\right) \right) =\boldsymbol{\mathcal O}\left(N^{-(\alpha_t+l)}\right)
\end{equation}
(strongly when $n=N$) $t$-locally uniformly. Finally, were deduce from \eqref{ae6} and deformation of $\Sigma_{\boldsymbol R}$ technique that
\begin{equation}
\label{ae10}
\sum_{k=0}^{l-1}\mathcal{C}\left( \left(\mathcal{C}^k_{\boldsymbol\Delta}(\boldsymbol I)- \mathcal{C}^k_{\boldsymbol\Delta_l}(\boldsymbol I)\right) \boldsymbol\Delta_l \right) =\boldsymbol{\mathcal O}\left(N^{-(2\alpha_t+l)}\right)
\end{equation}
(strongly when $n=N$) $t$-locally uniformly. Estimates \eqref{ae8}--\eqref{ae10} imply \eqref{ae7}. Now, to derive \eqref{ae1}, it only remains to notice that $\mathcal C^k_{\boldsymbol\Delta_l}(\boldsymbol I)\boldsymbol\Delta_l$ has an asymptotic expansion in powers of $N^{-\alpha_t}$ whose coefficients up to the order $l$ do not depend on $l$ or $n,N$ individually.

\subsection{Recurrence coefficients}

It follows from \eqref{cm4} and \eqref{cm5} that
\[
\left\{
\begin{array}{lll}
\ga_n^2(t,N) & = & h_n(t,N)/h_{n-1}(t,N), \medskip \\
\beta_n(t,N) & = & (P_n)_{n-1} - (P_{n+1})_n,
\end{array}
\right.
\]
where we write $P_n(z;t,N) = z^n + \sum_{k=0}^{n-1}(P_n)_kz^k$. Hence, we get from \eqref{rh1} that
\[
\boldsymbol Y(z)z^{-n\sigma_3} = \boldsymbol I + \frac1z\left( \begin{matrix} (P_n)_{n-1} & -\frac{h_n}{2\pi i} \smallskip \\ -\frac{2\pi i}{h_{n-1}} & * \end{matrix} \right) + \boldsymbol{\mathcal{O}}\left(\frac1{z^2}\right) =: \boldsymbol I + \frac{\boldsymbol Y_1(n;t,N)}z + \boldsymbol{\mathcal{O}}\left(\frac1{z^2}\right).
\]
Therefore,
\begin{equation}
\label{ae11}
\left\{
\begin{array}{lll}
\ga_n^2(t,N) & = & \big[\boldsymbol Y_1(n;t,N)\big]_{12}\big[\boldsymbol Y_1(n;t,N)\big]_{21}, \medskip \\
\beta_n(t,N) & = & \big[\boldsymbol Y_1(n;t,N)\big]_{11} - \big[\boldsymbol Y_1(n+1;t,N)\big]_{11}.
\end{array}
\right.
\end{equation}
We deduce from \eqref{rh33}, \eqref{oc12}, \eqref{ts6}, and the definition of $F$ that
\begin{multline*}
\boldsymbol Y(z)z^{-n\sigma_3} = \left(\frac{b-a}4\right)^{(n-N)\sigma_3}e^{-N\ell_t^*\sigma_3/2} \left( \boldsymbol I + \frac{\boldsymbol R_1(n;t,N) + \boldsymbol M_1(t)}z + \right. \\
\left. + \frac{nG_1(t)\sigma_3 + (N-n)D_1(t)\sigma_3}z + \boldsymbol{\mathcal{O}}\left(\frac1{z^2}\right) \right)e^{N\ell_t^*\sigma_3/2}\left(\frac{b-a}4\right)^{(N-n)\sigma_3}, 
\end{multline*}
where we write
\[
\left\{
\begin{array}{rll}
e^{g(z;t)} & = & z+ G_1(t) + \mathcal O\big(z^{-1}\big), \medskip \\
D(z;t) & = & 1+ z^{-1}D_1(t) +  \mathcal O\big(z^{-2}\big), \medskip \\
\boldsymbol K(z) & = & \boldsymbol I + z^{-1}\boldsymbol K_1(n;t,N) + \boldsymbol{\mathcal{O}}\big(z^{-2}\big), \quad \boldsymbol K\in\{\boldsymbol M,\boldsymbol R\}.
\end{array}
\right.
\]
In fact, it follows from \eqref{oc9} and \eqref{Dt-series} that $D_1(t)=1/4x^2(t)$, and analogously we deduce from \eqref{oc12} and \eqref{oc13} that $G_1(t)=-x(t)+1/4x^2(t)$. Further, using \eqref{rh10}, \eqref{oc6}, and \eqref{ts6} we see that 
\[
\boldsymbol M_1(t) = \left( \begin{matrix} 0 & -1/\sqrt{2x(t)} \smallskip \\ 1/\sqrt{2x(t)} & 0 \end{matrix} \right).
\]
Therefore \eqref{ae11} can be rewritten as
\begin{equation}
\label{ae12}
\left\{
\begin{array}{lll}
\gamma_n^2(t,N) &=& -1/(2x(t)) + \big([\boldsymbol R_1]_{12} - [\boldsymbol R_1]_{21} \big)/\sqrt{2x(t)} + [\boldsymbol R_1]_{12}[\boldsymbol R_1]_{21} \medskip \\
\beta_n(t,N) &=& x(t) + \big[\boldsymbol R_1(n;t,N) - \boldsymbol R_1(n+1;t,N)\big]_{11}.
\end{array}
\right.
\end{equation}
Hence, \eqref{oc3} follows from \eqref{ae1} and \eqref{ae12}. Moreover, as $\gamma_N^2(t,N)$ is expressed only through $\boldsymbol R_1(N;t,N)$, its expansion is strongly $t$-locally uniform.

To see the analyticity of $G_k(t;n-N)$ and $B_k(t;n-N)$ in $O_\mathsf{one-cut}$, we need to examine the dependence of $\boldsymbol R_1$ on $t$. To this end, write
\begin{equation}
\label{ae16}
\boldsymbol R_1 = -\frac1{2\pi i}\oint_{\cup\partial U_e} \left(\sum_{k=0}^{l-1}\mathcal C^k_{\boldsymbol\Delta_l}(\boldsymbol I)\boldsymbol \Delta_l\right)(s) ds + \boldsymbol{\mathcal{O}}\left(N^{-(l+1)}\right),
\end{equation}
which we can do according to \eqref{ae7}, where $\boldsymbol\Delta_l$ is given by \eqref{ae5}. Notice that on each $\partial U_e$ the function $\boldsymbol\Delta_l$ is a trace on $\partial U_e$ of a meromorphic matrix-valued function in $U_e$ with a single pole at $e$. Indeed, it is clear that $\boldsymbol\Delta_l$ is holomorphic in $U_e\setminus J_t$ and on $J_t$ it holds that
\begin{eqnarray*}
\boldsymbol\Delta_{l+} & = & N^{-\alpha_e}\boldsymbol M_- F_-^{(n-N)\sigma_3} \left(\begin{matrix} 0 & 1 \\ -1 & 0 \end{matrix}\right) \left(\sum_{k=0}^{l-1} \boldsymbol P_{e,k+}N^{-k}\right) F_+^{(N-n)\sigma_3}\boldsymbol M_+^{-1} \\
 & = & N^{-\alpha_e}\boldsymbol M_- F_-^{(n-N)\sigma_3} \left(\sum_{k=0}^{l-1} \boldsymbol P_{e,k-}N^{-k}\right)\left(\begin{matrix} 0 & 1 \\ -1 & 0 \end{matrix}\right) F_+^{(N-n)\sigma_3}\boldsymbol M_+^{-1} = \boldsymbol\Delta_{l-},
\end{eqnarray*}
where the first and the last equalities follow from \hyperref[rhm]{\rhm}(b) and \eqref{rh9} while the second can be verified by using the explicit expressions \eqref{rh15} and \eqref{g8}, \eqref{rh17} and \eqref{g12}, \eqref{rh23} and \eqref{g15}. Hence, $\boldsymbol \Delta_l$ is indeed meromorphic in each $U_e$ with a pole at $e$. From this, it is easy to see that $\mathcal C^k_{\boldsymbol\Delta_l}(\boldsymbol I)\boldsymbol\Delta_l$ is a trace on $\partial U_e$ of a meromorphic matrix-valued function in $U_e$ with a single pole at $e$. In particular, the integral in \eqref{ae16} does not depend on the radii of the disks $U_e$. Thus, its $t$-dependence comes only from the points $a,b,c$ and the conformal maps constructed in \eqref{g10}, \eqref{g12}, \eqref{g15}, and \eqref{g16}. It follows from Proposition~\ref{prop:x} that $a(t)$, $b(t)$, $c(t)$ are holomorphic functions of $t\in O_\mathsf{one-cut}$ with holomorphic continuations across each of the arcs $C_\mathsf{split}$, $C_\mathsf{birth}^a$, and $C_\mathsf{birth}^b$. The conformal maps $(-\phi_a)^{2/3}$ and $(-\phi_b)^{2/3}$, see \eqref{g10} and \eqref{g2}, have the same type of dependence on $t$. Thus, we indeed see that the functions $\boldsymbol R_1$ are analytic functions of $t\in O_\mathsf{one-cut}$.
 
Let now $n=N$. The first claim of \eqref{oc10} was derived in \cite[Corollary~4.2]{BD2}. Observe that to show the second claim, it is enough to prove that
\begin{multline}
\label{ae15}
\boldsymbol R_1 = \sum_{1\leq 2j+1\leq l-1}\left(\begin{matrix} -p_{2j+1} & q_{2j+1} \\ q_{2j+1} & p_{2j+1} \end{matrix}\right)N^{-2j-1}  + \sum_{2\leq 2j\leq l-1}\left(\begin{matrix} p_{2j} & q_{2j} \\ -q_{2j} & p_{2j} \end{matrix}\right)N^{-2j} + \\ + \boldsymbol{\mathcal O}\left(N^{-(l+1)}\right)
\end{multline}
for some constants $p_k,q_k$, as then it obviously holds that $[\boldsymbol R_1]_{12} - [\boldsymbol R_1]_{21}$ and $[\boldsymbol R_1]_{12}[\boldsymbol R_1]_{21}$ have asymptotic expansions only in even powers of $N^{-1}$. Using \eqref{rh15} and \eqref{rh10}, it is tedious but straightforward to verify that the expansion for $\boldsymbol\Delta_l$ has exactly the same form as the right-hand side of \eqref{ae15} (without $\boldsymbol{\mathcal O}$-term), where
\[
\left\{
\begin{array}{lll}
p_{2j} & \mapsto & \big(s_{2j}+t_{2j}\big)(-\phi_e/2)^{-2j}, \medskip \\
q_{2j} & \mapsto & \mathrm i\varepsilon_e\big(s_{2j}-t_{2j}\big)(-\phi_e/2)^{-2j}, \medskip \\
p_{2j+1} & \mapsto & 2\mathrm iAB\big(s_{2j+1}+t_{2j+1} + \varepsilon_e (s_{2j+1}-t_{2j+1})\big)(-\phi_e/2)^{-2j-1}, \medskip \\
q_{2j+1} & \mapsto & -2AB \big(\varepsilon_e (s_{2j+1}-t_{2j+1}) + s_{2j+1}+t_{2j+1}\big)(-\phi_e/2)^{-2j-1},
\end{array}
\right.
\]
on $\partial U_e$. Clearly, $\mathcal C_{\boldsymbol \Delta_l}(\boldsymbol I)=\mathcal C_-(\boldsymbol\Delta_l)$ also has the same form as the right-hand side of \eqref{ae15}. Another boring computation shows that the product $\mathcal C_{\boldsymbol \Delta_l}(\boldsymbol I)\boldsymbol\Delta_l$ has the same form as well. By induction, we get that all the summands under the integral sign in \eqref{ae16} have this form, from which \eqref{ae15} clearly follows.

\subsection{String Equations}

To prove Theorem~\ref{RecCoef1}, we need to introduce {\it discrete string equations} (see, e.g., \cite{BL}):
\[
\left\{\;
\begin{aligned}
{}&\ga_n[V'(\boldsymbol Q)]_{n,n-1}=\frac{n}{N}\,,\\
{}&[V'(\boldsymbol Q)]_{nn}=0,
\end{aligned}
\right.
\]
where $[\boldsymbol A]_{nm}$ is the $(n,m)$-th element of the matrix $\boldsymbol A$,
\[
\boldsymbol Q=\begin{pmatrix}
\be_0 & \ga_1 & 0 & 0 &\dots \\
\ga_1 &\be_1 & \ga_2 & 0 & \dots \\ 
0 & \ga_2 & \be_2 & \ga_3 & \dots \\
0 & 0 & \ga_3 & \be_3 &  \dots \\
\vdots & \vdots & \vdots & \vdots & \ddots
\end{pmatrix},
\]
and $\ga_n$ and $\be_n$ are recurrence coefficients \eqref{cm4} for polynomials satisfying orthogonality relations \eqref{cm3} with respect to a potential $V$. For the potential $V$ as in \eqref{cm2}, the discrete string equations become
\[
\left\{\;
\begin{aligned}
{}&\ga_n\big[-\boldsymbol Q^2+t\boldsymbol I\big]_{nn-1}=\frac{n}{N}\,,\\
{}&\big[-\boldsymbol Q^2+t\boldsymbol I\big]_{nn}=0,
\end{aligned}
\right.
\]
where, this time, $\boldsymbol I$ is the semi-infinite identity matrix. This gives the equations
\begin{equation}
\label{ae17}
\left\{\;
\begin{aligned}
{}&\ga_n^2(\be_{n-1}+\be_n)=-\frac{n}{N}\,,\\
{}&\ga_{n+1}^2+\ga_n^2+\be_n^2=t.
\end{aligned}
\right.
\end{equation}

To prove \eqref{oc16} using \eqref{ae17}, let us set
\[
v:=\left(\frac {n+1/2}N\right)^{-2/3} \quad \text{and} \quad u_*:=\left(\frac nN\right)^{-2/3}=v\left(1-\frac{v^{3/2}}{2N}\right)^{-2/3}.
\]
Then for all $N$ large, the value $\widehat B_k(t,u_*)$ can be computed as a series
\[
\widehat B_k(t,u_*) = \sum_{i=0}^\infty \frac{\partial^i B_k}{\partial u^i}(t,v)v^i\left(\left(1-\frac{v^{3/2}}{2N}\right)^{-2/3}-1\right)^i.
\]
Using the Taylor expansion of $(1-x)^{-2/3}$ at the origin, we can rewrite the above expression as a series in powers of $N$ with coefficients that are holomorphic in $\mathcal N$ functions. Thus, \eqref{oc15} can be equivalently written as
\[
\beta_n(t,N) \sim \sum_{k=0}^\infty \widetilde B_k(t,v)N^{-k},
\]
where the functions $\widetilde B_k(t,v)$ are holomorphic in $\mathcal N$. Hence, to prove \eqref{oc16} we need to show that $\widetilde B_{2j-1}(t,v)\equiv 0$ in $\mathcal N$.  To this end, let us represent each $\widetilde B_k(t,v)$ as a series in powers of $N^{-1}$:
\[
\widetilde B_k(t,v) = \sum_{j=0}^\infty \widetilde B_{k,j}(t,u_*)N^{-j}, \quad v=\left(\frac {n+1/2}N\right)^{-2/3}=u_*\left(1+\frac{u_*^{3/2}}{2N}\right)^{-2/3}.
\]
Notice also that
\begin{equation}
\label{ae18}
\widetilde B_{k,0}(t,u)=\widetilde B_k(t,u) \quad \text{and} \quad \widetilde B_{0,0}(t,u)=x(tu)/\sqrt u.
\end{equation}
Then it also holds that
\[
\widetilde B_k(t,w) = \sum_{j=0}^\infty \widetilde B_{k,j}(t,u_*)(-N)^{-j}, \quad w=\left(\frac {n-1/2}N\right)^{-2/3}=u_*\left(1-\frac{u_*^{3/2}}{2N}\right)^{-2/3}.
\]
Hence, we get that
\begin{equation}
\label{ae19}
\beta_n(t,N) + \beta_{n-1}(t,N) \sim 2\sum_{k=0}^\infty\left(\sum_{j=0}^\infty \widetilde B_{k,2j}(t,u_*)N^{-2j} \right) N^{-k},
\end{equation}
with the expansion valid locally uniformly in both variables. Then the constant term in the expansion of $\ga_n^2(\be_{n-1}+\be_n)$ is equal to 
\[
- \frac{2\widetilde B_{0,0}(t,u_*)}{2u_*x(u_*t)} = - u_*^{-3/2} = -\frac nN
\]
by \eqref{oc15} and \eqref{ae18}. Thus, it follows from the first relation in \eqref{ae17} that the rest of the terms in the expansion of $\ga_n^2(\be_{n-1}+\be_n)$ must be equal to zero. The $N^{-1}$-term is given by
\begin{equation}
\label{ae20}
-\frac{\widetilde B_{2l+1,0}(t,u_*)}{u_*x(u_*t)} = -\frac{\widetilde B_{2l+1}(t,u_*)}{u_*x(u_*t)}, \quad l=0,
\end{equation}
by \eqref{oc15}, \eqref{ae19}, and \eqref{ae18}. This implies that $\widetilde B_1(t,u_*)=0$. As we can vary $n$ and $N$, the last equality holds on a set with a limit point in $\mathcal N_t$. Hence, $\widetilde B_1(t,\cdot)\equiv0$ by holomorphy in the second variable.  Assuming that $\widetilde B_{2l-1}(t,\cdot)\equiv 0$ for all $l\leq L$, we get from \eqref{oc15}, \eqref{ae19}, and \eqref{ae18} that the $N^{-(2L+1)}$-term in the expansion of $\ga_n^2(\be_{n-1}+\be_n)$ is given by \eqref{ae20} with $l=L$. Previous argument yields that $\widetilde B_{2L+1}(t,\cdot)\equiv 0$ and the desired claim now follows from the principle of mathematical induction.

\subsection{Free energy}

In \cite[Proposition~5.1]{BD1}, it was shown that the free energy $F_N(t)$ and the recurrence coefficient $\gamma_N(t,N)$ satisfy Toda equation \eqref{cm6} for all real $t>t_\mathsf{cr}$. It was further shown in \cite{BD1} that
\begin{equation}
\label{ae21}
F_N(t) = \frac23 t^{3/2} - \frac14\log(4t) + \int_\infty^t\int_\infty^\tau\left(\ga_N^2(\sigma,N) - \frac 1{2\sqrt\sigma}-\frac1{4\sigma^2}\right)\mathrm d\sigma\mathrm d\tau,
\end{equation}
where the integrals are taken along positive reals. It was also proved in \cite{BD1} that an asymptotic expansion for $F_N(t)$ can be obtained by simply plugging the asymptotic expansion for $\ga_N^2(t,N)$ into \eqref{ae21} and integrating term by term; that is, \eqref{oc1} is valid uniformly on closed subsets of $(t_\mathsf{cr},\infty)$, where the functions $F^{(2k)}(t)$ can be computed via the following equations:
\begin{equation}
\label{ae23}
\left\{
\begin{array}{lll}
F^{(0)}(t) & = & \displaystyle \frac23 t^{3/2} - \frac14\log(4t) + \int_\infty^t\int_\infty^\tau\left(-\frac1{2x(\sigma)} - \frac 1{2\sqrt\sigma}-\frac1{4\sigma^2}\right)\mathrm d\sigma\mathrm d\tau, \medskip \\
F^{(2k)}(t) & = & \displaystyle  \int_\infty^t\int_\infty^\tau G_{2k}(\sigma;0) \mathrm d\sigma\mathrm d\tau, \quad k\geq1,
\end{array}
\right.
\end{equation}
(the integrals in \eqref{ae23} are well defined as it was shown that $G_{2k}(t;0)=\mathcal O\big(t^{-7/2}\big)$ and $x(t) = -\sqrt t + \frac1{2t} + \mathcal O\big(t^{-5/2}\big)$ uniformly as $O_\mathsf{one-cut}\ni t\to\infty$).

Since $Z_N(t)$ is an entire function of $t$, the free energy $F_N(t)$ is a meromorphic function of the parameter $t$. Hence, Toda equation \eqref{cm6} extends to the entire parameter plane. Recall that $\ga_N^2(t,N)$ are holomorphic functions of the parameter $t$ on each closed subsets of $O_\mathsf{one-cut}$ satisfying \eqref{condT} for all $N$ large enough (depending on the subset). Hence, using \eqref{ts4} and some algebra, we can rewrite \eqref{ae21} as
\[
F_N(t)  =  1 - \frac23 x^3(t) - \frac12\log\big(-2x(t)\big) + \int_\infty^t\int_\infty^\tau\left(\ga_N^2(\sigma,N) + \frac{7x^\prime(\sigma) + 2 \sigma x^{\prime\prime}(\sigma)}6 \right)\mathrm d\sigma\mathrm d\tau,
\]
where $x^\prime(t)$ is the derivative of $x(t)$ with respect to $t$. Hence, for any closed subset $T\subset O_\mathsf{one-cut}$ satisfying \eqref{condT} there exists a constant $N(T)$ such that the functions $F_N(t)$ are holomorphic on $T$. Similarly, we see that the functions $F^{(2k)}(t)$ are in fact holomorphic in $O_\mathsf{one-cut}$ and can be holomorphically extended across $C_\mathsf{split}$, $C_\mathsf{birth}^a$, and $C_\mathsf{birth}^b$. This, in particular, gives formula \eqref{oc1a}. Thus,
\[
F_N(t) - \sum_{k=0}^{K-1} F^{(2k)}(t)N^{-2k} =  \int_\infty^t\int_\infty^\tau\left(\ga_N^2(\sigma,N) - \sum_{k=0}^{K-1} G_{2k}(t;0)N^{-2k}\right)\mathrm d\sigma\mathrm d\tau = \mathcal O\big(N^{-2K}\big)
\]
uniformly on closed subsets of $O_\mathsf{one-cut}$ satisfying \eqref{condT}, which implies that asymptotic expansion \eqref{oc1} does indeed hold in $O_\mathsf{one-cut}$ as claimed.


\begin{thebibliography}{999}

\bibitem{AMAM1} G. \'Alvarez, L. Mart\'inez Alonso, and E. Medina. 
Determination of $S$-curves with applications to the theory of non--hermitian orthogonal polynomials. 
{\it J. Stat. Mech. (2013) P06006.}

\bibitem{AMAM2} G. \'Alvarez, L. Mart\'inez Alonso, and E. Medina. 
Phase structure and asymptotic zero densities of orthogonal polynomials in the cubic model.
{\it J. Comput. Appl. Math}, {\bf 284}, (2015), 10--25.

\bibitem{BE} M. Berg\`{e}re and B. Eynard. Universal scaling limits and matrix models, and $(p,q)$ Liouville gravity. {\it arXiv:0909.0854v1 [math-ph]}.

\bibitem{BIZ}
D. Bessis, X. Itzykson, and J. B. Zuber. Quantum Field Theory Techniques in Graphical Enumeration. {\it Adv. Appl. Math.} {\bf 1}
(1980), 109-157.

\bibitem{BIPZ}
E. Br\'{e}zin, C. Itzykson, G. Parisi, and J.-B. Zuber. Planar diagrams. {\it Commun. Math. Phys.} {\bf  59} (1978), 35--51.

\bibitem{BD1}
P. M. Bleher and A. Dea\~no.
Topological expansion in the cubic random matrix model. 
{\it Int. Math. Res. Not. {\bf 12} (2013), 2699--2755.}

\bibitem{BD2} P. M. Bleher and A. Dea\~no.
Painlev\'{e} I double scaling limit in the cubic matrix model.
arXiv:1310.3768.

\bibitem{BI} P. M. Bleher and A. R. Its.
Asymptotic of the partition function of a random matrix model.
{\it Annales de l'Institute Fourier, {\bf 55} no. 6 (2005), 1943--2000.}

\bibitem{BL} P. Bleher and K. Liechty.
Random Matrices and the Six-Vertex Model.
CRM Monograph Series, AMS, 2013.

\bibitem{Bout} P. Boutroux. 
Recherches sur les transcendants de M. Painlev\'e et l'\'etude asymptotique des \'equations diff\'erentielles du second ordre. 
{\it Ann. Sci. de l'\'E.N.S. {\bf 30} (1913), no. 3, 255--375.}

\bibitem{CHT} O. Costin, M. Huang, and S. Tanveer.
Proof of the Dubrovin conjecture and analysis of the tritronqu\'ee solutions of $P_I$. 
{\it Duke Math. J. {\bf 163} (2014), no. 4, 665–704.}


\bibitem{Dav1}
F. David. Phases of the large $N$ matrix model and non--perturbative effects in $2D$ gravity. {\it Nucl. Phys. B {\bf 348} (1991), 507--524.}

\bibitem{Dav2}
F. David. Non--perturbative effects in matrix models and vacua of two dimensional gravity {\it Phys. Lett. B {\bf 302} (1993), 403--410.}

\bibitem{Deift} P. Deift. 
Orthogonal Polynomials and Random Matrices: A Riemann--Hilbert Approach. 
American Mathematical Society, 2000.

\bibitem{DKMVZ} P. Deift, T. Kriecherbauer, K. T.--R. McLaughlin, S. Venakides and X. Zhou.
Uniform asymptotics for polynomials orthogonal with respect to varying exponential weights and applications to universality questions in random matrix theory. 
{\it Comm. Pure Appl. Math. {\bf 52}  (1999), 1335--1425.}

\bibitem{DKMVZ2} P. Deift, T. Kriecherbauer, K. T.--R. McLaughlin, S. Venakides and X. Zhou.
Strong asymptotics of orthogonal polynomials with respect to exponential weights. 
{\it Comm. Pure Appl. Math. {\bf 52}  (1999), 1491--1552.}

\bibitem{DiFrancesco} P. Di Francesco. Matrix model combinatorics: Applications to folding 
and coloring. {\it Random Matrices and Their Applications (P. M. Bleher and A. R. Its, eds.). 
Mathematical Sciences Research Institute Publications, Vol. 40, 2001.}

\bibitem{DLMF}
NIST Digital Library of Mathematical Functions. 
http://dlmf.nist.gov/, Release 1.0.6 of 2013-05-06. 
Online companion to \cite{OLBC10}.

\bibitem{DZ} P. Deift and X. Zhou.
A steepest descent method for oscillatory Riemann--Hilbert problems. Asymptotics for the MKdV equation. 
{\it Ann. Math. {\bf 137}  (1993), 295--368.}

\bibitem{DK} M. Duits and A.B.J. Kuijlaars.
Painlev\'e I asymptotics for orthogonal polynomials with respect to a varying quartic weight. 
{\it Nonlinearity {\bf 19} (2006), 2211--2245.}

\bibitem{Eyn} B. Eynard. Counting Surfaces,  CRM Aisenstadt Chair lectures, Progress in Mathematical Physics {\bf 70},
Birkh\"a user, Basel, 2016. 

\bibitem{EMcL}
N.M. Ercolani and K.T-R. McLaughlin. Asymptotics of the partition function for random matrices via Riemann--Hilbert techniques, and applications to graphical enumeration. {\it Int. Math. Res. Not. 14 (2003), 755--820.}

\bibitem{EMcLP}
N. M. Ercolani, K. T.--R. McLaughlin, and V. U. Pierce. Random matrices, graphical enumeration and the continuum limit of the Toda lattices, {\it Comm. Math. Phys. {\bf 278} (2008) 31--81.}

\bibitem{FIK}
A.S. Fokas, A.R. Its, and A.V. Kitaev. 
The isomonodromy approach to matrix models in 2D quantum gravity. 
{\it Comm. Math. Phys. {\bf 147} (1992), 395--430.}

\bibitem{Forrester} P. Forrester. 
\newblock {\em Log-Gases and Random Matrices}, volume 34 of {\em The London Mathematical Society 
Monographs Series}.
\newblock Princeton University Press, 2010.

\bibitem{GRakh87} A.A. Gonchar and E.A. Rakhmanov.
Equilibrium distributions and the degree of rational approximation of  analytic functions.
{\it Mat. Sb. {\bf 134(176)} (1987), no. 3, 306--352, 447}
English transl. in {\it Math. USSR-Sb. {\bf 62} (1989), no 2, 305--348.}

\bibitem{HKL} D. Huybrechs, A. Kuijlaars, and N. Lejon.
Zero distribution of complex orthogonal polynomials with respect to exponential weights. 
{\it J. Approx. Theory,  {\bf 184} (2014), 28--54.}

\bibitem{Jenkins} J. Jenkins. 
Univalent functions and conformal maps. 
Springer, 1965

\bibitem{Kap} A. A. Kapaev. 
Quasi-linear {S}tokes phenomenon for the {P}ainlev\'e first equation.
{\it J. Phys. A {\bf 37} (2004), 11149--11167.} 

\bibitem{KS} A. Kuijlaars and G. Silva. 
S-curves in Polynomial External Fields.
{\it J. Approx. Theory, {\bf 191} (2015), 1--37.}

\bibitem{Mulase} M. Mulase. 
Lectures on the asymptotic expansion of a Hermitian matrix integral.
{\it Lecture Notes in Physics 502, 91--134 (H. Aratyn et al., eds). Springer Verlag, 1998.}

\bibitem{OLBC10} F. W. J. Olver, D. W. Lozier, R. F. Boisvert, and C. W. Clark, editors. 
NIST Handbook of Mathematical Functions. 
Cambridge University Press, New York, NY, 2010. 
Print companion to \cite{DLMF}.

\bibitem{Pommerenke} Ch. Pommerenke. 
Univalent Functions, 
Vandenhoeck \& Ruprecht, G\"ottingen, 1975.

\bibitem{SaffTotik}
E.B. Saff and V.~Totik.
\newblock {\em Logarithmic Potentials with External Fields}, volume 316 of {\em
  Grundlehren der Math. Wissenschaften}.
\newblock Springer-Verlag, Berlin, 1997.

\bibitem{St85} H.~Stahl.
Extremal domains associated with an analytic function. {I, II}.
{\it Complex Variables Theory Appl. {\bf 4} (1985), 311--324, 325--338.}

\bibitem{St85b} H.~Stahl.
Structure of extremal domains associated with an analytic function.
{\it Complex Variables Theory Appl. {\bf 4} (1985), 339--356.}

\bibitem{St86} H.~Stahl.
Orthogonal polynomials with complex valued weight function. {I, II}.
{\it Constr. Approx. {\bf 2} (1986), no. 3,225--240, 241--251.}

\bibitem{Strebel}  K. Strebel. 
Quadratic differentials. 
Springer, 1984.

\bibitem{Zvonkin} A. Zvonkin. Matrix Integrals and Map Enumeration:
An Accessible Introduction. {\it Mathl. Comput. Modelling {\bf 26, 8-10} (1997), 281--304.}



\end{thebibliography}
\end{document}